\documentclass{aa}
\usepackage{graphicx}
\usepackage{amsmath}
\usepackage[utf8]{inputenc}

\usepackage{multirow}
\usepackage{minipage}

\usepackage{geometry}
\usepackage{pdflscape}

\pdfoutput=1

\newcommand\kms{km$\,$s$^{-1}$}
\newcommand\Msol{M$_{\odot}$}
\newcommand\Lsol{L$_{\odot}$}

\DeclareMathOperator\erf{erf}

\title{The AMIGA sample of isolated galaxies}
\subtitle{XIII. The HI content of an almost ``nurture free'' sample}
\titlerunning{The HI properties of the AMIGA sample}
\author{M.~G.~Jones\inst{\ref{inst1}}\thanks{E-mail: mjones@iaa.es} \and D.~Espada\inst{\ref{inst2},\ref{inst10}} \and L.~Verdes-Montenegro\inst{\ref{inst1}} \and  W.~K.~Huchtmeier\inst{\ref{inst3}} \and U.~Lisenfeld\inst{\ref{inst4},\ref{inst5}} \and S.~Leon\inst{\ref{inst6}} \and J.~Sulentic\inst{\ref{inst1}} \and J.~Sabater\inst{\ref{inst8}} \and D.~E.~Jones\inst{\ref{inst9},\ref{inst11}} \and S.~Sanchez\inst{\ref{inst1}} \and J.~Garrido\inst{\ref{inst1}} }
\authorrunning{M.~G.~Jones et al.}

\date{\today}

\institute{
Instituto de Astrof\'{i}sica de Andaluc\'{i}a (CSIC), Apdo. 3004, 18008 Granada, Spain\label{inst1}
\and
National Astronomical Observatory of Japan (NAOJ), 2-21-1 Osawa, Mitaka, Tokyo 181-8588, Japan\label{inst2}
\and
The Graduate University for Advanced Studies (SOKENDAI), 2-21-1 Osawa, Mitaka, Tokyo, 181-0015, Japan\label{inst10}
\and
Max-Planck-Institut fuer Radioastronomie, Postfach 2024, D-53010 Bonn, Germany\label{inst3}
\and
Departamento de F\'{i}sica Te\'{o}rica y del Cosmos, Universidad de Granada, 18071 Granada, Spain\label{inst4}
\and
Instituto Carlos I  de F\'{i}sica Te\'{o}rica y Computacional, Universidad de Granada, Granada, Spain\label{inst5}
\and
Joint ALMA Observatory – ESO, Av. Alonso de Córdova, 3104 Santiago, Chile\label{inst6}
\and
Institute for Astronomy, University of Edinburgh, EH9 3HJ, Edinburgh, UK\label{inst8}
\and
SAMSI, 19 T.W. Alexander Drive, P.O. Box 110207, Research Triangle Park, NC 27709, USA\label{inst9}
\and
Department of Statistical Sciences, Duke University, P.O. Box 90251, Durham, NC 27708, USA\label{inst11}
}

\abstract{
    We present the largest catalogue of HI single dish observations of isolated galaxies to date, as part of the multi-wavelength compilation being performed by the AMIGA project (Analysis of the interstellar Medium in Isolated GAlaxies). Despite numerous studies of the HI content of galaxies, no revision focused on the HI scaling relations of the most isolated $L_{*}$ galaxies has been made since \citet{Haynes_1984}. 
}{
    The AMIGA sample has been demonstrated to be almost ``nurture free'', therefore, by creating scaling relations for the HI content of these galaxies we will define a metric of HI normalcy in the absence of interactions. 
}{
    The catalogue comprises of our own HI observations with Arecibo, Effelsberg, Nan\c{c}ay and GBT, and spectra collected from the literature. In total we have measurements or constraints on the HI masses of 844 galaxies from the Catalogue of Isolated Galaxies (CIG). The multi-wavelength AMIGA dataset includes a revision of the B-band luminosities ($L_{\mathrm{B}}$), optical diameters ($D_{25}$), morphologies, and isolation. Due to the large size of the catalogue, these revisions permit cuts to be made to ensure isolation and a high level of completeness, which was not previously possible. With this refined dataset we fit HI scaling relations based on luminosity, optical diameter and morphology. Our regression model incorporates all the data, including upper limits, and accounts for uncertainties in both variables, as well as distance uncertainties.
}{
    The scaling relation of HI mass with $D_{25}$ is in good agreement with that of \citet{Haynes_1984}, but our relation with $L_{\mathrm{B}}$ is considerably steeper. This disagreement is attributed to the large uncertainties in the luminosities, which introduce a bias when fitting with ordinary least squares regression (as was done in previous works), and the different morphology distributions of the samples. We find that the main effect of morphology on the $D_{25}$-relation is to increase the intercept towards later types, while for the $L_{\mathrm{B}}$-relation it is to flatten the slope. These trends were not evident in previous works due to the small number of detected early-type galaxies. Applying our relations to HI detected galaxies in the Virgo cluster we find that although the typical HI-deficiency is only $\sim$0.3 dex, the tail of the distribution extends over an order of magnitude beyond that of the AMIGA sample. These results are in general agreement with previous studies of HI-deficiency in the Virgo cluster.
}{
    The HI scaling relations of the AMIGA sample define an up-to-date metric of the HI content of almost ``nurture free'' galaxies. These relations allow the expected HI mass, in the absence of interactions, of an individual galaxy to be predicted to within 0.25 dex (for typical measurement uncertainties). These relations are thus suitable for use as statistical measures of the impact of interactions on the neutral gas content of galaxies. 
}

\begin{document}

\maketitle

\section{Introduction}

\defcitealias{Haynes_1984}{HG84}

Galaxies in and around high density environments such as clusters and compact groups undergo an array of environmental processes that impact their morphological type, gas content, and star formation rate. The effects of tidal forces and ram pressure stripping are ubiquitous in clusters \citep[e.g.][]{Kenney_2004,Lucero_2005,Chung_2009,Abramson_2011}, and the impact of the former is detectable even for galaxy pairs by the elevation of their star formation rates \citep[e.g.][]{Patton_2013}, or in extreme cases by stellar or gaseous tidal tails.

HI is one of the most sensitive components of the ISM (interstellar medium) to environmental effects as it typically extends approximately twice as far as the stellar disc. HI-rich galaxies with close neighbours are frequently seen to have HI tails and bridges extending well beyond any detectable stellar component, in some cases $\sim$500 kpc long \citep{Serra_2013,Leisman_2016,Hess_2017}. In clusters ram pressure stripping depletes galaxies of much of their HI reservoir leaving the majority of them HI-deficient. Furthermore, the rate of gas-poor spirals increases towards the centre of clusters \citep{Haynes_1984b,Solanes_2001,Lah_2009,Chung_2009}, and perhaps even in groups \citep{Hess_2013,Odekon_2016,Brown_2017}. One of the most extreme examples of environment is compact groups, small groups (4-10 members) with number densities comparable to cluster cores \citep{Hickson_1982,Hickson_1992}. These groups are found to be HI-deficient, and while they show evidence of highly effective stripping events \citep{Verdes-Montenegro_2001,Rasmussen_2008,Borthakur_2015,Walker_2016}, their formation and evolution are not yet understood, in particular the fate of atomic gas. This highlights the need for up-to-date benchmark of the HI content of undisturbed galaxies to act as a fair reference with which to compare.

In order to understand the impact of environmental effects on a galaxy's HI content in a statistical sense, rather than on a system by system basis, a predictor of the expected HI content for a given galaxy is required to act as a baseline. This predictor must be calibrated by HI observations of galaxies with as minimal impact from interactions and environmental effects as possible to ensure that the baseline represents the HI content of unperturbed systems. The AMIGA project \citep[Analysis of the interstellar Medium of Isolated GAlaxies,][]{Verdes_Montenegro_2005} is an in depth study of isolated galaxies from a starting sample of 1050 CIG galaxies \citep[Catalogue of Isolated Galaxies,][]{Karachentseva_1973}. AMIGA was initially focused on studying the ISM in isolated galaxies, but as well as collecting a rich, multi-wavelength dataset has made numerous refinements to the quantification of the isolation and environment of these galaxies and their properties in the radio, infrared, and optical. These quantifications have demonstrated AMIGA to be an almost ``nurture free'' sample with galaxies that have been isolated for 3 Gyr on average \citep{Verdes_Montenegro_2005}, and have properties that are distinct even from those of field galaxies. Thus, AMIGA constitutes an ideal sample for calibration of a predictor of HI content.

While interferometric 21 cm observations can provide spatially resolved maps of the HI emission of a galaxy, they generally have poorer surface brightness sensitivity than single dish observations and can introduce bias due to scale dependent attenuation of features. Therefore, as the global properties of a system's HI, including its mass and basic kinematics, can be found from its 21 cm spectral profile alone, and because single dish spectra are both more plentiful in the literature and require shorter observations, they represent the best way to measure the total HI content in this case.

As the optical properties of galaxies are thought to be less impacted, or at least impacted on a longer timescale, than HI properties, the optical luminosity and optical diameter are typically used as proxies for the HI mass. \citet{Haynes_1984} (hereafter HG84) performed the seminal study of the HI properties of isolated galaxies using 324 Arecibo spectra of CIG galaxies to calibrate their predictors of HI mass. These scaling relations are still widely used today to measure the quantity ``HI-deficiency'':
\begin{equation}
\label{eqn:HI_def}
\mathrm{DEF} = \log{M_{\mathrm{HI}}^{\mathrm{exp}}/\mathrm{M}_{\odot}} - \log{M_{\mathrm{HI}}^{\mathrm{obs}}/\mathrm{M}_{\odot}},
\end{equation}
where $M_{\mathrm{HI}}^{\mathrm{exp}}$ is the expected HI mass based on a predictor, and $M_{\mathrm{HI}}^{\mathrm{obs}}$ is the observed HI mass. This definition of HI-deficiency means that galaxies with positive $\mathrm{DEF}$ are poor in HI relative to what is expected.

\citet{Solanes_1996} extended the work of \citetalias{Haynes_1984} by assessing the correlation between galaxy size and HI mass for 532 field galaxies in the Pisces-Perseus region. As that region contains a chain of clusters the Sa-Sc spirals in their sample were selected to have low projected neighbour densities to ensure they were not cluster members, however, almost none of these galaxies would be considered isolated by the AMIGA criteria (see appendix \ref{Solanes_iso_comp}). Hence, it is important to note that here `field' and `isolated' are two quantitatively separate categories. \citet{Solanes_2001} then used the predictor calibrated in \citet{Solanes_1996} to measure the HI-deficiency of galaxies in 18 nearby clusters, and mapped the HI-deficiency across the Virgo region.

More recently \citet{Toribio_2011a,Toribio_2011} used ALFALFA \citep[Arecibo Legacy Fast ALFA survey,][]{Giovanelli_2005,Haynes_2011} to perform a principal component analysis of the HI and optical properties of 1624 field galaxies in low density environments (selected with weaker criteria than AMIGA's) within the ALFALFA footprint in the direction of Virgo. Unlike the previous works (and this paper) the ALFALFA survey provides a blind HI-selected, rather than optically-selected, sample which means that the relations calculated by \citet{Toribio_2011} are optimal for the average HI-rich galaxy, however, this excludes parts of the population such as isolated early-type galaxies that are HI-poor and thus not detectable by ALFALFA.

\citet{Denes_2014} used HIPASS \citep[HI Parkes All Sky Survey,][]{Barnes_2001,Meyer_2004} and a compilation of optical and infrared properties to construct scaling relations of HI-selected galaxies. Their scaling relations were constructed from the HIPASS galaxies, excluding the highest 30\% in neighbour density (out to the 7th optically-selected neighbouring galaxy). This sample contains many more galaxies than the previous samples, but this is a direct consequence of weaker isolation criteria. With these relations it was confirmed that HI-deficiency is seen to correlate with the densest environments.

Finally, \citet{Bradford_2015} used a combination of ALFALFA data and their own HI observations to fit scaling relations between stellar masses (estimates from the NASA Sloan Atlas) and HI masses of isolated galaxies. This work focused on low-mass galaxies (mostly below the mass range covered by AMIGA) and therefore chose to define isolation as a minimum separation of 1.5 Mpc from a massive (potential) host galaxy. This definition was expanded to include non-dwarf galaxies, allowing the relations to be extended to higher masses. However, the sample suffers from incompleteness at higher masses and defining a consistent metric of isolation for both dwarf and $L_{*}$ galaxies is a challenge.

All of these related works, with the exception of \citetalias{Haynes_1984}, are based on samples with significant numbers of field galaxies, not truly isolated galaxies. Therefore, they do not necessarily represent a galaxy population that has been without interactions for an extended period \citep[the average AMIGA galaxy has been without substantial interaction for 3 Gyr,][]{Verdes_Montenegro_2005}, and thus are not appropriate to act as the baseline for the expected HI content of galaxies in the absence of interactions.

Another growing use for HI scaling relations is in HI spectral line stacking experiments. With the imminent arrival of SKA precursor and pathfinder facilities the redshift range of HI galaxy surveys will be pushed to order unity through the use of stacking. HI scaling relations can be used to estimate the contribution of source confusion to such stacks \citep{Delhaize_2013,Jones_2016,Elson_2016}, and to act as a comparison for the average properties of the stacked galaxies. Although these applications can both be (and likely will be) fulfilled by comparison with simulations, HI scaling relations offer not only an additional method that does not depend on the veracity of simulations, but also a method that can rapidly provide estimates with a minimal investment of computation time.

In this paper we use a collection of 844 spectra of CIG galaxies, both from the literature and AMIGA's own observations, to measure a new baseline for the HI content of highly isolated galaxies. This measure has not been updated (for the most isolated galaxies) since \citetalias{Haynes_1984}. Our larger sample of isolated galaxies with HI observations, combined with the ancillary dataset AMIGA has collected and characterised, allows us to make cuts to ensure both isolation and a high level of completeness, while still retaining a large enough sample to perform a statistical analysis. Furthermore, the regression model used here is more sophisticated than in previous works. It accounts for measurement uncertainties in all quantities (including the source distances), and incorporates the information contained in the upper limits. The retention of upper limits also means that our science sample covers the range of morphologies in a much more representative manner than \citetalias{Haynes_1984}, as early types tend to be undetected in HI. The new baseline of HI content of the most isolated galaxies that we calculate here, will allow studies of the atomic gas in galaxies in terms of ``nature versus nurture'', and for very gas-deficient systems will provide an up-to-date estimate of how much has been lost.

The paper is arranged as follows: in the next section we describe the AMIGA sample and the compiled optical properties, section \ref{sec:HI_data} details our HI observations and the HI data compiled from the literature, in section \ref{sec:data_red} we describe how that data was uniformly reduced, section \ref{sec:analysis} presents our regression model and the results of our analysis, and in section \ref{sec:discuss} we discuss these results before summarising in section \ref{sec:summary}.

\section{Sample}

The AMIGA \citep{Verdes_Montenegro_2005} sample is drawn from the CIG \citep{Karachentseva_1973}, which includes 1051 isolated galaxies \citep[although CIG 781 has since been shown to be a globular cluster,][]{Leon_2003}. AMIGA is an ongoing project to study the ISM of these galaxies and has observed and compiled a multi-wavelength database covering the optical, H$\alpha$, NIR, FIR, radio continuum, as well as HI and CO lines. AMIGA has made substantial contributions to updating and qualifying this catalogue of sources. \citet{Leon_2003} used SExtractor and DSS (Digitized Sky Survey) to redefine the source positions of the CIG. Mostly these updated positions agreed within a few arcsec of the original position, but in certain cases there were deviations of over half an arcmin. \citet{Verdes_Montenegro_2005} evaluated the completeness of the AMIGA sample using the $V/V_{\mathrm{max}}$ test, finding it to be 80-95\% complete for objects with B-band magnitudes brighter than 15.0. \citet{Verley_2007b,Verley_2007a} measured the degree of isolation of the galaxies in this sample, estimating both the local number density and the strength of the tidal forces exerted by any neighbours. Criteria for both of these parameters were then chosen with the goal of removing any galaxies from the sample that could have their evolution impacted by the presence of neighbours. The isolation criteria were revised again in \citet{Argudo_Fernandez_2013} based on SDSS DR9 images and spectroscopy. However, because AMIGA is an all sky sample, restricting it to the SDSS footprint excludes much of the collected HI data. Therefore, we choose not to use this most recent revision and show in appendix \ref{sec:isolation_regression} that our results are mostly consistent with those of this more restricted sample.

The AMIGA sample has also been demonstrated to be the sample of galaxies with the lowest levels of all properties that are enhanced by interaction. \citet{Lisenfeld_2007} found that the FIR luminosity of AMIGA galaxies falls over 0.2 dex below that of a random sample of galaxies (selected without constraints on environment), while the ratio of FIR to B-band luminosity is more than 0.1 dex lower, suggesting that the star formation rate (SFR) in an average galaxy is enhanced relative to that of an AMIGA galaxy. \citet{Lisenfeld_2011} observed CO in 173 AMIGA galaxies and found them to be 0.2-0.3 dex poorer in molecular gas than interacting galaxies. The galaxies in the AMIGA sample are also radio-quiet, with most radio emission emanating from mild SF in the disc, and have a very low AGN-fraction as evidenced by the lack of excess ($<1.5\%$ of sources) above the radio continuum-FIR correlation \citep{Leon_2008,Sabater_2008}, although, curiously there is still a non-negligible fraction showing optical nuclear activity \citep{Sabater_2012}.
Finally, \citet{Espada_2011} used a high signal-to-noise and velocity resolution subset of the HI dataset of this paper to show that AMIGA has the lowest level of HI-asymmetry of any galaxy sample. This body of evidence confirms the assertion that AMIGA is an excellent example of a ``nurture free'' galaxy sample which can act as the baseline control sample for studying the properties of non-interacting galaxies.

\subsection{Optical properties}

The optical properties of the sample were mostly taken directly from the AMIGA 2012 data release \citep{Fernandez_Lorenzo_2012} or compiled from HyperLeda\footnote{\url{http://leda.univ-lyon1.fr/}} \citep{Makarov_2014}. Here we briefly describe the parameters used. For a full description consult the referenced articles.

\subsubsection{Optical positions}

The optical positions of the CIG were updated by \citet{Leon_2003} who used DSS images and SExtractor. These new positions are used in this work to make corrections to observations that pointed at slightly incorrect locations.

\subsubsection{Apparent magnitudes}

The B-band magnitudes from the AMIGA 2012 release were compiled from HyperLeda and the standard corrections were applied to give the corrected magnitude as:
\begin{equation}
B_{\mathrm{c}} = B - A_{\mathrm{g}} - A_{\mathrm{i}} - A_{\mathrm{K}},
\end{equation}
where $B$ is the observed B-band magnitude, $A_{\mathrm{g}}$ is the Galactic extinction, $A_{\mathrm{i}}$ in the galaxy's internal extinction, and $A_{\mathrm{K}}$ is the K-correction. $A_{\mathrm{g}}$ was taken directly from HyperLeda, as was $A_{\mathrm{i}}$, except that it used the revised AMIGA morphologies. The K-correction was updated in this work to reflect the latest available heliocentric velocities of the sources (see section \ref{sec:dists}).

\subsubsection{B-band luminosity}

A physical property of the galaxy is required to act as a predictor of the HI mass, therefore, the corrected B-band apparent magnitudes must be converted to a luminosity. The luminosity is calculated in terms of the Sun's bolometric luminosity. We use the Sun's bolometric absolute magnitude, $M_{\mathrm{bol,\odot}} = 4.88$ \citep[as in][]{Lisenfeld_2011,Fernandez_Lorenzo_2012}, and the equation:
\begin{equation}
\label{eqn:luminosity}
\log{L_{\mathrm{B}}\,h_{70}^{2} / L_{\odot}} = 10 + 2\log{D \,h_{70} / \mathrm{Mpc}} + 0.4(M_{\mathrm{bol,\odot}} - B_{\mathrm{c}}),
\end{equation}
where $D$ is the calculated distance to the source.

\subsubsection{Morphologies}

\begin{figure}
\centering
    \includegraphics[width=\columnwidth]{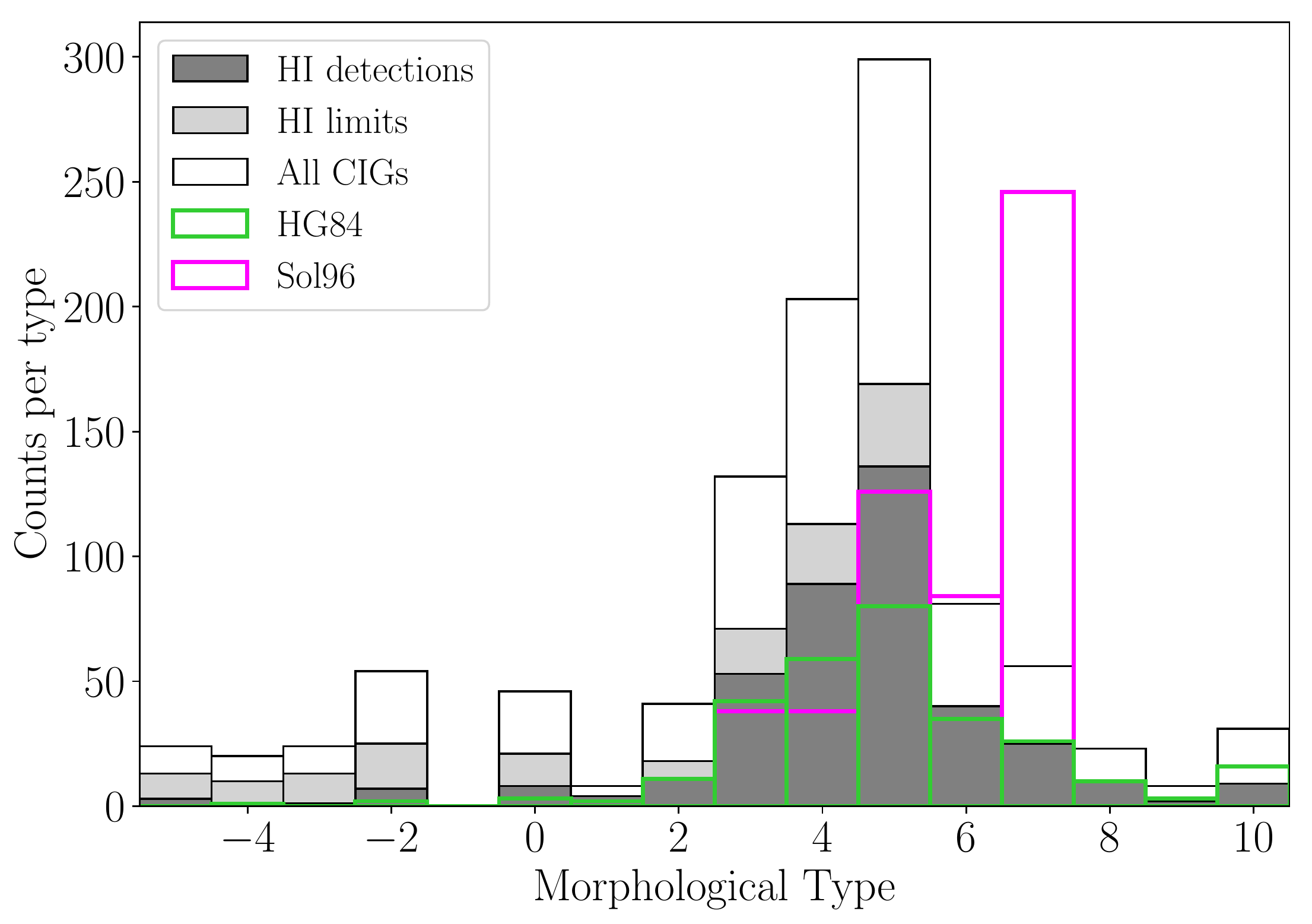}
    \caption{Morphology distributions of galaxy samples used in this work and the works with which we compare. The morphological types used for the AMIGA and \citetalias{Haynes_1984} samples are from the AMIGA database, while those for \citet{Solanes_1996} are taken from the original article. The dark grey bars (detections) combined with the light grey bars (marginals and non-detections) make up the AMIGA HI science sample, which has a median type of 4 (Sbc). For comparison the white bars show the full CIG sample. The pink bars shown the field sample of \citet{Solanes_1996}, which has a median type of 6 (Scd), and the green bars the CIG-based sample of \citetalias{Haynes_1984}, which has a median type of 5 (Sc).}
    \label{fig:morph_dist}
\end{figure}

Morphologies given in \citet{Fernandez_Lorenzo_2012} were used for this work. These morphologies are mostly based on SDSS images or AMIGA's own optical images, for a much smaller number of sources the morphologies are from the original AMIGA revision of morphologies \citep{Sulentic_2006} based on POSS II images \citep[Second Palomar Observatory Sky Survey][]{Reid_1991}, or in cases where no images were available the morphologies were taken from NED (NASA/IPAC Extragalactic Database) or HyperLeda. The numerical scale follows the RC3 system.

The morphology distributions of all the CIG and the AMIGA HI science sample (see section \ref{sec:cuts}) are shown in Figure \ref{fig:morph_dist} along with the other galaxy samples with which we compare results (see section \ref{sec:comp_rels}).

\subsubsection{Optical diameters, axis ratios, and inclinations}
\label{sec:opt_diam_incl}

The major axis optical diameters at 25 mag/arcsec$^{2}$ in B-band ($D_{25}$), axial ratios ($R_{25}$), and inclinations were taken directly from \citet{Fernandez_Lorenzo_2012}. $D_{25}$ and $R_{25}$ were compiled from HyperLeda in that work, whereas inclinations were estimated using AMIGA's morphologies (but otherwise following the HyperLeda methodology).

\subsubsection{Position angles}

As position angles had not been compiled as part of the AMIGA 2012 release they were compiled for this work from HyperLeda. These angles are required for the beam corrections of the Nan\c{c}ay telescope as its beam is non-circular.

\subsection{Distances \& velocities}
\label{sec:dists}

\begin{figure}
    \centering
    \includegraphics[width=\columnwidth]{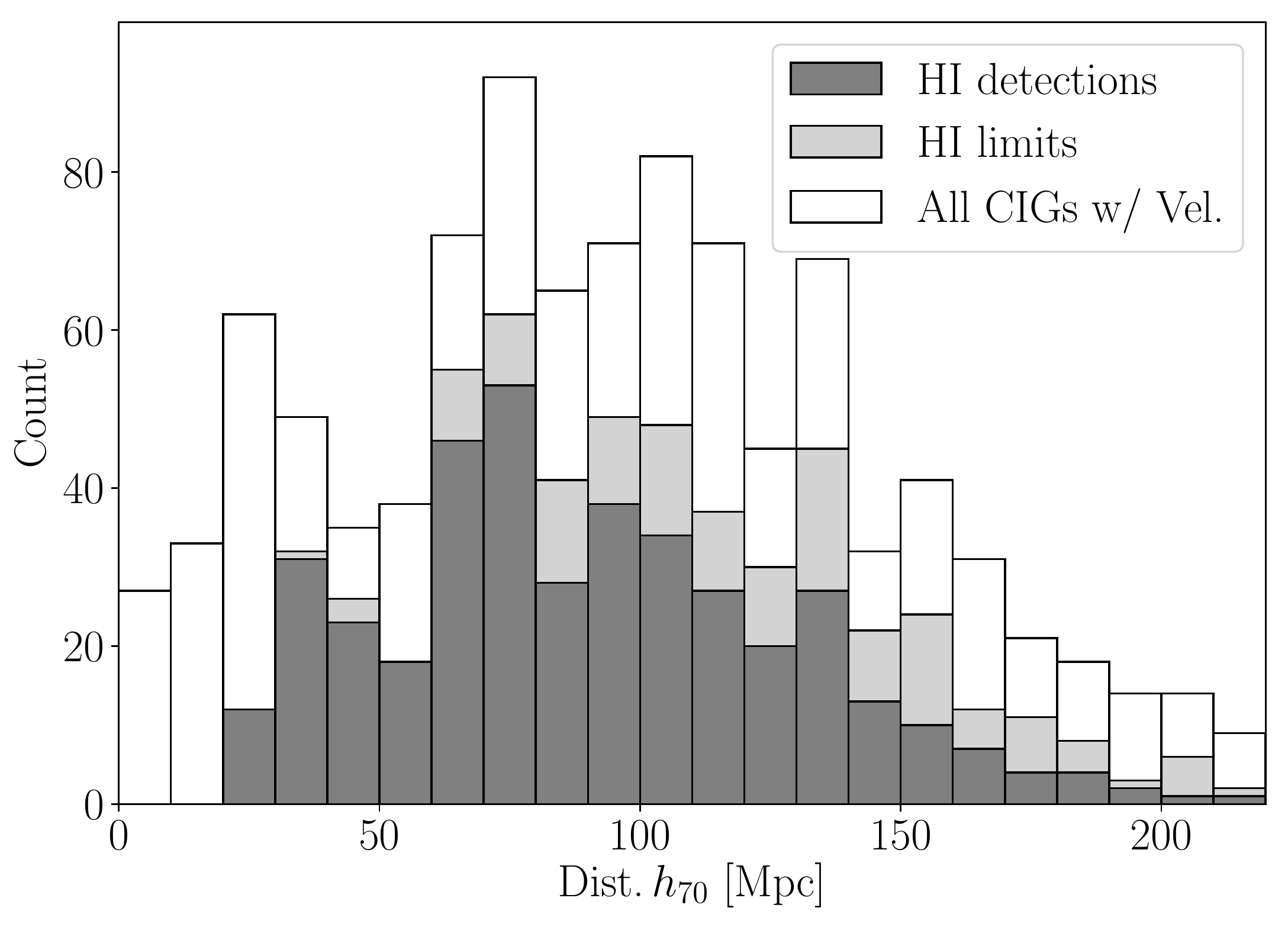}
    \caption{The calculated distances to all CIG galaxies with measurements of recession velocity (white). The subset of the CIG that is the AMIGA HI science sample (defined in section \ref{sec:cuts}) is shown in grey, the dark grey indicating detections and light grey the upper limits. Note that the science sample contains no sources with heliocentric velocities less than 1500 \kms \ due to the isolation requirements, as explained section \ref{sec:cuts}.}
    \label{fig:dist_dist}
\end{figure}

To calculate distances from heliocentric velocities we extended the method of previous AMIGA releases which used Hubble flow and velocities corrected for Local Group motion. We adopt the model of \citet{Mould_2000} which corrects for Local Group motion and then has separate attractor velocity fields for the Virgo cluster, the Shapley supercluster, and the Great Attractor. Each of these attractors is modelled as a spherical overdensity with symmetric infall. For a full description of the model refer to the original reference. The resulting distances are shown in figure \ref{fig:dist_dist}. $H_{0}$ is assumed to be 70 $\mathrm{km\,s^{-1}\,Mpc^{-1}}$ throughout this paper.\footnote{It should be noted that this is an update from previous AMIGA papers, which used $H_{0} = 75 \, \mathrm{km\,s^{-1}\,Mpc^{-1}}$, as both WMAP and Planck results now support a lower value of $H_{0}$ \citep{Hinshaw_2013,Ade_2016}.}

Comparing the distances to sources in common with ALFALFA we find that the Mould-model distances are systematically higher than ALFALFA distances by about 3 Mpc. The scatter between the two methods is also about 3 Mpc, but the deviations are highly correlated with position on the sky, as would be expected because the positions and velocity fields of the attractors are different in the two methods \citep[the ALFALFA flow model is described in][]{Masters_2005}.

Although no sources with heliocentric velocities less than 1500 \kms \ are used in the final regression analysis, the distances to sources with $V_{\mathrm{helio}} < 1000 \; \mathrm{km\,s^{-1}}$ were replaced by literature values from primary and secondary distance indicators \citep[as in][]{Verdes_Montenegro_2005}, with the exception of CIGs 506, 657, 711, 748, and 753 for which no such distance estimates exist. Finally, the errors in the distances were estimated by assuming a normal distribution of galaxy peculiar velocities of width 200 \kms \ and a Gaussian uncertainty in $H_{0}$ of 2 $\mathrm{km\,s^{-1}\,Mpc^{-1}}$. The Mould-model distances were then recalculated 10000 times, with each iteration having a randomly drawn Hubble constant and a random selection of peculiar velocities for all the sources. The calculated distances to each galaxy were fit with a normal distribution and its standard deviation taken as the uncertainty in the distance. The uncertainty for sources with redshift independent distance measurements was assumed to be 10\%, however, these low redshift sources are not used in the final regression sample. It should also be noted that the heliocentric velocity used for the distance determination was not necessarily the systemic HI velocity calculated in this work, but instead the best available velocity in the AMIGA dataset (see appendix \ref{app:data_tab} for more information).

\section{HI data}
\label{sec:HI_data}

The 844 HI spectra compiled in this paper are from both the literature and our own observation, in approximately equal quantities. At the outset of the project, spectra of CIG galaxies were identified in the literature and all the remaining sources were observed where possible. From a starting sample of 1050 targets (the CIG) spectra of a total of 897 were compiled or observed (although not all observations resulted in usable data). In the cases where we used existing observations we required that the spectra were published (or made available to us) rather than just the spectral parameters. This requirement meant that all the spectral parameters of this compilation could be extracted using the same fitting method, allowing a highly uniform HI database of isolated galaxies to be created.

\subsection{HI spectra from the literature}

HI spectra were compiled from the literature using NED and the original articles. In most cases the spectra had been compiled (and digitised where necessary) by NED, however, for a small number (8) of spectra only the published plots were available and we performed the digitisation ourselves.\footnote{The digitisation was performed by hand using WebPlotDigitizer v3.11 (\url{http://arohatgi.info/WebPlotDigitizer/app/}).} A complete list of the 26 original references and the total number of spectra taken from each is shown in Table \ref{tab:refs}.

\begin{table*}
\centering
\caption{Literature spectra used in this compilation}
\label{tab:refs}
\begin{tabular}{l c c c}
\hline\hline
Reference & No. of spectra & Telescope(s)\tablefootmark{$\dagger$} & Reference code\\ \hline
\citet{Springob_2005}          & 238               & AOL, AOG, G91, G43, NRT             & Sp05\\
\citet{Haynes_1984}          & 100               & AOL             & HG84\\
\citet{Meyer_2004}          & 15               & HIP             & Me04\\
KLUN\tablefootmark{$\ddagger$} & 14               & NRT             & KLUN\\
\citet{Tifft_1988}          & 9               & G91             & TC88\\
\citet{Hewitt_1983}          & 6               & AOF             & He83\\
\citet{Courtois_2009}          & 4               & GBT             & Co09\\
\citet{Haynes_1980}          & 3               & AOL             & HG80\\
\citet{Bicay_1986}          & 3               & AOL             & Bi85\\
\citet{Bothun_1985}          & 2               & AOL             & Bo85\\
\citet{Lewis_1985}          & 2               & AOL             & Le85\\
\citet{Lu_1993}          & 2               & AOL             & Lu85\\
\citet{Masters_2014}          & 2               & GBT             & Ma14\\
\citet{Richter_1987}          & 2               & G91             & RH87\\
\citet{Theureau_1998}          & 2               & NRT             & Th98\\
\citet{Balkowski_1981}          & 1               & NRT             & BC81\\
\citet{Haynes_1991}          & 1               & G91             & HG91\\
\citet{Haynes_2011}          & 1               & AOG             & Ha11\\
\citet{Huchtmeier_1995}          & 1               & ERT             & Hu95\\
\citet{Lewis_1983}          & 1               & AOL             & Le93\\
\citet{Mirabel_1988}          & 1               & AOL             & MS88\\
\citet{Rubin_1976}          & 1               & G91             & Ru76\\
\citet{Schneider_1992}          & 1               & G91             & Sc92\\
\citet{Staveley-Smith_1987}          & 1               & JBL             & SD87\\
\citet{Theureau_2005}          & 1               & NRT             & Th05\\
\citet{vanDriel_1995}          & 1               & AOL             & vD95\\ \hline
\end{tabular}
\tablefoot{
\tablefoottext{$\dagger$}{These are the telescopes which we use spectra from, but the original references may also contain observations with other telescope. The telescope codes are described in Table \ref{tab:beam_widths}.}
\tablefoottext{$\ddagger$}{The Kinematics of the Local Universe (KLUN) is a long term project with the data compiled in many papers \citep{Bottinelli_1992,Bottinelli_1993,Theureau_1998,Theureau_2005,Theureau_2007}.}
}
\end{table*}

\subsection{HI observations}

The AMIGA team performed HI observations of 488 CIG galaxies with the Arecibo, Effelsberg (ERT), Green Bank (GBT), and Nan\c{c}ay (NRT) radio telescopes. A full summary of these observations is displayed in Table \ref{tab:obs} and here we outline the observing strategy used at each facility. All targets were observed using a total power switching mode (ON-OFF) at all telescopes and both polarisations were averaged together.

\begin{table*}
\centering
\caption{Summary of AMIGA's HI observations of CIG galaxies}
\label{tab:obs}
\begin{tabular}{c c c c c c}
\hline\hline
Telescope    & Date        & Resolution ($\mathrm{km\,s^{-1}}$) & Bandwidth ($\mathrm{km\,s^{-1}}$) & Boards $\times$ Channels    &  Detection Rate\\ \hline
Arecibo      & 2002        & 0.67,2.66                           & 1400,5550                          & 2$\times$2048 & 70\%              \\ 
Effelsberg   & 2002 - 2004 & 5.24                                & 1200                               & 4$\times$256 & 67\%               \\
Green Bank   & 2002 - 2003 & 1.20,2.50                           & 1200,2500                          & 2$\times$1024 & 94\%              \\
Nan\c{c}ay & 2002 - 2005 & 2.57                                & 10550,2600                         & 2$\times$4096,4$\times$2048 & 30\%\\ \hline
\end{tabular}
\tablefoot{
Description of columns: 1) telescope name, 2) date when observations were conducted, 3) spectral resolution given as the approximate velocity width of a channel, 4) approximate velocity range of the full bandwidth, 5) number of boards and the number of channels on each board, 6) source detection rate. In several cases multiple observing modes were used, these are separated with a comma in the table.
}
\end{table*}

\subsubsection{Arecibo}

A total of 34 CIG galaxies were observed with the Arecibo 305 m telescope using its Gregorian optics system and L-band wide receiver. The autocorrelator was configured either in a high or a low resolution mode, corresponding to a bandwidths of approximately 1400 or 5550 \kms, depending on whether the source was of known or unknown redshift. Total integration times were about 30 minutes per galaxy and the system temperature was approximately 30 K.

\subsubsection{Effelsberg}

Observations of 186 galaxies were performed with the Effelsberg radio telescope. Most of these targets were selected because they fall outside of Arecibo's declination range and therefore generally have declinations above 37$^{\circ}$ or below $-1^{\circ}$. Observations were performed in 10 minute ON-OFF pairs with a total bandwidth of 6.25 MHz across 256 channels, giving a typical channel width of $\sim$5 \kms \ over a range of about 1200 \kms. The system temperature was about 30 K.

\subsubsection{Green Bank}

A total of 51 CIG galaxies were observed with the GBT. Integration times of between 10 and 60 minutes were used for ON-OFF pairs of targets below 10000 \kms. Bandwidths of 5 or 10 MHz were used depending on the expected emission strength and width. The system temperature was approximately 20 K.

\subsubsection{Nan\c{c}ay}

During a total of 600 hours we observed 277 CIG galaxies. Sixty of these suffered from strong interference or severe baseline problems and had to be discarded. For sources of unknown redshift a total bandwidth of 50 MHz was used giving a velocity range of approximately 10500 \kms, which was centred at 7000 \kms \ to try to maximise the probability of detecting the target's HI emission (as we anticipated that targets at very low velocities would have already been detected). For sources of known redshift a narrowed bandwidth of 12.5 MHz ($\sim$2500 \kms) was used. The best system temperatures (at dec of 15$^{\circ}$) was about 35 K.

\subsection{Selection of spectra}

Of the 488 CIG galaxies observed by the AMIGA team 429 are included in the final sample, along with 415 spectra from the literature. Our own observations were omitted in cases where there is still no known redshift (24 CIG galaxies) of an undetected source, or a redshift was obtained after our observations and it revealed that the source would not have been (completely) within the observed bandwidth (29 CIG galaxies). Without knowing the HI emission of a target should fall within the bandwidth, an upper limit of the flux cannot be confidently estimated. A small number of our observations were discarded because a literature spectrum was deemed preferable to our own spectrum (6 CIG galaxies).

In cases where there were multiple spectra with detections of the same target the preferred spectrum was selected by hand. As the comparison was performed by a person it did not follow an exact algorithm, but considered the following factors:
\begin{itemize}
\item The rms noise of the spectrum.
\item The telescope beam size relative to the size of the optical disc of the target galaxy.
\item Spectral resolution.
\item Other problems such as RFI, unstable baselines, and proximity to the edge of the bandpass.
\end{itemize}
Generally the first two of these were the most important. When the angular size of the optical disc was comparable to the telescope beam, the observation with the largest beam was almost always preferred, even at the expense of some signal-to-noise. The rationale behind this choice is that it is better to incur a larger random error due to increased noise in the spectrum, than a larger systematic error due to flux residing outside the primary beam. In cases where beam size was unimportant, generally more recent and higher spectral resolution spectra were favoured. In the case of non-detections the spectrum with the lowest rms noise was favoured.

As much as was possible ALFALFA spectra were avoided (only one ALFALFA spectrum is used the final sample) such that an independent comparison could be made between the observed flux scales of our dataset and those CIG galaxies with ALFALFA spectra. This choice did not decrease the quality of our database because the rms noises in the overlapping spectra were typically similar to those from ALFALFA and no other telescope used had a beam size smaller than Arecibo.

\section{HI data reduction}
\label{sec:data_red}

The HI single dish spectra of a total of 844 CIG galaxies were obtained through our own observations or compiled from the literature. The AMIGA collaboration observed 488 CIG galaxies with the Arecibo 305 m telescope, the Effelsberg radio telescope, the Green Bank telescope, and the Nan\c{c}ay radio telescope. The 415 spectra obtained from the literature predominantly came from \citet{Springob_2005} and \citetalias{Haynes_1984} (see Table \ref{tab:refs} for the complete list of sources). For many of the literature observations the original spectra were unavailable in digital format and were substituted for with the digitised spectra from NED. In addition, we digitised 8 spectra ourselves.

\subsection{Determination of spectral and source parameters}

The baselines of our own observations were fit with low order polynomials and the rms noise was estimated in an emission free region of each spectrum. The same procedure was applied to the literature spectra which were published without the baselines removed. All spectra were inspected by eye (and smoothed as necessary) to determine if there was a likely detection, or potential marginal detection, of the CIG galaxy. The spectra without a detection were retained to be used as upper limits only if the source had an existing redshift that fell in the observed bandpass. Upper limits on the source HI mass was estimated for the marginal and non-detected sources as described in section \ref{sec:HI_lims}. 
A threshold for the upper limits of 5-$\sigma$ was chosen because this is approximately when there is a transition from a mix of detections and marginals (identified by eye), to solely marginals.

The source parameters were extracted from the spectra with either detections or marginal detections, using our own implementation of the \citet{Springob_2005} method to fit HI spectral profiles. This method was selected because it does not require a parametric form of the profiles to be assumed, but is found to be more resilient in cases of low S/N compared to using the observed datapoints themselves to define the source properties \citep[e.g.][]{Fouque_1990}. Here we provide a brief description of the method. For a complete description refer to the original article \citep{Springob_2005}. 

The Springob-method assumes that HI profiles are double-horned in shape and begins by finding the peak flux density on the left and right edges of the profile. The two edges are then fit with straight lines using the datapoints between 15\% and 85\% of the peak flux density of that side of the profile minus the rms noise. The velocity width is then measured as the separation between the 50\% levels (of the peak flux density minus rms) of the left and right sides (each calculated separately), while the centre velocity is taken as the mean of the velocities at the left and right 50\% levels. Finally, the integrated flux is calculated by summing the flux density in the channels between the two zero points of the lines fitted to the left and right sides of the profile (and then multiplying by the mean channel width of the summed channels). The error in the integrated flux is estimated using the empirical relation
\begin{equation}
\label{eqn:Sint_err}
\sigma_{S_{\mathrm{int}}} = 2 \sigma_{\mathrm{rms}} \sqrt{1.4 W_{50} \delta v},
\end{equation}
where $\sigma_{\mathrm{rms}}$ is the rms noise, $W_{50}$ is the velocity width at the 50\% level, and $\delta v$ is the spectrum channel width in \kms.
As the Springob-method assumes that the spectral profile is double horned in shape, it loses some of its objectivity when applied to profiles with only a single peak or cases where the highest point in the profile is not in either of the horns. These cases are flagged in the data reduction process, but generally were found to give similar results to the Fouqu{\'e}-method. For spectra with the highest peak not falling in either horn, the peak signal-to-noise was adjusted after the initial fitting to reflex the true profile peak height. 

A complication with the Springob-method is that there must be at least 3 spectral channels with flux between the 15\% and 85\% levels within each horn in order for a straight line to be fitted. While \citet{Springob_2005} mostly had high resolution spectra, preventing this from being a serious concern, a number of the compiled spectra are from older observations with relatively poor spectral resolution ($> 15$ \kms). If a straight line could not be fit due to there being too few points within the relevant interval then additional points were linearly interpolated between the true datapoints for the purposes of fitting only (these spectra were also flagged to indicate this had been done).

Finally, as some of the spectra required interpolation or were not double horned we decided not to use the uncertainties in the line fits to determine the errors in the widths. Instead we used the estimates of \citet{Fouque_1990}, which gives the uncertainty in the systemic velocity, $V_{50}$, as
\begin{equation}
\label{eqn:W50_err}
\sigma_{V_{50}} = \frac{4 \sqrt{1.2 n_{\mathrm{smo}} \delta v \, (W_{20} - W_{50})/2 } } {\mathrm{snr_{p}}},
\end{equation}
where $n_{\mathrm{smo}}$ is the number of channels the final spectrum was smoothed over, $\delta v$ is the channel width in \kms, $\mathrm{snr_{p}}$ is the peak signal-to-noise ratio, and $W_{20}$ is the velocity width in \kms \ calculated as described above except at the 20\% level. The uncertainty in the velocity width, $W_{50}$, is taken to be $\sqrt{2}$ times this value.

During this fitting process flags were also set if the method was thought to be potentially erroneous. This could occur if, for example, there was a substantial noise spike near the edge of the profile that obscured the true location of the profile edge. 

It was also determined by eye whether a given target was considered detected, marginally detected, or not detected. Upon review it was found that the marginal detections almost all corresponded to profiles with signal-to-noise ratios of less than 5. This was then adopted as the quantitative threshold for a detection and all profiles with a signal-to-noise of less than 5 were considered upper limits when deriving the scaling relations (see section \ref{sec:HI_lims}).

\subsection{HI flux and width corrections}
\label{sec:corrections}

\begin{figure}
    \centering
    \includegraphics[width=\columnwidth]{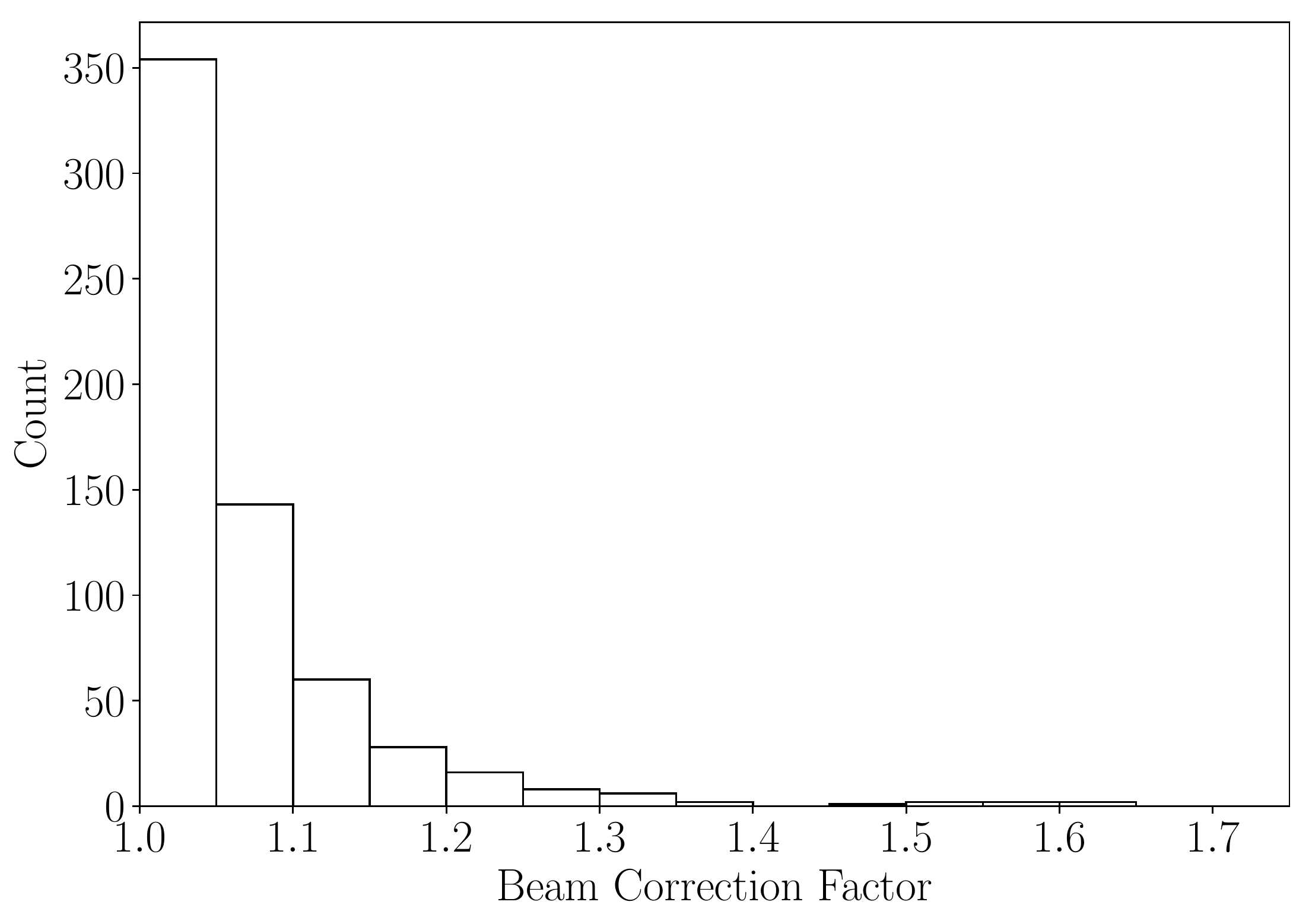}
    \caption{The distribution of beam correction factors for all galaxies detected in HI.}
    \label{fig:beam_corr}
\end{figure}

The HI integrated flux of a source can be suppressed below its true value by inaccurate pointing of the telescope, beam attenuation (if the angular size of emission is comparable to the telescope's beam), or both. Inaccurate pointing can be caused by errors in the input source catalogue or due to the intrinsic uncertainty in the telescope's pointing accuracy. The smaller the beam of the telescope the more severe both of these effects will be because the smaller the beam the greater the attenuation of the incoming signal for a given offset. 

\citet{Leon_2003} remeasured the optical positions of the CIG and found that there was a typical offset of 2'' (although in some cases were as large as 38''), while typical pointing uncertainties for radio telescopes are 5-15''. \citetalias{Haynes_1984} estimated that Arecibo's pointing uncertainty led to an average of 5\% decrease in flux in target sources. The decrease is likely to be even smaller for other telescopes as Arecibo was the largest used by approximately a factor of 3 in diameter. As the centre of the beam has the highest gain, any offset from the centre results in a decrease in the measured flux. Therefore, \citetalias{Haynes_1984} corrected for this effect by multiplying by a constant correction factor. The updated positions calculated in AMIGA reveal that many of the original observations in our compilation were not targeting the centre of the source, meaning that the random pointing uncertainty would not always act to decrease the observed flux. Therefore, we choose not to make a correction for this effect. However, the systematic effect caused by the incorrect target positions is corrected for in the beam attenuation correction, as explained below.

When observed in HI, nearby galaxies cannot typically be treated as point sources because their distribution of HI frequently extends to angular scales comparable to the size of a radio telescope's beam \citep[e.g.][]{Shostak_1978}, this means that a correction must be applied for the beam filling factor, $f$, in order to get the corrected flux, $S_{\mathrm{c}} = f S_{\mathrm{obs}}$, where $S_{\mathrm{obs}}$ is the observed flux and $f$ is calculated as follows:
\begin{equation}
f = \frac{ \int \sigma_{\mathrm{HI}}(x,y) B(x,y) \, \mathrm{d}x\mathrm{d}y }{ \int \sigma_{ \mathrm{HI}}(x,y) \, \mathrm{d}x\mathrm{d}y }
\end{equation}
Here $x$ and $y$ are the angular Cartesian coordinates on the sky, $\sigma_{\mathrm{HI}}$ is the neutral hydrogen surface density distribution, and $B$ is the beam response pattern of the telescope. We followed the approach of \citet{Hewitt_1983} using a circular Gaussian beam and a circular double Gaussian for the HI surface density. The characteristic length of the first Gaussian component of the HI surface density is assumed to be $R_{1} = 0.65D_{25}$ (in B-band), the second Gaussian component has a magnitude of -0.6 times that of the first and a length scale of $0.23R_{1}$, in order to create the central HI hole. Finally, the whole distribution is compressed along one axis according to the inclination derived from the optical properties. The position angle is unimportant because the beam functions are circularly symmetric (with the exception of NRT, see below). Finally the centre of the distribution is offset by the difference between the revised position \citep{Leon_2003} and the target coordinates of the original observation.\footnote{No positional offset was made for CIGs 68, 543, and 561 because of suspected typographical errors in target coordinates listed in the original reference.} The value of $f$ is then calculated numerically.

In the case of spectra observed with NRT there is the additional complication that the beam response cannot be assumed to be circular. We therefore use a double Gaussian beam that has a HPBW of 20' in the North-South direction and 4' in the East-West direction. This asymmetric beam also means that the position angle of the source is, in theory, important. Source position angles were obtained from HyperLeda for all objects and used to rotate the model HI distribution relative to the assumed telescope beam (only in the case of NRT). It should be noted that the uncertainties in the position angle can be very large, with different measurements in HyperLeda frequently varying by over 10$^{\circ}$, however, given the large size of the NRT beam the impact that this is expected to have is minimal. The HPBWs assumed for other telescopes can be found in Table \ref{tab:beam_widths}. The distribution of beam corrections is shown in Figure \ref{fig:beam_corr}. Over 90\% of HI detections in this dataset have a beam correction factor of less than 20\%

\begin{table}
\centering
\caption{Telescope beam widths and codes}
\label{tab:beam_widths}
\begin{tabular}{l c c}
\hline\hline
Telescope           & \begin{tabular}{c}HPBW\\ (arcmin)\end{tabular} & Code \\ \hline
Arecibo (Gregorian) & 3.5                                                     & AOG  \\
Arecibo (Dual circular) & 3.3                                                     & AOL  \\
Arecibo (Flat)      & 3.9                                                     & AOF  \\
Effelsberg          & 8.8                                                     & ERT  \\
Green Bank 100 m    & 9.0                                                     & GBT  \\
Green Bank 300 ft   & 10                                                      & G91  \\
Green Bank 140 ft   & 21                                                      & G43  \\
Jodrell Bank        & 12                                                      & JBL  \\
Nan\c{c}ay          & 4 $\times$ 20\tablefootmark{$\dagger$}              & NRT  \\
Parkes              & 13                                                      & HIP  \\ \hline
\end{tabular}
\tablefoot{
Description of columns: 1) telescope name, 2) the telescope half power beam widths, 3 ) a 3 character code to identify each telescope.\\
\tablefoottext{$\dagger$}{The N-S extent of the Nan\c{c}ay beam changes with elevation, however, as this dimension is always much larger than any of our galaxies this term can be safely neglected in the beam correction factor.}
}
\end{table}

The velocity widths of all sources were corrected following the methodology of \citet{Springob_2005}. The first correction to the velocity width is for instrumental broadening, $c_{\mathrm{inst}}$. This is calculated following the empirical expressions given in \citet{Springob_2005} equations 3, 5-7, and their table 2. We replace the channel width with the channel width times $n_{\mathrm{smo}}-2$ (the expressions assume the spectra have already been Hanning smoothed across 2 channels). The next correction is for broadening due to the cosmological expansion. This term is simply $c_\mathrm{cosmo} = (1+z_{50})^{-1}$, where $z_{50}$ is the heliocentric redshift measured at the 50\% level. The instrumental effects have to be corrected first because corrections should be applied in the reverse order of how they impacted the originally emitted spectrum, starting with the impact of the instrument, then the expansion of the Universe, and finally the properties of the source itself. We do not make any of the third type of corrections (e.g. inclination and turbulent motions) as the velocity widths are not part of our statistical analysis.

\subsection{HI masses}

\begin{figure}
    \centering
    \includegraphics[width=\columnwidth]{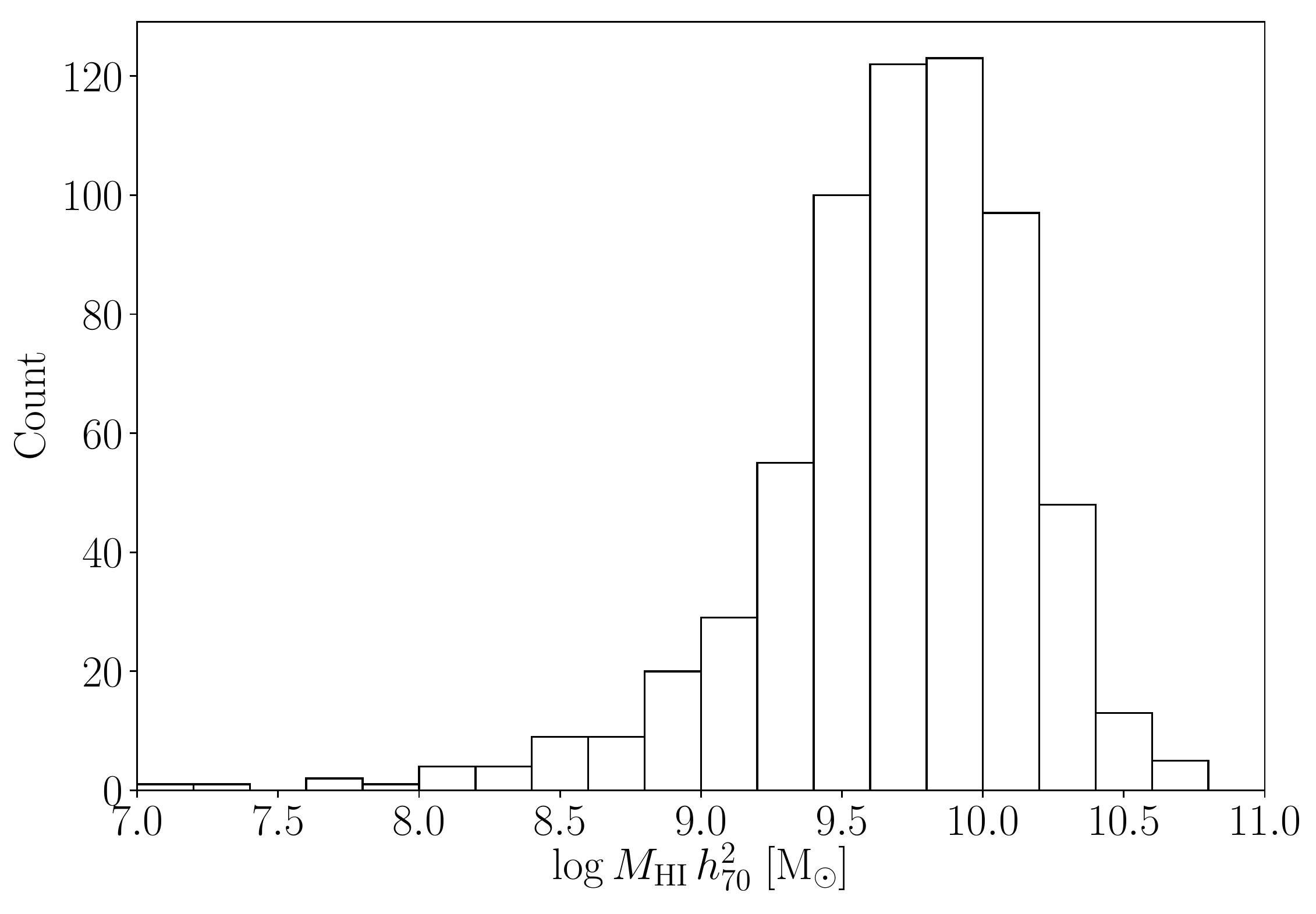}
    \caption{The distribution of HI masses for all sources detected in HI.}
    \label{fig:HI_masses}
\end{figure}

With the measurements of the HI fluxes and beam correction factors we use the normal equation to calculate the HI masses of the detected sources,
\begin{equation}
    \label{eqn:HI_mass}
    \frac{M_{\mathrm{HI}} \, h_{70}^{2}}{\mathrm{M_{\odot}}} = 235600 \times \frac{S_{\mathrm{c}}}{\mathrm{Jy \, kms^{-1}}} \left( \frac{D\,h_{70}}{\mathrm{Mpc}} \right)^{2}
\end{equation}
where $D$ is the estimated distance to the source in Mpc. The distribution of HI masses of all HI detections in shown in figure \ref{fig:HI_masses}.

\subsection{HI mass upper limits}
\label{sec:HI_lims}

As a means to make a fair comparison of the sensitivity of all spectra the parameter $\sigma_{\mathrm{rms},10}$ was calculated for all spectra; the rms noise if the spectra all had 10 \kms \ channel widths.
\begin{equation}
\label{eqn:rms10}
\sigma_{\mathrm{rms},10} = \sigma_{\mathrm{rms}} \sqrt{ \frac{n_{\mathrm{smo}} \delta v}{10 \; \mathrm{km\,s^{-1}} } },
\end{equation}
where $\sigma_{\mathrm{rms}}$ is the spectrum's measured rms noise for its given channel width ($\delta v$) and smoothing ($n_{\mathrm{smo}}$). The integrated signal-to-noise ratio of all detections and marginal detections was calculated using a similar approach to ALFALFA \citep{Giovanelli_2005}
\begin{equation}
    \label{eqn:snr}
    \mathrm{S/N} = \frac{1000 \times S_{\mathrm{int}}}{\sigma_{\mathrm{rms},10} \sqrt{W_{50} \times 10 \, \mathrm{km\,s^{-1}}}},
\end{equation}
where $S_{\mathrm{int}}$ is the integrated flux in Jy \kms, and the rms noise in 10 \kms \ channels ($\sigma_{\mathrm{rms},10}$) is in mJy. A maximum value of 300 \kms \ was set for $W_{50}$ (that is, widths above 300 \kms \ were set to 300 \kms \ for this calculation only) because, as confirmed by \citep{Haynes_2011}, beyond this point smoothing the profile no longer results in the same improvement of signal-to-noise.

All spectra with $\mathrm{S/N}$ less that 5 were treated as upper limits. The distinction between non-detections and marginal detections is that marginal detections were originally identified by eye as marginal detections or detections (but have $\mathrm{S/N} < 5$), whereas in the case of non-detections, no HI emission in the appropriate velocity range was identified. As 5-$\sigma$ is the threshold we have set to separate detections from upper limits, we will use 5-$\sigma$ upper limits on the HI mass for those sources not considered bona fide detections.

To calculate these upper limits the spectral profile of the source was assumed to be rectangular, with a flux density of $5\sigma_{\mathrm{rms},10}$. The velocity widths of each source was estimated from the B-band Tully-Fisher relation (TFR). We used the relation for field galaxies calculated by \citet{Torres-Flores+2010}, which converted to our unit system is
\begin{equation}
    \log L_{\mathrm{B}}\,h_{70}^{2}/\mathrm{L_{\odot}} = 2.94 \log 2 v_{\mathrm{max}}/\mathrm{km\,s^{-1}} + 2.45,
\end{equation}
where $v_{\mathrm{max}}$ is the maximum rotation velocity of the galaxy's rotation curve. We assume that the velocity width is $W_{\mathrm{TFR}} = 2 v_{\mathrm{max}} (1+z) \sin i$, where $i$ is the inclination (see section \ref{sec:opt_diam_incl}). A minimum width of 100 \kms \ was set because less than 5\% of our final detection sample has widths this narrow and narrower widths make sources more likely to be detected. Finally, the distance to each source was used as calculated in section \ref{sec:dists}, giving the upper limits on the HI mass as
\begin{equation}
\frac{M^{\mathrm{max}}_\mathrm{HI}\,h_{70}^{2}}{\mathrm{M_{\odot}}} = 1.178 \times 10^{3} \; \frac{\sigma_{\mathrm{rms},10}}{\mathrm{mJy}} \left( \frac{W_{\mathrm{TFR}}}{\mathrm{km\,s^{-1}}} \right) \left( \frac{D\,h_{70}}{\mathrm{Mpc}} \right)^{2}.
\end{equation}

Using widths based on the TFR steepens the final scaling relations that we calculate by about 5\% compared to assuming a constant width. However, because for a given sensitivity per channel the flux (mass) limit of a non-detection grows with its velocity width, assuming a constant width introduces a non-physical dependence between $L_{\mathrm{B}}$ and the limit on the HI mass. Instead, using the TFR to determine the widths introduces the natural relation between $L_{\mathrm{B}}$ and the widths into the upper limits of the HI mass.

\subsection{Comparison with ALFALFA integrated fluxes and velocity widths}
\label{sec:alfa_comp}

\begin{figure}
    \centering
    \includegraphics[width=\columnwidth]{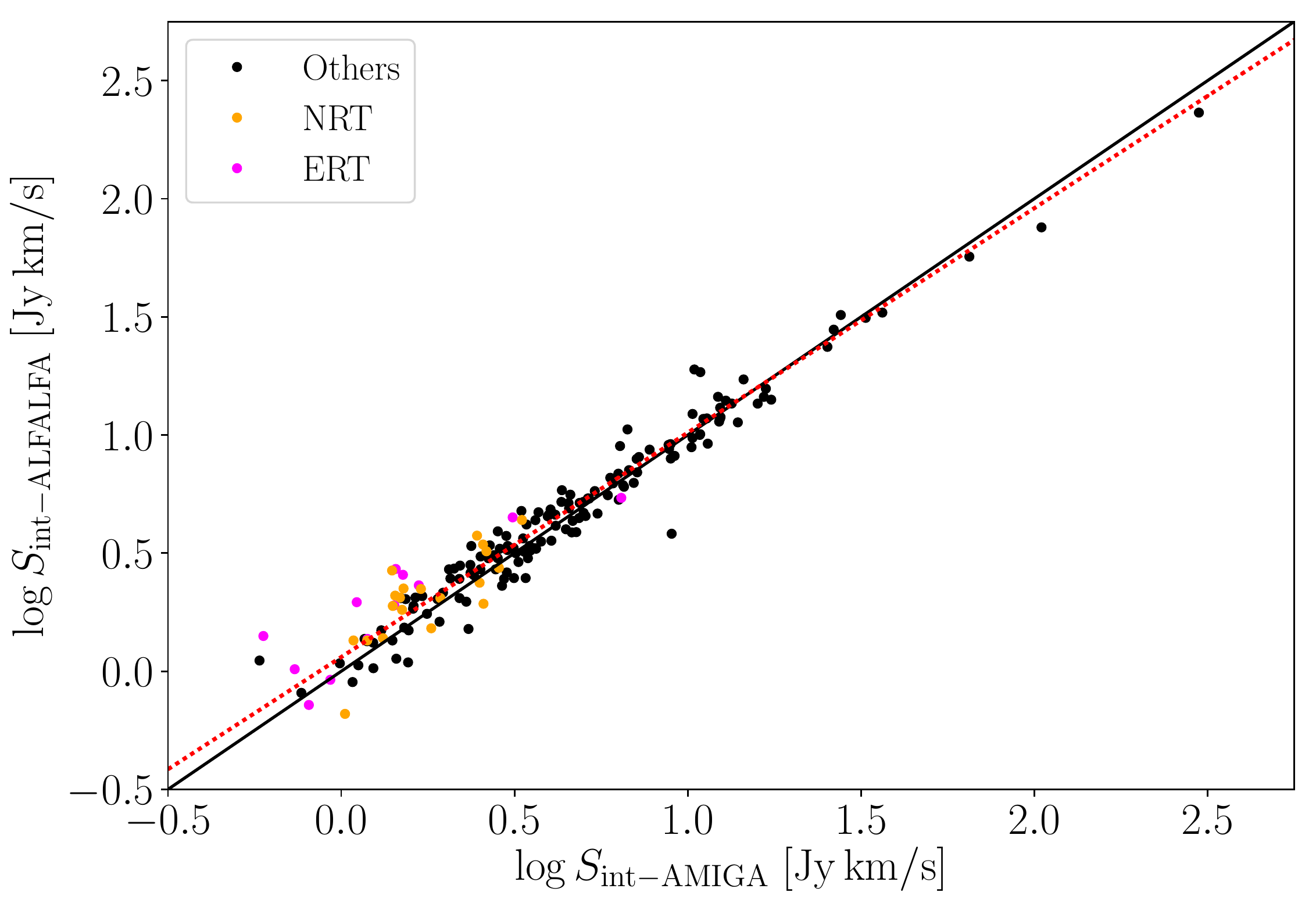}
    \caption{The comparison between ALFALFA and AMIGA measurements of HI integrated flux. The orange and pink points show detections from NRT and ERT respectively, and the black points show detections from all other telescopes. The thin black line indicates equality, while the dotted red line shows the best fit to all the points. Statistical error bars are not shown as for the majority of the points these are comparable in size to the points themselves, indicating that absolute calibration is the cause of most of the scatter and offset.}
    \label{fig:flux_comp}
\end{figure}

\begin{figure}
    \centering
    \includegraphics[width=\columnwidth]{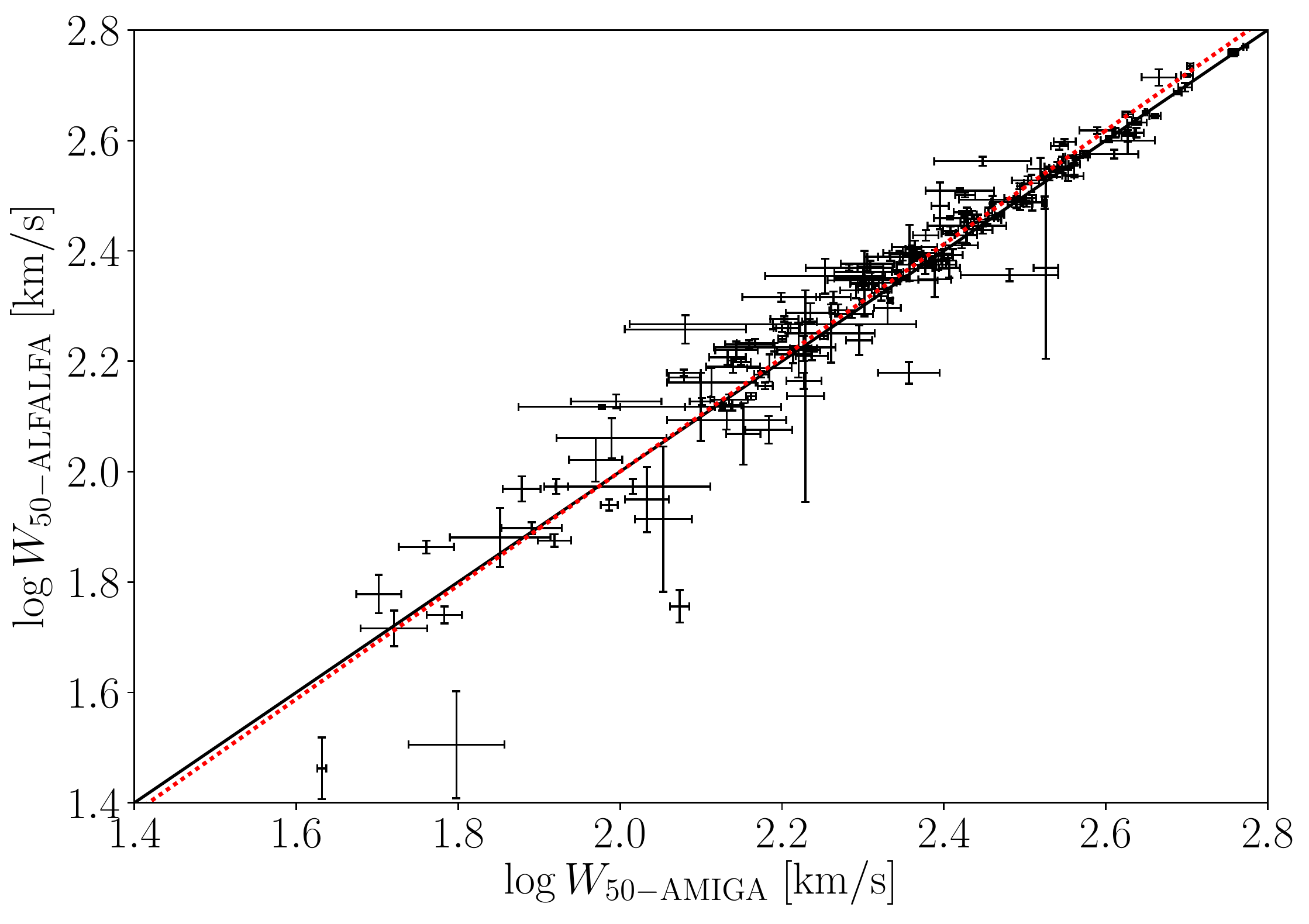}
    \caption{The comparison between ALFALFA and AMIGA measurements of HI profile widths at the 50\% level. The black line indicates equality and the dotted red line the best fit to the data. The highly outlying points are either from low signal-to-noise detections, or sources where the profile shapes in ALFALFA and AMIGA have differences for unknown reasons.}
    \label{fig:width_comp}
\end{figure}

As only one of our compilation of spectra came from ALFALFA we can use the ALFALFA catalogue\footnote{Here we used the 70\% ALFALFA catalogue which is available at \url{http://egg.astro.cornell.edu/alfalfa/data/index.php}} as a means to compare and verify our corrected integrated flux and velocity width measurements. The two catalogues were cross matched for agreement within 30'' and 200 \kms, using the optical counterpart positions and the HI recession velocities given in the ALFALFA catalogue. To estimate how likely a mismatch was with this automated procedure we integrated the ALFALFA correlation function \citep{Papastergis_2013,Jones_2015} over the match volume to determine how many interlopers are expected. As essentially all our detections are above $\log M_{\mathrm{HI}}/\mathrm{M_{\odot}} = 8$, this was set as the minimum mass for a believable mismatch. This gave the chance of a mismatch as less than 1\%. Therefore, we consider all automated matches to be correct. The comparison of the flux and velocity widths are shown in Figures \ref{fig:flux_comp} and \ref{fig:width_comp}.

It appears that there is very good qualitative agreement between the two datasets. Indeed the relation between the velocity widths is $\log W_{50 - \mathrm{ALFALFA}} = 1.03 \log W_{50 - \mathrm{AMIGA}} - 0.06$. However, in the case of the fluxes the best fit line is at a small, but significant angle to the 1:1 line ($\log S_{\mathrm{int} - \mathrm{ALFALFA}} = 0.95 \log S_{\mathrm{int - \mathrm{AMIGA}}} + 0.06$), indicating that there is a systematic disagreement of up to 20\% (at the lowest and highest fluxes) in the flux between the AMIGA and ALFALFA measurements.\footnote{The deviation from unity of the slopes for both the flux and width comparison may be a similar magnitude, however, it is important to remember that the flux measurements span almost 3 orders of magnitude whereas the width measurements span only 1.}

Discrepancies at the highest fluxes are not surprising as these large and bright sources are often extended beyond the Arecibo beam, which can cause complications in determining the flux, especially with a multi-beam receiver such as ALFA (Arecibo L-band Feed Array). However, these sources were not found to be the main cause of the offset gradient. Instead sources observed with NRT and ERT were found to have systematically low integrated fluxes compared to ALFALFA, with a mean offset of $\sim$0.2 dex. While the most obvious explanation for such an offset might be the beam correction factor, as the Arecibo beam is much smaller than both the NRT and ERT beams, all of the matched NRT and ERT sources have optical diameters of 1 arcmin or less, meaning their beam correction factors in ALFALFA would be approximately 10\% or less and thus cannot explain the offset.

A similar discrepancy was noticed before by the NIBLES team in \cite{vanDriel_2016}. They attributed this to a difference between single and multi-beam detectors, but our dataset does not appear to support this interpretation because the integrated fluxes measured from the 145 single beam spectra in our sample (excluding NRT and ERT) that overlap with ALFALFA are in good agreement with those of ALFALFA, despite it being a multi-beam survey. Furthermore, the NIBLES comparison was performed only against very high signal-to-noise sources in ALFALFA, which are not a representative sample of all the ALFALFA sources.

Even though NRT and ERT only contribute $\sim$20\% of the overlapping measurements, removing these data from the fit more than halved the magnitude of the deviation. Further investigation showed an apparent frequency (or redshift) dependence in the ratio of the ALFALFA fluxes to the AMIGA fluxes obtained with NRT and ERT. However, this trend had a poor correlation and although it could be an indication of a gain calibration issue, we were unable to identify the root cause of the apparent offset. Therefore, no correction was made to the NRT and ERT data to bring the flux scales in line with ALFALFA and the rest of our dataset, but we note that applying such a correction would steepen the final scaling relations that we calculate by a few percent. The implications of this choice for the final scaling relations are described in Appendix \ref{sec:flux_disc}.

The remaining scatter around the best fit line was measured along the length of the line and took values in the range 0.1-0.15 dex, with a mean of 0.12 dex across all the data. This is a better estimate of the uncertainty in the flux than the statistical error found during spectral fitting because for most sources uncertainties in the absolute calibration of the telescope dominate over the statistical uncertainty in a given spectrum. Therefore this value (0.12 dex) was set as the minimum possible uncertainty in the integrate flux and, later on, the HI mass.

\section{Analysis}
\label{sec:analysis}

In this section we present our fundamental results, the HI scaling relations, but first we describe the selection of the final science sample, explain our regression model and discuss how the problems associated with previous regression methods used to fit HI scaling relations have been addressed.

\subsection{Completeness \& isolation}
\label{sec:cuts}

\begin{table*}
\centering
\caption{Sample size after each successive cut}
\label{tab:samp_size}
\begin{tabular}{l|ccc|c}
\hline\hline
Cumulative cuts  & Detections & Marginals & Non-detections & Total \\ \hline
No cuts          & 625        & 18        & 201            &  844  \\
Completeness     & 566        & 17        & 145            &  728  \\
Isolation        & 427        & 16        & 129            &  572  \\
Profile quality  & 399        & 16        & 129            &  544  \\ \hline
\end{tabular}
\end{table*}

The ancillary data collected by the AMIGA team allows cuts to be made to the sample to ensure that the final scaling relations are fit to only galaxies with quantified isolation and a sample that is highly complete. Due to the substantially larger size of this dataset (compared to \citetalias{Haynes_1984}), even after these significant cuts have been made there still remain sufficient sources to perform a statistical analysis.

The completeness of the CIG was assessed by \citet{Verdes_Montenegro_2005} using a $V/V_{\mathrm{max}}$ test and found to be 80-95\% complete below a B-band magnitude of 15. The magnitudes of the AMIGA sample were revised by \citet{Fernandez_Lorenzo_2012}, which shifted this cut to a magnitude of 15.3. This threshold is applied to our HI sample which removes approximately 15\% of the sources.

Next, isolation was ensured by following the recommended cuts of \citet{Verley_2007b}. The dimensionless local number density, $\eta_{\mathrm{k}}$, calculated by the distance to the 5th neighbour, is cut at a maximum value of 2.4. The $Q$ parameter, which signifies the strength of the tidal forces exerted by neighbours relative to the binding strength of the galaxy, is cut at a maximum value of -2, corresponding to an external tidal force of 1\% of the galaxy's internal forces. Neighbour density is frequently used alone to define isolation, but these two parameters are complementary because strong tidal forces can be caused by just one very nearby neighbour, without significantly impacting $\eta_\mathrm{k}$. With both of these isolation criteria set the sample is ensured to be quite distinct to samples in higher density environments. It should also be noted that all sources with heliocentric velocities below 1500 \kms \ are removed in this step because it is extremely difficult to accurately quantify isolation for such nearby sources \citep{Verley_2007b}. Hence, this cut also has the effect of removing any dwarf galaxies that were in the CIG, as these are only sufficiently bright when they are relatively nearby. Therefore, the relations calculated in this paper are not applicable to dwarf galaxies as there are none in our science sample.

Finally, any sources which had flags set during the spectral fitting procedure to indicate the spectral parameters are potentially spurious were also removed, which reduced the remaining detections by 5\%. This leaves a final sample of 544 CIG galaxies (399 detections, 16 marginal detections, and 129 non-detections in HI) that we will refer to as the AMIGA HI science sample. This sample is used in all the following analysis unless explicitly stated otherwise. The exact sample size after each of the cuts explained above is shown in Table \ref{tab:samp_size}.

Applying the same isolation and completeness cuts described above to the full CIG leaves 618 galaxies. Therefore, although many galaxies have been cut from the HI sample there are still detections or upper limits on the HI content of almost 90\% of the full isolated and complete sample.

\subsection{Regression model}
\label{sec:reg_mod}

The data are expected to exhibit a good positive correlation between, for example, $\log M_{\mathrm{HI}}$ and $\log L_{\mathrm{B}}$. However, this correlation most likely has a significant amount of intrinsic scatter due to covariates, such as galaxy morphology. In addition, the data contain errors in both the independent and dependent variables, and censorship of the dependent variable is common due to non-detection. Finally, the part of the errors that originate from the distance uncertainty is the same for both variables, making their errors correlated. Each of these properties of the dataset can erroneously impact the final regression line if not accounted for in the regression method.

The simplest methods, such as ordinary least squares (OLS), account only for scatter in the dependent variable, but can be straightforwardly extended to include the uncertainty in the measurements of the dependent variable. Therefore, both of these aspects of the data are usually modelled in the astronomy literature. All the works that we compare with used either the OLS method \citep{Haynes_1984,Solanes_1996} or the OLS-bisector method \citep{Denes_2014}.

Measurement uncertainty in the independent variable is less straightforward to account for than uncertainty in the dependent variable, and is therefore frequently neglected. This is known as the ``errors-in-variables'' problem in statistics. Failing to account for these errors leads to a biasing of the regression line gradient (towards a flatter slope). Many methods also do not allow the incorporation of upper limits. However, upper limits can contain information about all the parameters of the regression fit and so simply ignoring them can make the results dependent on the sensitivity of the observations, or result in less precise estimates of the regression parameters than obtainable with the upper limits included. Finally, in the presence of correlated errors, standard regression methods can produce misleading results because they do not account for the fact that the measurements of the variables are not independent.

While there are many methods available in the literature to fit regression lines, they tend to be aimed at addressing a subset of these issues, but all are anticipated to be potentially important effects in this case. Therefore, we construct a parametric model designed for this particular situation and estimate the regression parameters by maximising the likelihood of the observed data given the model.

Assume that the data follow a linear trend with intrinsic scatter $\sigma_{\xi}$:
\begin{equation}
\label{eqn:reg_mod}
y^{*}_{i} = \beta_{0} + \beta_{1}x^{*}_{i} + \xi,
\end{equation}
where a star denotes the true value (as opposed to the observed value), $i$ indicates simply the $i$th data point, and $\beta_{0}$ and $\beta_{1}$ are the regression coefficients that we wish to determine. Here we also use the notation that the greek letter $\xi$ is a random variable and $\sigma_{\xi}$ is its standard deviation about a zero mean. This notation is also used for other random variables in this section. We have also taken care to consistently use the phrase ``intrinsic scatter'' to refer to estimates of $\sigma_{\xi}$ for the various relations calculated in this paper. Some of the scatter in the data is due to the measurement uncertainties (which can be large). Estimates of the measurement uncertainties for each datapoint are included in the method described below, which permits the fitting of an estimate of $\sigma_{\xi}$, that is, the scatter intrinsic to the physical relation that is not accounted for by measurement uncertainty.

The independent variable is assumed to have a Gaussian measurement error, $\sigma_{\eta_{i}}$, and a Gaussian error due to the distance uncertainty\footnote{Note that $\sigma_{\delta}$ is twice the estimated uncertainty in the log distance because luminosity and mass both scale with the square of the distance.}, $\sigma_{\delta_{i}}$, such that $x^{\mathrm{obs}}_{i} = x^{*}_{i} + \eta_{i} + \delta_{i}$, where $x^{\mathrm{obs}}_{i}$ is the observed value of the $i$th data point. Similarly the dependent variable is assumed to have a Gaussian measurement error $\sigma_{\epsilon_{i}}$, giving $y^{\mathrm{obs}}_{i} = y^{*}_{i} + \epsilon_{i} + \delta_{i}$, where $\delta_{i}$ takes exactly the same value as in the previous equation. This means that the errors in the $x$- and $y$-directions are correlated, even though $\eta_{i}$ and $\epsilon_{i}$ are independent.

Due to this correlation the errors in the $x$- and $y$-directions cannot be modelled as two independent normal distributions and instead are treated as a bivariate normal with covariance matrix
\begin{equation}
\Sigma_{i} =
  \left[ {\begin{array}{cc}
   \sigma_{x_{i}}^{2} & \rho_{i}\sigma_{x_{i}}\sigma_{y_{i}} \\       \rho_{i}\sigma_{x_{i}}\sigma_{y_{i}} & \sigma_{y_{i}}^{2} \\      \end{array} } \right]
\end{equation}
where $\sigma_{x_{i}}^{2} = {\sigma_{\eta_{i}}}^{2} + {\sigma_{\delta_{i}}}^{2}$, $\sigma_{y_{i}}^{2} = \sigma_{\xi}^{2} + {\sigma_{\epsilon_{i}}}^{2} + {\sigma_{\delta_{i}}}^{2}$, and $\rho_{i} = {\sigma_{\delta_{i}}}^{2}/\sigma_{x_{i}}\sigma_{y_{i}}$.

First consider only sources which are successfully detected. The observed independent variables will be normally distributed about their true values as indicated by $\sigma_{x_{i}}$ and the dependent variable will be normally distributed above and below the true regression line according to $\sigma_{y_{i}}$, giving the likelihood of the detected data as
\begin{eqnarray}
\mathcal{L}_{\mathrm{det}} &=& \prod\limits_{i} \frac{1}{2 \pi \sigma_{x_{i}} \sigma_{y_{i}} \sqrt{1 - \rho_{i}^{2}} } \exp \left( \frac{-1}{2(1-\rho_{i}^{2})} \right. \nonumber \\
&&\times \left. \left[ \frac{(x^{\mathrm{obs}}_{i}-x^{*}_{i})^{2}}{\sigma_{x_{i}}^{2}} + \frac{(y^{\mathrm{obs}}_{i}-\beta_{1}x^{*}_{i}-\beta_{0})^{2}}{\sigma_{y_{i}}^{2}} \right. \right. \nonumber \\
&&\left. \left. - \frac{2 \rho_{i} (x^{\mathrm{obs}}_{i}-x^{*}_{i}) (y^{\mathrm{obs}}_{i}-\beta_{1}x^{*}_{i}-\beta_{0})}{\sigma_{x_{i}}\sigma_{y_{i}}} \right] \right) .
\end{eqnarray}
When only considering the detected sources this is the likelihood that should be maximised by finding the optimal values of $\beta_{0}$, $\beta_{1}$, and $\sigma_{\xi}$. This method also treats the true values of the observations of $x$ as parameters and these are also found in the optimisation, but are discarded. In a more sophisticated treatment, such as a Bayesian hierarchical model, our prior knowledge of the intrinsic distribution of $x^{*}$ could be included rather than treating these as free parameters.

In the case where non-detections (or marginal detections) are also included, a different likelihood is required because $y^{\mathrm{obs}}_{j}$ is unknown, there is only an upper limit on its value (here the indices have been changed to $j$ rather than $i$ to prevent confusion between detections and non-detections.). We assume that the unobserved values of the HI mass ($y$) follow the same conditional distribution at each value of $x$ as the detections do, therefore, the appropriate weighting of each value of $y$ for the upper limits is obtained by integrating the likelihood above over all possible values of $y^{\mathrm{obs}}_{j}$. The values at which the non-detections become censored (the HI masses of the upper limits) are near random because they depend on which telescope the source was observed with, how long for, its distance, and its assumed velocity width. However, our assumption is somewhat uncertain because there is likely some morphological dependence on whether or not a source is detected, and the morphology distribution is not constant across all diameters and luminosities. This means that on some level our assumption is probably invalid. However, given the scope of our dataset this is a necessary simplification to proceed (although we explore the dependence on morphology in section \ref{sec:morph_rel}). When setting the upper limit for this integration we also make the simplifying assumption that it is absolute, i.e. that it is unaffected by the measurement and distance uncertainties. 
The upper limits are calculated at a level 5 times the rms noise in the spectra. The possibility that a signal at this level has been missed in our reduction process is remote. Furthermore, the fractional uncertainty in the distances is significantly less than 1 for all sources. Therefore, we are confident that the true HI mass of these sources falls below the stated limits.
With these assumptions the likelihood for the non-detections becomes
\begin{eqnarray}
\mathcal{L}_{\mathrm{lim}} &=& \prod\limits_{j} \frac{1}{2\sqrt{2 \pi} {\sigma_{x_{j}}}} \exp{\left( \frac{-(x^{\mathrm{obs}}_{j}-x^{*}_{j})^{2}}{2 {\sigma^{2}_{x_{j}}}} \right)} \nonumber \\ 
&&\times \left[ 1 - \erf \left( \frac{1}{\sqrt{2 \left( 1-\rho^{2}_{j} \right) }} \right. \right. \nonumber \\
&&\times \left. \left. \left( \frac{(x^{\mathrm{obs}}_{j}-x^{*}_{j})\rho_{j}}{\sigma_{x_{j}}} - \frac{y^{\mathrm{up}}_{j} - \beta_{0} - \beta_{1} x^{*}_{j}}{\sigma_{y_{j}}} \right) \right) \right] 
\end{eqnarray}
where $y^{\mathrm{up}}_{j}$ is the upper limit for the $j$th non-detection. When calculating $\sigma_{y_{j}}$ we no longer have a measurement error ($\sigma_{\epsilon_{j}}$) because no signal was detected, however, in place of $\sigma_{\epsilon_{j}}$ we use the scatter found in the calibration of the flux scales of our spectra (0.12 dex), which in practice was the relevant value for essentially all the detections as well. Finally, when performing the actual maximisation, $\log \mathcal{L}_{\mathrm{det}}$ for all the detections is added to $\log \mathcal{L}_{\mathrm{lim}}$ for all the non-detections to give the complete log-likelihood. 

Error estimates for each of the regression parameters can be calculated via the jackknife method. To jackknife a sample each datapoint is removed one at a time and the remaining $N-1$ datapoints are used to calculate the fit. The variance of each parameter can then be estimated by summing the squared deviations from the mean parameter value (across all $N$ jackknife samples) and weighting by $(N-1)/N$.

\subsection{HI scaling relations}
\label{sec:HI_rels}

\begin{figure}
    \centering
    \includegraphics[width=\columnwidth]{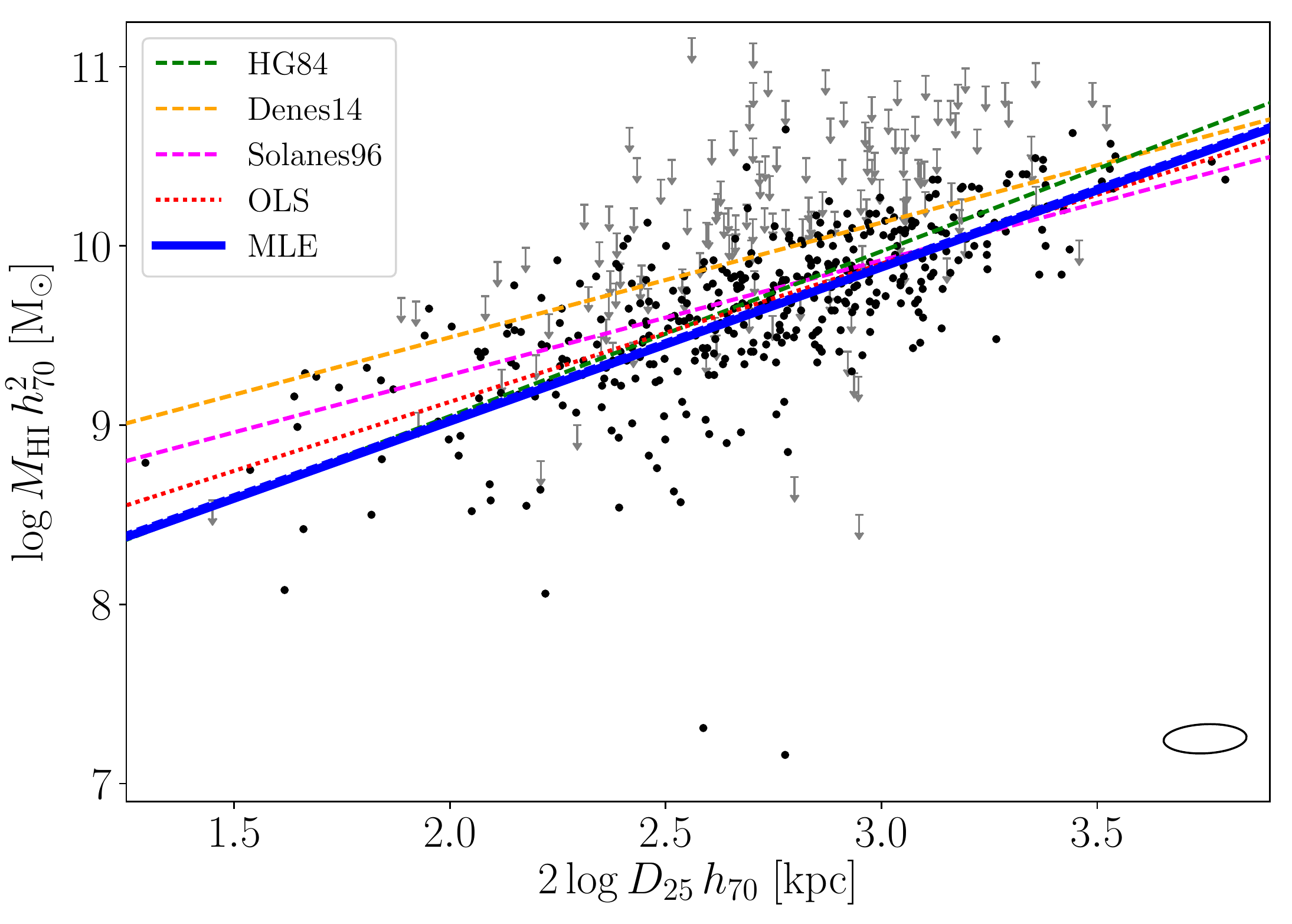}
    \caption{The scatter plot of the HI mass of AMIGA galaxies as a function of their optical diameters ($D_{25}$ in B-band). The black points indicate sources detected in HI while grey arrows indicate upper limits. The typical 1-$\sigma$ error ellipse of the data points is shown in the bottom right corner. The heavy blue lines show the regression fits of this work. The solid line corresponds to the full regression model including upper limits, while the dashed line is for the same model but only including detections (mostly hidden behind the solid line). The red dotted line is the ordinary least squares fit for the detections only. The green, purple, and orange dashed lines are from \citetalias{Haynes_1984}, \citet{Solanes_1996}, and \citet{Denes_2014} respectively.}
    \label{fig:D25_rel}
\end{figure}

\begin{figure}
    \centering
    \includegraphics[width=\columnwidth]{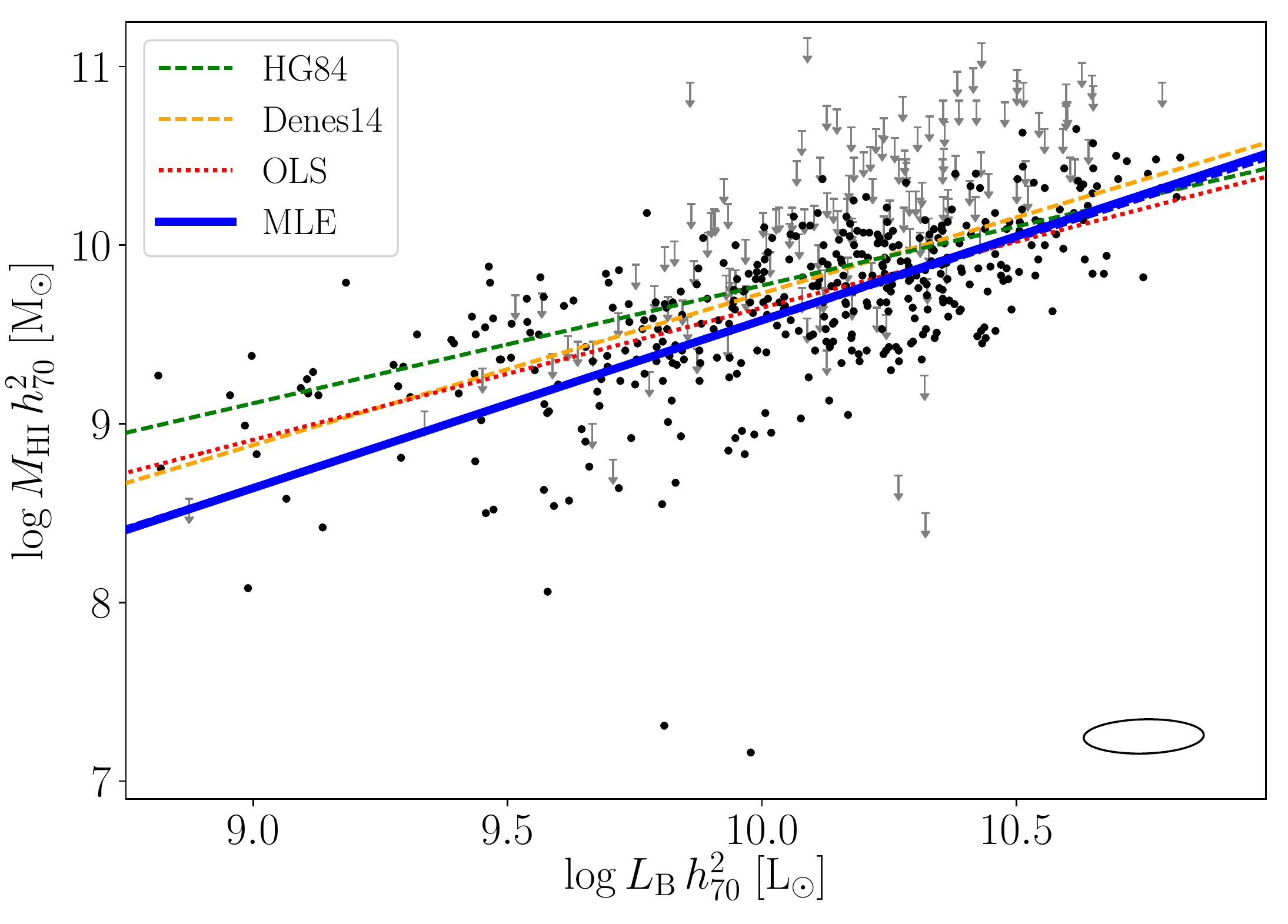}
    \caption{The scatter plot of the HI mass of AMIGA galaxies as a function of their optical luminosities (in B-band). The black points indicate sources detected in HI while grey arrows indicate upper limits. The typical 1-$\sigma$ error ellipse of the data points is shown in the bottom right corner. The heavy blue lines show the regression fits of this work. The solid line corresponds to the full regression model including upper limits, while the dashed line is for the same model but only including detections (mostly hidden behind the solid line). The red dotted line is the ordinary least squares fit for the detections only. The green and orange dashed lines are from \citetalias{Haynes_1984} and \citet{Denes_2014}.}
    \label{fig:LB_rel}
\end{figure}

We selected $\log D_{25}$ and $\log L_{\mathrm{B}}$ to use as predictors of HI content because these have the strongest correlations with $\log M_{\mathrm{HI}}$ out of all of the available observational properties. The correlation coefficient between $\log D_{25}$ and $\log M_{\mathrm{HI}}$ (detections only) is 0.73, and 0.69 between $\log L_{\mathrm{B}}$ and $\log M_{\mathrm{HI}}$. This is consistent with previous studies, which have generally found the optical diameter to be the best predictor of HI mass.

The regression model described in the previous subsection was fit to the AMIGA HI science sample (described in section \ref{sec:cuts}) and is shown by the blue lines in Figures \ref{fig:D25_rel} and \ref{fig:LB_rel}, and the coefficients are given in Tables \ref{tab:D25_rel} and \ref{tab:LB_rel}. For the purposes of comparison our regression model is fit both to all the data (shown by the solid blue line), including upper limits, and to just the detections (dashed blue line). An ordinary least squares (OLS) fit to the detections is also shown by the dotted red line. In both plots the OLS fit has a shallower gradient than those corresponding to our regression model. The reason for this is that the independent variable has considerable uncertainties (as shown by the typical error bars) which are not accounted for in the OLS method, causing an underestimate of the gradient. These plots also show fits from \citet{Haynes_1984}, \citet{Solanes_1996}  and \citet{Denes_2014} for comparison, which are discussed in detail in section \ref{sec:comp_rels}.

Tables \ref{tab:D25_rel} and \ref{tab:LB_rel} do not include values for the intrinsic scatter of the OLS fits because for this method only the total scatter about the relation can be calculated, which is 0.28 and 0.30 for the $D_{25}$ and $L_{\mathrm{B}}$ relations respectively. The corresponding values for our maximum likelihood method are 0.21 and 0.20, respectively. These are considerably smaller because our method accounts for the measurement uncertainties, excluding them from the scatter estimates, which are thus estimates only of the intrinsic scatter in the physical relation, not the total scatter.

The five exceptionally low HI-mass sources (two limits and three detections) that fall well below the main trend were excluded from the fitting process. To identify which points to exclude an iterative 3-$\sigma$ rejection algorithm was used. The relations were first fit using all the data, and the points and limits that were not consistent within $3\sigma_{\xi}$ of the fitted relation were removed and the relation was fit again. This process was iterated until the fit remained unchanged.

All of the removed sources fall well below the relation. There are no strongly outlying detections above the relation, and although there are many limits well above the main relation, as these are upper limits they are still consistent with it. In total five sources were removed (from all subsequent fits): CIGs 13, 68, 358, 609, and 1042. These sources do not appear to follow the assumptions of the regression model and therefore should not be fit with it. CIGs 13, 358, and 1042 are all early types, so their low HI content is not particularly surprising, also the photometry of CIG 1042 is highly uncertain due to a bright foreground star. However, CIGs 68 and 609 are types Sab and Sc, respectively, and thus would normally be expected to be quite HI-rich.

The general action of the upper limits is to modestly improve the precision of the estimates of the regression parameters (see Tables \ref{tab:D25_rel} and \ref{tab:LB_rel}). In this dataset the detections are numerous and cover the full ranges of both $D_{25}$ and $L_{\mathrm{B}}$, which allows the regression parameters to be determined reasonably precisely with the detections alone. The upper limits are distributed in both $D_{25}$ and $L_{\mathrm{B}}$ in a similar way to the detections and the majority lie well above the relations, so they have minimal impact on the regression parameters.

An alternative fitting method was also considered where each term in the likelihood was weighted by $1/V_\mathrm{max}$, similarly to in \citet{Solanes_1996}. This produced the relations $\log M_{\mathrm{HI}} = 0.86 \times 2 \log D_{25}/\mathrm{kpc} + 7.30$ and $\log M_{\mathrm{HI}} = 0.92 \log L_{\mathrm{B}}/\mathrm{L_{\odot}} + 0.37$, both with intrinsic scatters of 0.21 dex. The gradient and intercept parameters are easily within 1-$\sigma$ of the full HI science sample MLE values without the $1/V_\mathrm{max}$ weighting. Therefore, we do not to use $1/V_\mathrm{max}$ in the rest of this paper because it does no appear to have any significant effect.

\begin{table*}
\centering
\caption{Regression fits between $2 \log D_{25}/\mathrm{kpc}$ and $\log M_{\mathrm{HI}}/\mathrm{M_{\odot}}$}
\label{tab:D25_rel}
\begin{tabular}{c c c c c}
\hline\hline
Method & Sample     & Gradient & Intercept & Intrinsic Scatter (dex) \\ \hline
MLE    & All        & 0.86 $\pm$ 0.04     & 7.30 $\pm$ 0.12      & 0.21 $\pm$ 0.01    \\ 
MLE    & Detections & 0.86 $\pm$ 0.04     & 7.32 $\pm$ 0.13      & 0.21 $\pm$ 0.01    \\ 
OLS    & Detections & 0.77 $\pm$ 0.04     & 7.59 $\pm$ 0.10      & -       \\ \hline
\end{tabular}
\tablefoot{MLE indicates the maximum likelihood estimator described in section \ref{sec:reg_mod}, and OLS is the ordinary least squares method. Error estimates were made using jackknife resampling, with the exception of the ordinary least squares fit, which used the standard error estimates.}
\end{table*}

\begin{table*}
\centering
\caption{Regression fits between $\log L_{\mathrm{B}}/\mathrm{L_{\odot}}$ and $\log M_{\mathrm{HI}}/\mathrm{M_{\odot}}$}
\label{tab:LB_rel}
\begin{tabular}{c c c c c}
\hline\hline
Method & Sample     & Gradient & Intercept & Intrinsic Scatter (dex) \\ \hline
MLE    & All        & 0.94 $\pm$ 0.08     & 0.18 $\pm$ 0.80      & 0.20 $\pm$ 0.02    \\ 
MLE    & Detections & 0.92 $\pm$ 0.09     & 0.37 $\pm$ 0.90      & 0.21 $\pm$ 0.02    \\ 
OLS    & Detections & 0.74 $\pm$ 0.04     & 2.25 $\pm$ 0.40      & -       \\ \hline
\end{tabular}
\tablefoot{MLE indicates the maximum likelihood estimator described in section \ref{sec:reg_mod}, and OLS is the ordinary least squares method. Error estimates were made using jackknife resampling, with the exception of the ordinary least squares fit, which used the standard error estimates.}
\end{table*}

The residuals of the relations were compared against various other properties to look for any residual correlations. The residuals of the $D_{25}$-relation showed no correlation with $L_{\mathrm{B}}$ and vice versa, indicating that both of these parameters are a proxy for the same underlying property of the galaxy, its mass. These two sets of residuals both had correlation coefficients of 0.02. There was also minimal residual correlation found with FIR luminosity \citep{Lisenfeld_2007}, with the correlation coefficients being 0.15 and -0.11 for the $D_{25}$ and $L_{\mathrm{B}}$ relations respectively.

There was a slightly stronger suggestion of a residual correlation with morphology, with both relations producing residual correlation coefficients with morphological type of about 0.3.
This residual correlation indicates that some of the intrinsic scatter in the relation is due to differences in morphological type.

\subsection{HI scaling relations for different morphologies}
\label{sec:morph_rel}

\begin{figure}
    \centering
    \includegraphics[width=\columnwidth]{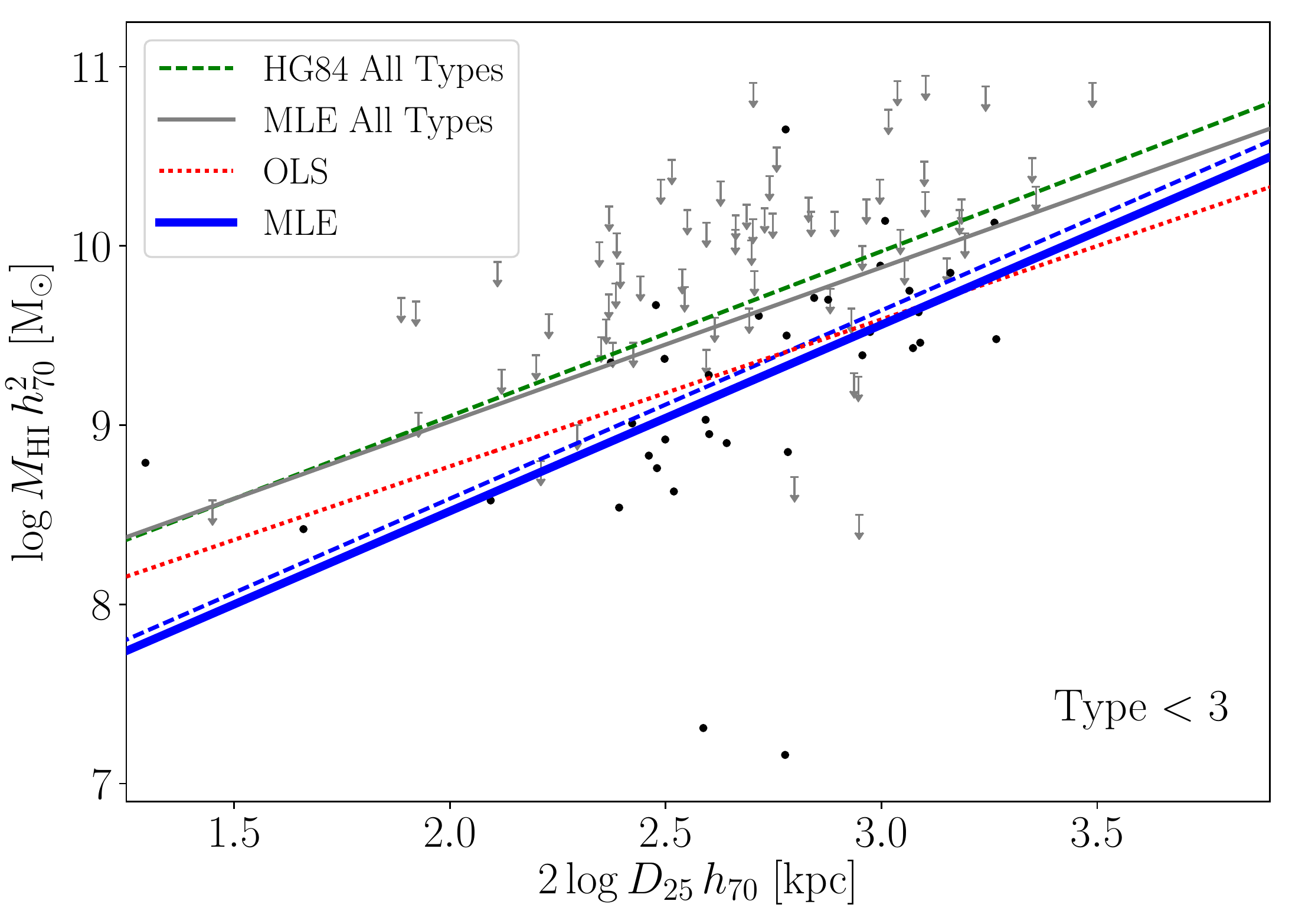}
    \includegraphics[width=\columnwidth]{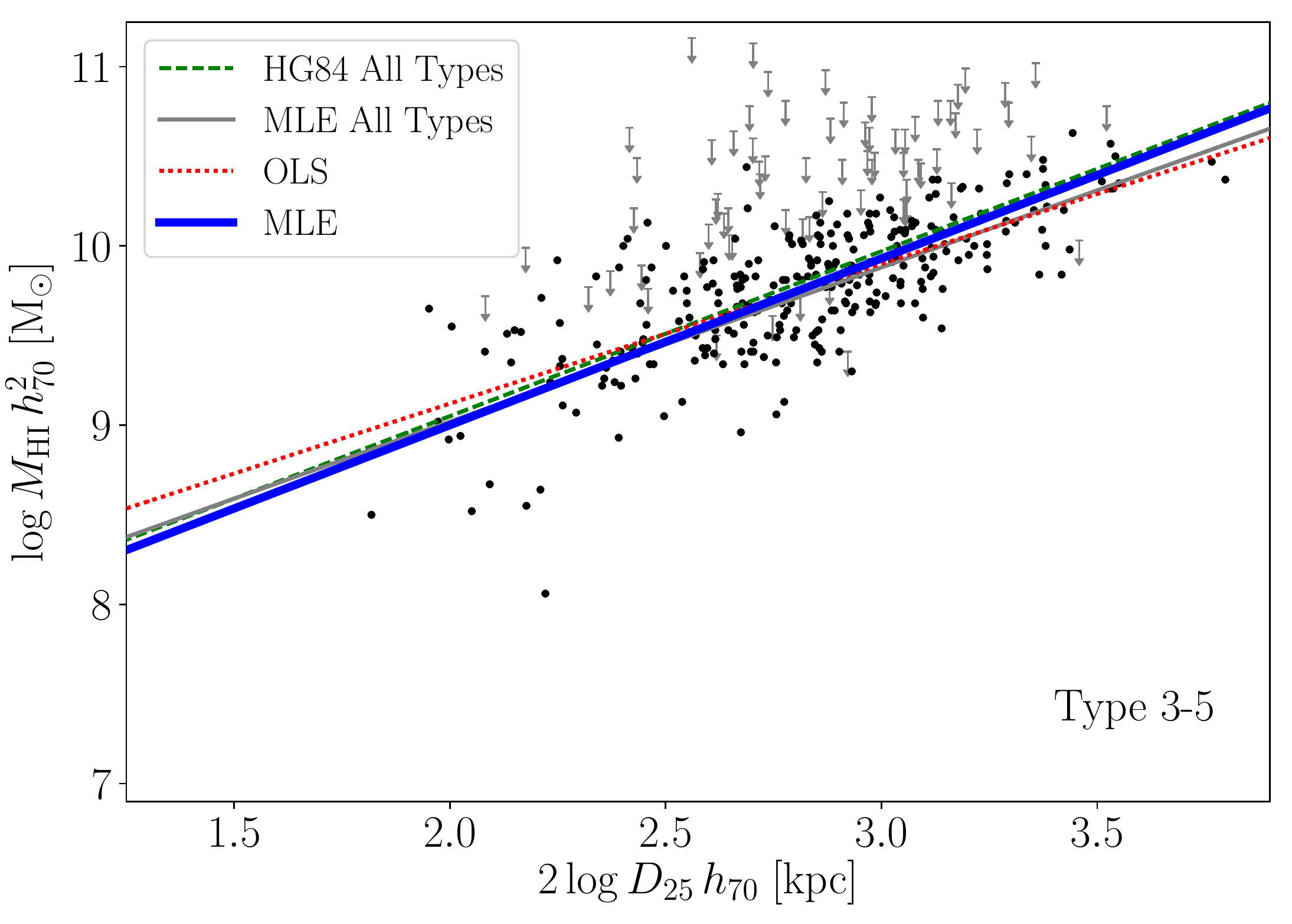}
    \includegraphics[width=\columnwidth]{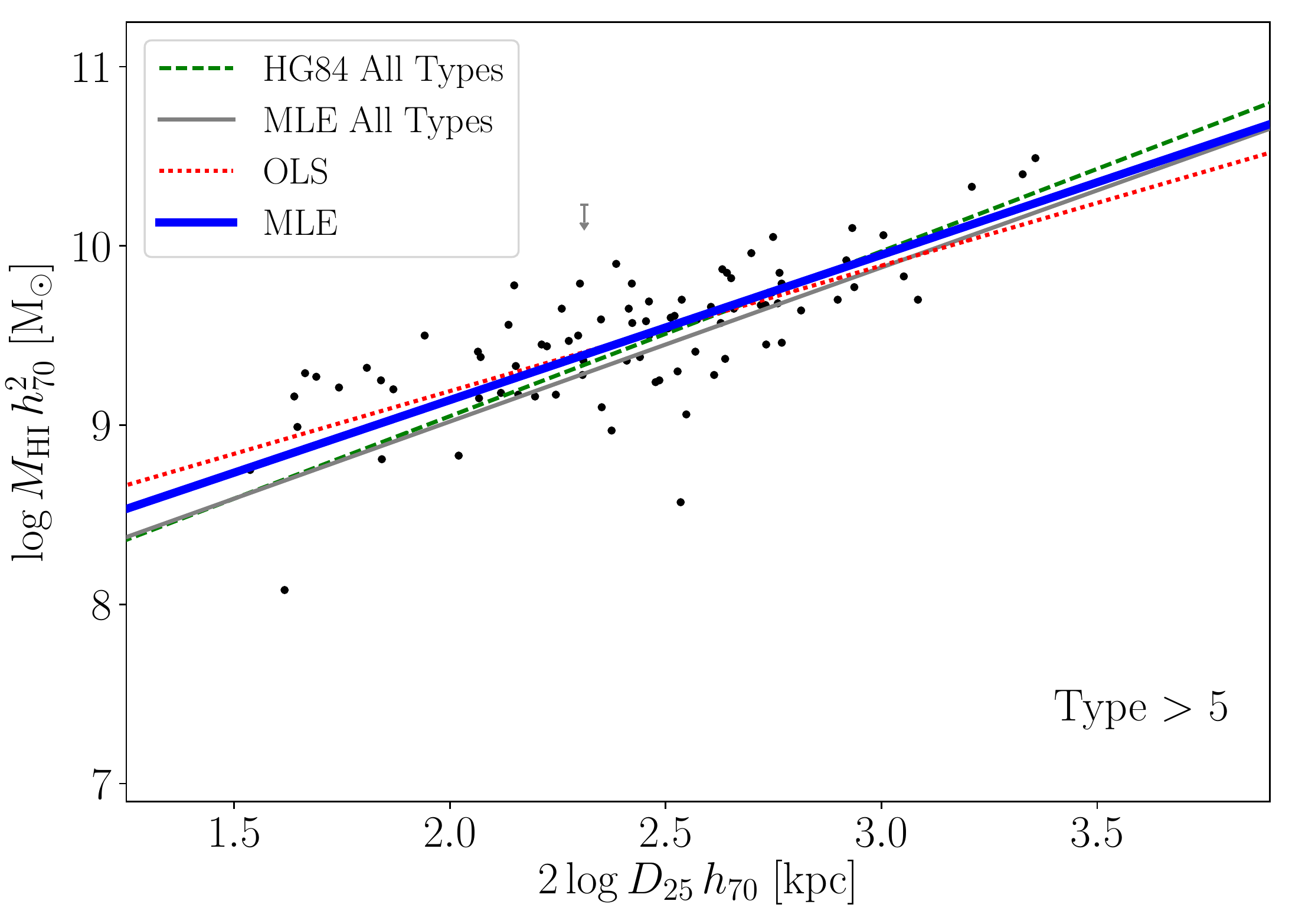}
    \caption{Scaling relations with $D_{25}$ split by morphological type; early and intermediate types (top), late types (middle), very late types (bottom).}
    \label{fig:morph_rel_D25}
\end{figure}

\begin{figure}
    \centering
    \includegraphics[width=\columnwidth]{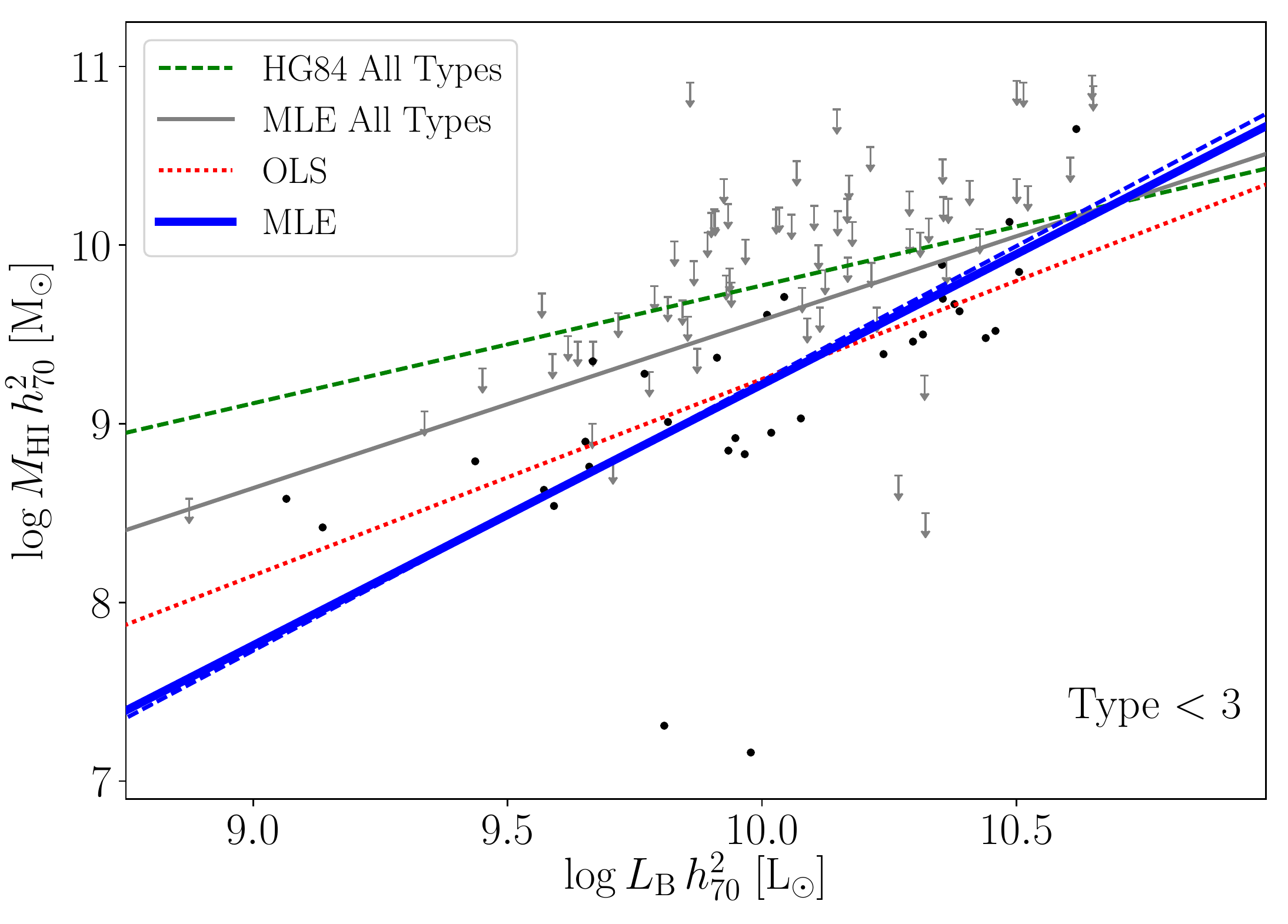}
    \includegraphics[width=\columnwidth]{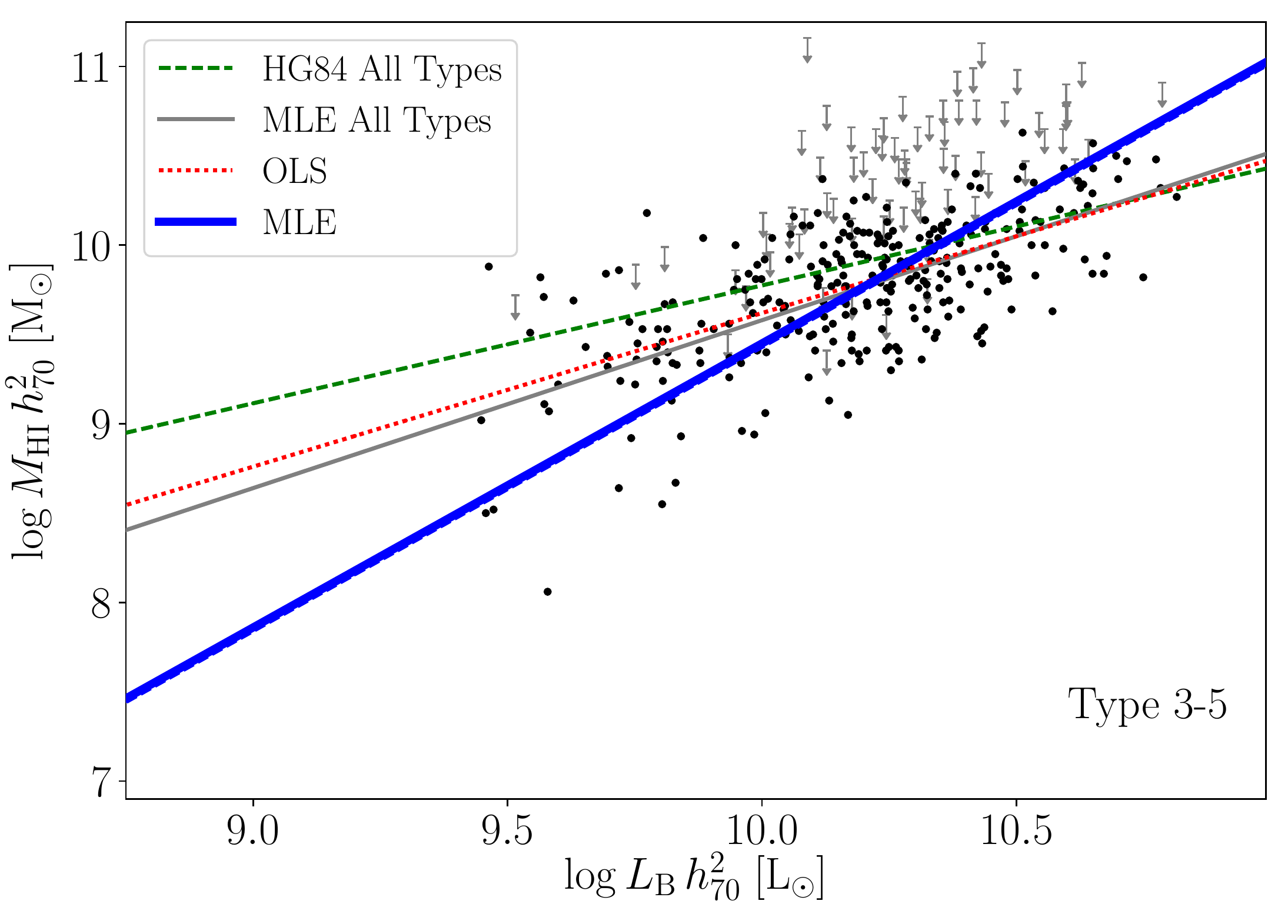}
    \includegraphics[width=\columnwidth]{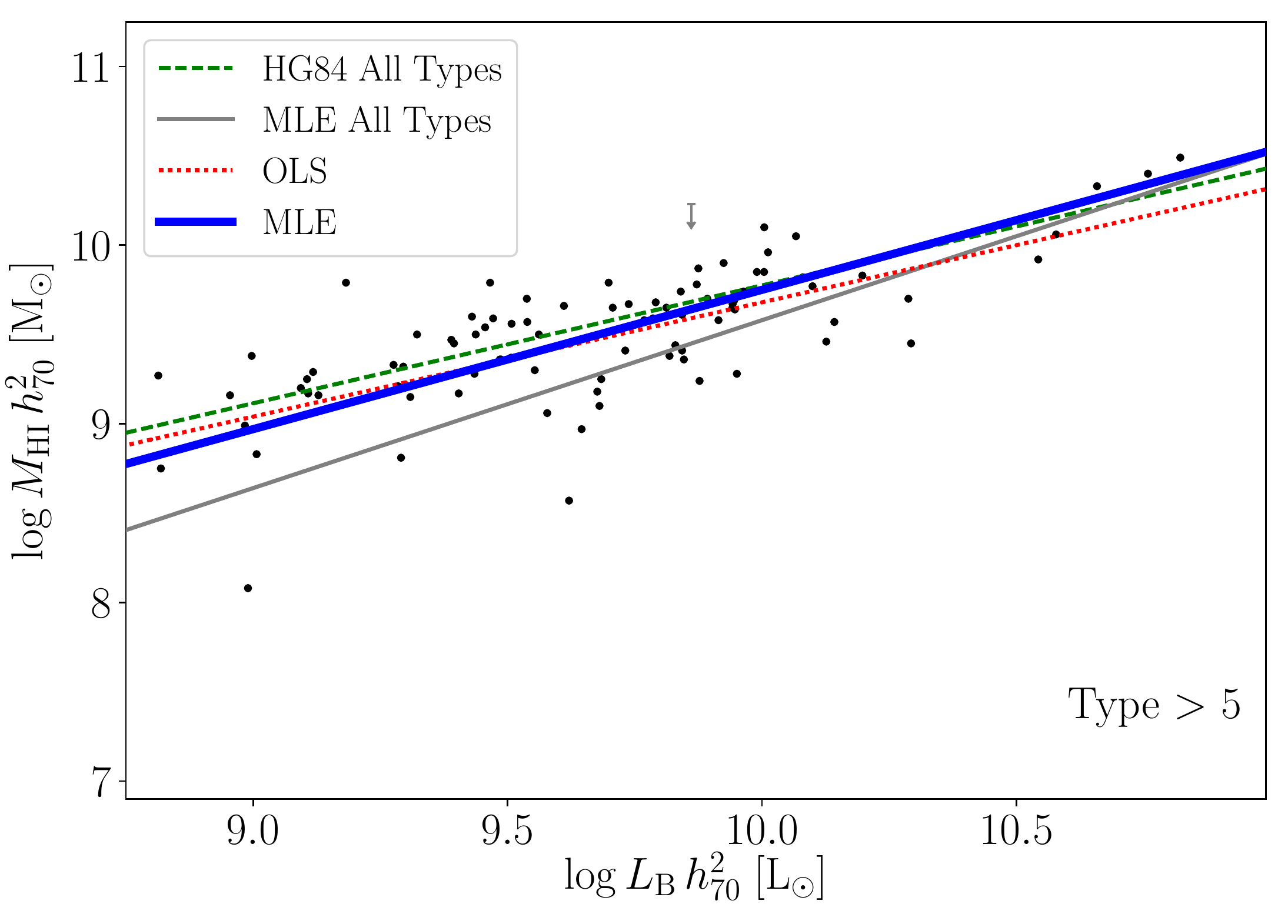}
    \caption{Scaling relations with $L_{\mathrm{B}}$ split by morphological type; early and intermediate types (top), late types (middle), very late types (bottom).}
    \label{fig:morph_rel_LB}
\end{figure}

As morphology is a categorical variable it cannot be included in the regression model in the same manner that the numerical variables are. Ideally scaling relations would be fit individually for each morphological type, however, the currently existing sample of well isolated galaxies simply is not large enough to permit this approach. Therefore, we have split the sample into three bins of morphology that roughly correspond to early and intermediate types ($T<3$, earlier than Sb), the main portion of AMIGA ($3 \leq T \leq 5$, i.e. late types from Sb to Sc), and very late types ($T > 5$, later than Sc). These relations are shown in Figures \ref{fig:morph_rel_D25} and \ref{fig:morph_rel_LB} along with the relation of the full sample (thin grey line) and the \citetalias{Haynes_1984} relation of their full sample (green dashed line). The coefficients of the regression lines are shown in Tables \ref{tab:morph_rel_D25} and \ref{tab:morph_rel_LB}.

As morphological type goes from early to late the gradient of the $D_{25}$-relation (Figure \ref{fig:morph_rel_D25}) changes very little, but the intercept increases by about 1 dex. This change is not unexpected because later types are usually found to be more HI-rich than earlier types. For the $L_{\mathrm{B}}$-relation the effect of morphology is quite different, with the most striking change being the gradient, which varies from $\sim$1.5 for early types and late types, to 0.78 for very late types. The physical interpretation of this is that low-luminosity, late-type galaxies are more HI-rich than low-luminosity early types, which is again consistent with what would be expected.

It should also be noted that although the uncertainties in the intercepts of the relations are extremely large, the intercept and gradient uncertainties are over 99\% negatively correlated (from our jackknife estimates), indicating that the uncertainties in the gradient and intercept should not be considered independently. The translational uncertainties in the $y$-position of the relation lines are considerably smaller than the quoted uncertainties in the intercepts, as these only represent the uncertainties at the origin of the $x$-axis, which lies far outside the range of the data (in both cases). This effect is much more pronounced for the $L_{\mathrm{B}}$-relation, than the $D_{25}$-relation, because the data are considerably further from the origin of the $x$-axis in the chosen units, which causes a greater lever arm effect.

\begin{table*}[t]
\centering
\caption{Relations with $D_{25}$ split by morphological type}
\label{tab:morph_rel_D25}
\begin{tabular}{c c c c c}
\hline\hline
Type & Gradient & Intercept & \begin{tabular}{c}Intrinsic Scatter\\(dex)\end{tabular} & \begin{tabular}{c}Correlation\\coefficient\end{tabular} \\ \hline
<3   & 1.04 $\pm$ 0.21     & 6.44 $\pm$ 0.59      & 0.27 $\pm$ 0.08    & 0.67\\
3-5  & 0.93 $\pm$ 0.06     & 7.14 $\pm$ 0.18      & 0.16 $\pm$ 0.02    & 0.74\\
>5   & 0.81 $\pm$ 0.09     & 7.53 $\pm$ 0.24      & 0.17 $\pm$ 0.03    & 0.73\\ \hline
\end{tabular}
\tablefoot{The relations in this table were all calculated using the maximum likelihood estimate for the detections and upper limits combined. Errors were estimated via jackknifing the sample.}
\end{table*}

\begin{table*}[t]
\centering
\caption{Relations with $L_{\mathrm{B}}$ split by morphological type}
\label{tab:morph_rel_LB}
\begin{tabular}{c c c c c c c}
\hline\hline
Type & Gradient & Intercept & \begin{tabular}{c}Intrinsic Scatter\\(dex)\end{tabular} & \begin{tabular}{c}Correlation\\coefficient\end{tabular} \\ \hline
<3   & 1.46 $\pm$ 0.24     & -5.38 $\pm$ 2.41     & 0.00 $\pm$ n/a     & 0.86\\
3-5  & 1.59 $\pm$ 0.10     & -6.45 $\pm$ 1.05     & 0.00 $\pm$ n/a    & 0.64\\
>5   & 0.78 $\pm$ 0.12     & 1.95 $\pm$ 1.17      & 0.15 $\pm$ 0.06    & 0.73\\ \hline
\end{tabular}
\tablefoot{
The relations in this table were all calculated using the maximum likelihood estimate for the detections and upper limits combined. Errors were estimated via jackknifing the sample. The intrinsic scatter has no error estimates for the first two samples as all jackknife iterations produced intrinsic scatter values less than 0.01.
}
\end{table*}

\section{Discussion}
\label{sec:discuss}

In this section we discuss the interpretation of our results, focusing on two particular aspects: the impact of morphology and comparison with samples in different environments. The morphological dependence is discussed based on the split relations calculated above, although the different morphology distributions of the previous works that we compare with are also discussed. The AMIGA sample represents the most isolated environment, and we compare this with both field and cluster environments. We leave the comparison with compact groups for another paper.

\subsection{Morphological dependence}

As shown in Figures \ref{fig:morph_rel_D25} and \ref{fig:morph_rel_LB} there is a definite dependence of the scaling relations on morphological type. This means that when comparing to a sample of mainly spiral galaxies (dominated by types 3-5, as is the population of isolated galaxies) the single relations shown in Tables \ref{tab:D25_rel} and \ref{tab:LB_rel} are the most appropriate scaling relations to use, but when considering a more morphologically diverse sample using just a single relation will result in biases for the early- and very late-type objects.

The three morphology bin relations (in both $D_{25}$ and $L_{\mathrm{B}}$) can be made into a piece-wise relation to predict the HI mass of galaxies of different morphological types based on either $D_{25}$ or $L_{\mathrm{B}}$.
These piece-wise relations reduce the correlation coefficient between morphological type and the residual HI mass from $\sim$0.3 to -0.05 for the $D_{25}$-relation, and to -0.03 for the $L_{\mathrm{B}}$-relation, indicating that the dependence on morphology has been markedly reduced.

Although these piece-wise relations are somewhat ad hoc because the bins were chosen purely based on the morphology distribution of our sample, we nevertheless recommend they be used when a sample contains early types or very late types to address the bias in the predicted HI content that would otherwise arise. With a substantially larger dataset it would be possible to derive a more robust correction based on fitting the relations for each type, but this is not presently possible for isolated galaxies.

Another point to note is the estimated intrinsic scatter of the relations. As is shown in Table \ref{tab:morph_rel_D25}, all the $D_{25}$-relations, regardless of morphology, have an estimated intrinsic scatter of 0.15-0.3 dex. However, the intrinsic scatter estimates for the $L_{\mathrm{B}}$-relations are all 0.15 dex or below, with the relations for early and late types being consistent with zero intrinsic scatter. While one should not over-interpret this result, it does suggest that if morphology was fully accounted for then the $L_{\mathrm{B}}$-relation may actually be intrinsically tighter than the $D_{25}$-relation. If true, this might imply that if the contribution of the bulge to the overall luminosity was removed (as bulge-to-disc ratio is a key property in defining morphological class) then the disc luminosity could actually be a better predictor of $M_\mathrm{HI}$ than the disc size.

The trends we observe with morphology are somewhat different than those of \citetalias{Haynes_1984}. While we find that the intercept of the $D_{25}$-relation increases with type, they find virtually no change. In the case of the $L_{\mathrm{B}}$-relation we see a definite flattening of the gradient for later types, which is again not evident in \citetalias{Haynes_1984}. When comparing the samples used in the two papers these discrepancies are not surprising because their sample, after non-detections were excluded, included very few early types (see Figure \ref{fig:morph_dist}) and would therefore struggle to illuminate these trends.

\subsection{Comparison with previous scaling relations}
\label{sec:comp_rels}

\begin{table*}
\centering
\caption{The gradients and intercepts of the comparison relations}
\label{tab:comp_rel}
\begin{tabular}{l|cc|cc|}
\hline\hline
\multirow{2}{*}{Reference} & \multicolumn{2}{|c|}{$2 \log D_{25}/\mathrm{kpc}$}             & \multicolumn{2}{|c|}{$\log L_{B}/\mathrm{L_{\odot}}$} \\ \cline{2-5} 
                           & Gradient & \multicolumn{1}{c|}{Intercept} & Gradient      & Intercept      \\ \hline
\citet{Haynes_1984}        & 0.92     & 7.21                           & 0.66          & 3.17\\
\citet{Solanes_1996}       & 0.64     & 8.00                           & -             & -  \\
\citet{Denes_2014}         & 0.64     & 8.21                           & 0.85          & 1.23\\ \hline
\end{tabular}
\end{table*}

Figures \ref{fig:D25_rel} and \ref{fig:LB_rel} include scaling relations from \citet{Haynes_1984}, \citet{Solanes_1996}, and \citet{Denes_2014} for comparison purposes.\footnote{We do not make a comparison with \citet{Toribio_2011} because their relations are based on SDSS r-band properties, which we were unable to reliably relate to our B-band properties.} These relations are based on HI datasets that span a range of low-density environments, but none as isolated as AMIGA. These relations had to be converted to the same unit system used here in order to facilitate a fair comparison. This means that the linear regression coefficients listed in Table \ref{tab:comp_rel} are not the same as in the original sources. Although we leave the details of these conversion for appendix \ref{sec:rel_conv}, it should be noted that this is an essential step, without which our interpretation of the comparisons would change.\footnote{This is also a note of caution that when using the relations calculated in this paper it is essential to ensure the definitions of luminosities, diameters, and morphological types used are equivalent to those of this work.} In particular, it is important to note that the optical properties of the \citetalias{Haynes_1984} sample were taken from the UGC \citep{Nilson_1973}, which means that they were measured by eye not digitally.

The \citetalias{Haynes_1984} sample, like ours, is based on the CIG and therefore the most similar comparison sample. However, despite coming from the same original catalogue that work did not have the refined information on isolation or completeness that AMIGA has, and thus is not quite as isolated or complete. In total they observed 324 CIG galaxies with the Arecibo telescope. Approximately 11\% of their sample was not detected or only marginally detected, this fraction was omitted from the fits of the final scaling relations. Applying our own isolation and completeness criteria to the sources in common between our catalogue and that of \citetalias{Haynes_1984} has very much the same effect as on our own full dataset. This indicates that, as is the case for the full CIG, $\sim$30\% of the \citetalias{Haynes_1984} galaxies were not well isolated according to our definition. The \citetalias{Haynes_1984} scaling relations were also fit just to detections using the OLS method, so would not have accounted for the bias due to the uncertainties in the dependent variable. 

In the $D_{25}$-relation our fit is very similar to that of \citetalias{Haynes_1984}, although our sample appears to be marginally less HI-rich. The $L_{\mathrm{B}}$-relations are more different, with our relation predicting that low mass galaxies are about 0.5 dex poorer in HI than the \citetalias{Haynes_1984} relation. The steeper gradient of our relation is likely in part due to the fact that the uncertainties in $L_{\mathrm{B}}$ are large (see error ellipse in Figure \ref{fig:LB_rel}) and the OLS fitting method used in \citetalias{Haynes_1984} does not account for this. However, this does not appear to be a complete explanation because although our own OLS fit (red dotted line) has a shallower gradient than the main relation (solid blue line), it is not as shallow as the \citetalias{Haynes_1984} relation.

The next relation which we compare to is from \citet{Solanes_1996}. This relation was calculated based on a sample of 532 field spiral galaxies in the direction of the Pisces-Perseus supercluster. A threshold neighbour density was set to ensure that galaxies associated with the many clusters in the region were not selected, but this sample is a sample of field galaxies, not a sample of isolated galaxies, and therefore cannot be considered nearly ``nurture free'' as AMIGA can (see appendix \ref{Solanes_iso_comp}). Their relation with $D_{25}$ is shallower than ours and the other relations, which might suggest that in the environment of this sample there is already a small amount of HI-deficiency for the largest galaxies. The relation also indicates that the average galaxy in the \citet{Solanes_1996} sample are considerably richer in HI than the AMIGA galaxies, and this is especially noticeable for the smallest galaxies.

The final relation that we compare with is also from a field sample, but in this case it is HI-selected rather than optically-selected. \citet{Denes_2014} used the HI detections of the HIPASS catalogue, excluding the galaxies in the 30\% densest environments, to derive scaling relations between optical magnitudes and sizes, and HI mass. This isolation cut essentially only excludes the HI detections that were in relatively high density regions, such as the edge of clusters. The most striking difference between these relations and ours is that the sample they are based on is clearly more HI-rich, which is most apparent in the $D_{25}$-relation (Figure \ref{fig:D25_rel}). This is not surprising as they were drawn from an HI-selected sample, however, it is important to remember that when using a scaling relation one should be conscious of the sample on which it was based and whether it is an appropriate sample to draw comparisons with.

Comparing all these relations against each other and our relations split by morphology illuminates some potential causes for their apparent differences. In Figure \ref{fig:D25_rel} both the \citet{Solanes_1996} and \citet{Denes_2014} relations have shallower slopes than our relation or that of \citetalias{Haynes_1984}. As we have seen that changes in morphology do not appear to strongly alter the slope of this relation, this may be an indication that the difference in this slope is caused by environment, with both the samples of isolated galaxies having a steeper slope than the field samples. While, unsurprisingly, the most HI-rich is the HIPASS sample (an HI-selected sample).

In the case of the $L_{\mathrm{B}}$-relation the \citetalias{Haynes_1984} slope is shallower than that of our relation, with low-mass galaxies being found to be much more HI-rich than we find. However, the average morphological type of the \citetalias{Haynes_1984} sample is later than our sample (see Figure \ref{fig:morph_dist}) and there are almost no sources with early-type morphologies, which would have the effect of flattening the slope of the relation (see Figure \ref{fig:morph_rel_LB}). Unfortunately, as we do not have the morphology distribution of the \citet{Denes_2014} sample we cannot tell if this fits with the same explanation, however, as the gradient is considerably flatter than any we measure, even for very late types, it may be that this is a consequence of the sample being HI-selected rather than being due to morphology.

To check this interpretation of the difference between the slopes of our $L_{\mathrm{B}}$-relation and that of \citetalias{Haynes_1984}, we made OLS fits (the method \citetalias{Haynes_1984} used) to both the \citetalias{Haynes_1984} sources in our dataset and fits directly to their original sample, using their values. The OLS fit to the \citetalias{Haynes_1984} detections in our sample (using our values) has a gradient of 0.64, almost identical the gradient of 0.66 found in the original work. In contrast the OLS fit to all our detections, which increases the number of sources of types S0 or earlier from 8 to 23, has a gradient of 0.79, which is substantially steeper. This suggests that the morphology distribution is indeed the cause of the discrepancy. However, a possible confounding factor is that our sample has no sources below 1500 \kms, which raises the possibility that it may be dwarf galaxies, rather than the morphology distribution, that are causing the \citetalias{Haynes_1984} fit to have a flatter slope. To test this we took the original \citetalias{Haynes_1984} dataset, removed all source with heliocentric velocities below 1500 \kms (46 sources), and then fit the OLS regression line again. Rather than steepening the fit, this flattened it further to 0.57, suggesting that the difference between the \citetalias{Haynes_1984} $L_{\mathrm{B}}$-relation and our own may even be underestimated.

In summary, we find that the \citetalias{Haynes_1984} $L_{\mathrm{B}}$-relation differs from the equivalent relation of this work, in part because our sample contains more detected early-type galaxies and includes even more through upper limits, and also because our fitting method accounts for uncertainties in $L_{\mathrm{B}}$. Both of these effects cause the gradient of the relation to steepen, which results in the final $L_{\mathrm{B}}$-relation being almost 50\% steeper than that of \citetalias{Haynes_1984}. In the case of the $D_{25}$-relation it appears to be quite robust to both of these effects (although the resilience to the latter is expected because the uncertainties are smaller) and our final fit parameters agree within the uncertainties with those of \citetalias{Haynes_1984} (see tables \ref{tab:D25_rel} and \ref{tab:comp_rel}). However, there does appear to be a marginal decrease in the HI-richness of our sample relative to theirs, which is likely explained by the increased number of early types in the sample.

\subsection{Broken scaling relations}

In recent years several works have found trends with breaks in them between galaxies' optical and HI properties \citep[e.g.][]{Catinella_2010,Huang_2012,Maddox_2015}, however, our relations do not appear to show evidence of any break. 

Using the HI-selected population of ALFALFA, \citet{Huang_2012} found that the scaling relation between stellar mass and HI mass has a slope change at a stellar mass of about $10^{9}$ \Msol, and this result was confirmed in \citet{Maddox_2015} via a similar analysis of the ALFALFA dataset. The fact that such a break in the scaling relation is not apparent in our dataset does not indicate a conflict with these results because the break is expected to occur at $M_{\ast} \sim 10^{9}$ \Msol \ and our dataset begins at $L_{\mathrm{B}} \sim 10^{9}$ \Lsol (although stellar mass and B-band luminosity are not directly equivalent they should be the same order of magnitude). This reiterates the point that the AMIGA science sample does not contain a population of dwarf galaxies and the relations of this paper are likely inappropriate for such a population.

GASS \citep[GALEX Arecibo SDSS Survey,][]{Catinella_2010} studied the scaling relation of gas fraction ($M_{\mathrm{HI}}/M_{\ast}$) with stellar mass surface density, of high stellar mass galaxies. They found that there is a break occurring at a stellar mass surface density of $10^{8.5} \; \mathrm{M_{\odot}\, kpc^{-2}}$, with galaxies of higher surface densities having sharply reduced HI content. Using our B-band properties as a very rough proxy for stellar mass surface density, we find that the typical AMIGA galaxy has surface density $4 L_{\mathrm{B}}/\pi D^{2}_{25} \sim 10^{7.5} \; \mathrm{L_{\odot}\, kpc^{-2}}$ and none are above $10^{8.5} \; \mathrm{L_{\odot}\, kpc^{-2}}$. Therefore, we again find that our sample does not extend into the range where this break is relevant.

\begin{figure*}
    \centering
    \includegraphics[width=\textwidth]{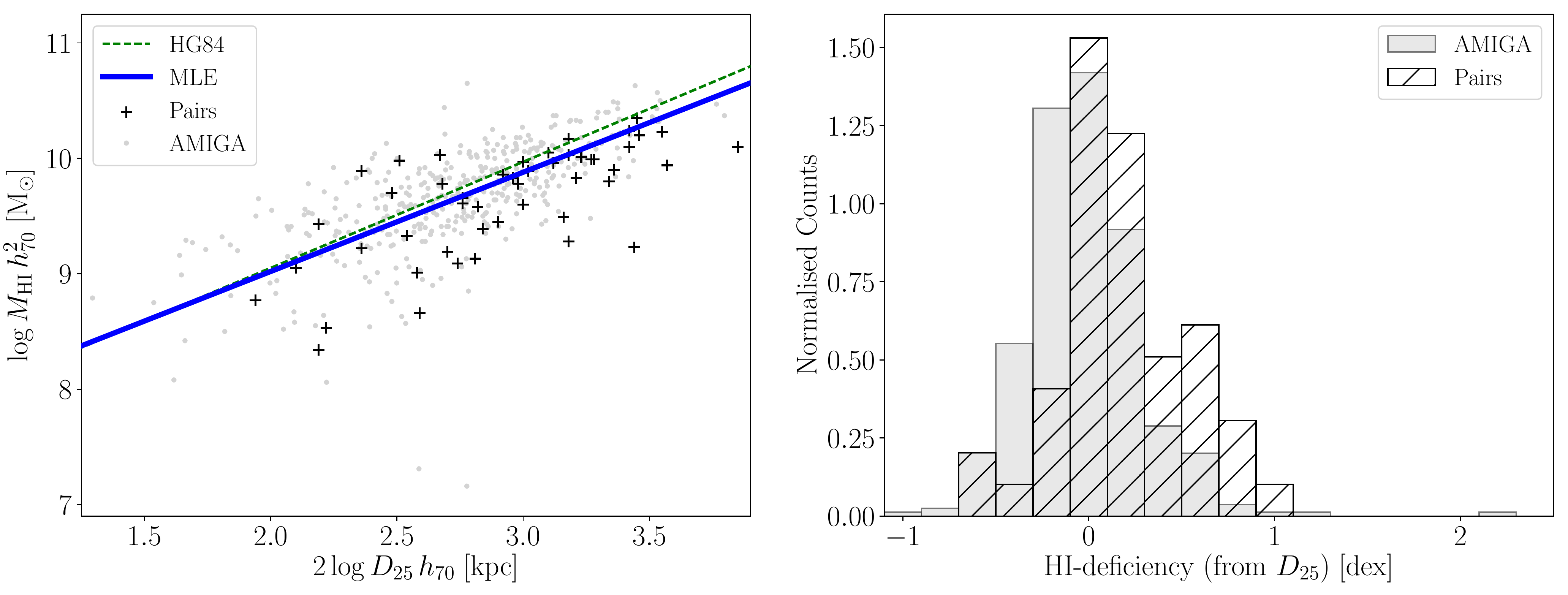}
    \caption{\textit{Left}: Scatter plot of the HI masses of isolated galaxy pairs as a function of their optical diameters \citep[data from][]{Zasov_1994}, shown with black crosses. The light grey points in the background are the HI detections of the AMIGA HI science sample. The solid blue line shows the MLE regression fit of this work and the dashed green line shows the \citetalias{Haynes_1984} relation. \textit{Right}: HI-deficiency of isolated galaxy pairs (diagonal hatching) compared to the AMIGA HI science sample (light grey). HI-deficiency here is calculated with the $D_{25}$-relation (without use of morphological type).}
    \label{fig:pairs_reg}
\end{figure*}

\begin{figure*}
    \centering
    \includegraphics[width=\textwidth]{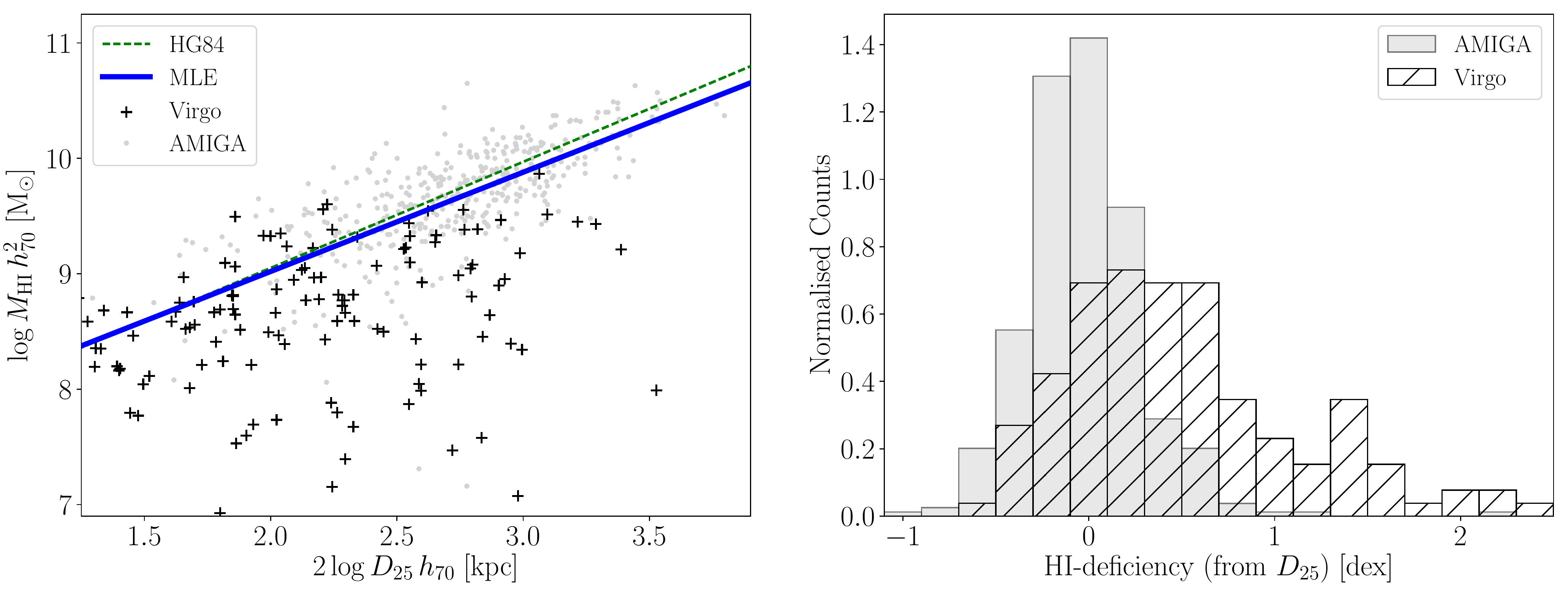}
    \includegraphics[width=\textwidth]{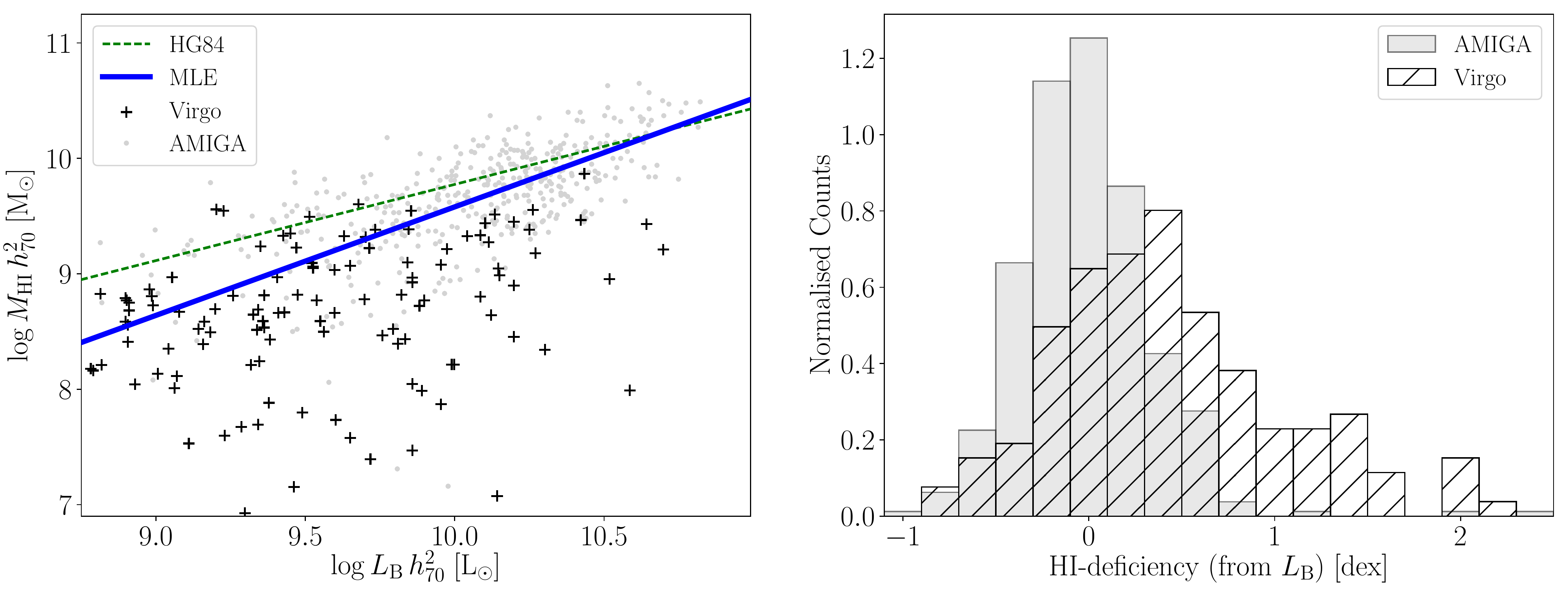}
    \caption{\textit{Left column}: Scatter plots of the HI masses of Virgo cluster galaxies from the VCC, shown with black crosses. The light grey points in the background are the HI detections of the AMIGA HI science sample. The solid blue line shows the MLE regression fit of this work and the dashed green line shows the \citetalias{Haynes_1984} relation. \textit{Right column}: HI-deficiency of Virgo cluster galaxies from the VCC (diagonal hatching) compared to the AMIGA HI science sample (light grey). \textit{Top row}: Here the samples are compared based on $D_{25}$. \textit{Bottom row}: Here the samples are compared based on $L_{\mathrm{B}}$.}
    \label{fig:HIdef_comp}
\end{figure*}

\subsection{Comparison with isolated pairs}
\label{sec:comp_pairs}

To facilitate a comparison with galaxies that are not entirely isolated, but also not field objects, we used the dataset of \citet{Zasov_1994} that was extracted from the Catalog of Isolated Pairs of Galaxies \citep[CPG,][]{Karachentsev_1972}. The Zasov sample was selected to be pairs consisting of one early and one late-type galaxy. This was done such that the detected HI emission (typically with a spatial resolution that cannot separate the two galaxies) can be assumed to originate entirely from one component, the late-type galaxy. While this assumption is not entirely correct as there are HI-rich early types \citep[e.g.][]{Serra_2012}, in the vast majority of cases the late-type galaxy is expected to contain orders of magnitude more HI than the early-type galaxy. With this assumption the optical properties of the late type in each pair can then be compared with the HI content, and in turn contrasted with the relations of isolated galaxies.

The left panel of Figure \ref{fig:pairs_reg} shows the data from \citet{Zasov_1994} for the isolated pairs compared to the data and regression fit of the AMIGA HI science sample, as well as the \citetalias{Haynes_1984} fit shown as in previous plots. The right panel shows the HI-deficiency of the galaxy pairs (calculated with the $D_{25}$-relation) along with the AMIGA HI-deficiencies. This comparison should be treated with caution because not only is the pairs dataset small, we were also only able to make a conversion for the different Hubble constants used, as there is no overlap with our dataset, so more detailed calibration (as was done for the previous comparisons) was not possible. Therefore, the reader should take under consideration that the exact positions of the points are somewhat uncertain in our unit system even though we compare them directly in Figure \ref{fig:pairs_reg}.

While the distribution of HI-deficiency of the isolated pairs is peaked at zero, the wing towards positive HI-deficiencies is more heavily populated than for AMIGA. This results in the mean HI-deficiency being 0.2 dex, indicating that a minor amount of HI has been removed from these galaxies. A correlation between the HI-deficiency of the pair and the pair separation was investigated, but none was evident. Instead it appears to be the largest galaxies that are causing the pairs distribution to be slightly HI-deficient, as is apparent in the left panel of Figure \ref{fig:pairs_reg} because most of the paired galaxies with diameters above $\sim$30 kpc fall below the regression line.

\subsection{HI-deficiency of Virgo cluster galaxies}

To contrast the HI content of the AMIGA galaxies with a cluster environment, Virgo cluster galaxies with HI measurements were obtained from HyperLeda. As the optical fluxes and diameters in our own compilation were collected from HyperLeda this ensured that at least the optical scales are directly comparable. The Virgo region is extremely complicated, so we also only selected sources which had redshift-independent distance measures placing them at less than 40 Mpc, and were identified as Virgo cluster members in the VCC \cite[Virgo Cluster Catalog,][]{Binggeli_1985}. These criteria produced a sample of 132 Virgo galaxies for comparison. Unfortunately the morphological types of these sources, with an equivalent definition was not available, and so they could only be compared against the global relations.

Figure \ref{fig:HIdef_comp} show the HI-deficiency of HI-detected galaxies in the Virgo cluster, with their deficiencies calculated by both the $D_{25}$ and $L_{\mathrm{B}}$ relations. We see that although the typical galaxy is only deficient by a factor of $\sim$2 (or 0.3 dex) the distribution of HI-deficiencies in Virgo is highly skewed, with the high HI-deficiency tail of the distribution extended approximately an order of magnitude beyond that of the AMIGA sample. Furthermore, it should be noted that because the selection criteria require the galaxies to be detected in HI, the true level of HI-deficiency is likely to be higher than is shown here. Curiously, however, the outlying isolated galaxies that were excluded from the regression analysis have HI-deficiencies very similar to the most extremely deficient Virgo galaxies that are still detected in HI.

These results are generally consistent with previous studies of HI-deficiency in the Virgo cluster \citep[e.g.][]{Huchtmeier_1989,Solanes_2001}, which also find a typical deficiencies to be a factor of $\sim$2, but also detected galaxies that are apparently missing over 90\% of their initial HI content. VIVA \citep[VLA Imaging survey of Virgo galaxies in Atomic gas,][]{Chung_2009} finds a mean HI-deficiency of about 0.5 dex, based on the $D_{25}$-relation of \citetalias{Haynes_1984}. This greater level of deficiency is likely because this sample was selected optically and then observed in HI, rather than requiring a prior HI detection (as we have here). However, the most extreme galaxies are also found to have HI-deficiencies of about 2 dex. Finally, \citet{Cortese_2011} demonstrated that galaxies in Virgo are also HI-deficient for their stellar masses, but that the level of deficiency (anti-)correlates with other factors, such as stellar surface density.

The left panels of Figure \ref{fig:HIdef_comp} also indicate an apparent advantage of our relations over those of \citetalias{Haynes_1984} (aside from those discussed earlier). It is straightforward to see that the \citetalias{Haynes_1984} $D_{25}$-relation would produce almost identical HI-deficiencies to our relation because the two lines fall in almost the same place in the top left panel. However, in the bottom left panel the \citetalias{Haynes_1984} $L_{\mathrm{B}}$-relation would measure significantly higher levels of HI-deficiency in the least luminous galaxies compared to their $D_{25}$-relation. This shows that the two relations we have calculated produce more self-consistent measures of HI-deficiencies.

\begin{figure}
    \centering
    \includegraphics[width=\columnwidth]{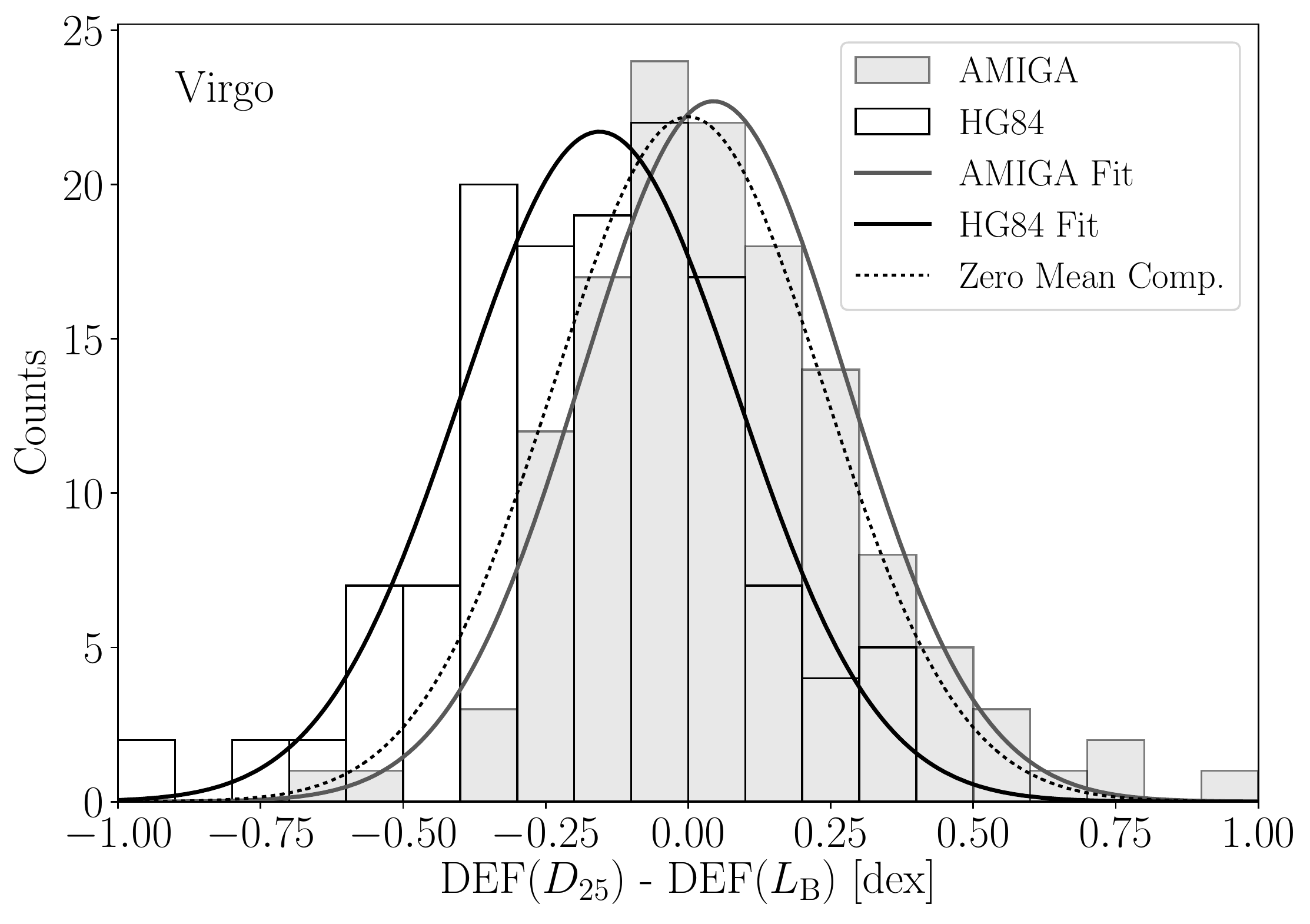}
    \caption{Comparison of the differences between the HI-deficiencies of Virgo cluster galaxies derived using the $D_{25}$ and $L_{\mathrm{B}}$ relations of this work (grey bars) and the differences using the \citetalias{Haynes_1984} relations (unfilled bars). The Gaussian fit to our values is shown with the grey solid line, the fit to the \citetalias{Haynes_1984} values is shown with the solid black line. A third Gaussian with zero mean and a width taken to be the average of the two fits, is shown for comparison by the dotted line.}
    \label{fig:HIdef_diff_comp}
\end{figure}

This result is shown more clearly in Figure \ref{fig:HIdef_diff_comp} which compares the difference between the HI-deficiency of each galaxy calculated using either the $D_{25}$ or the $L_{\mathrm{B}}$-relation, for both our relations and those of \citetalias{Haynes_1984}. The \citetalias{Haynes_1984} values are clearly offset to the left indicating that their $L_{\mathrm{B}}$-relation produces larger estimates of HI-deficiency than their $D_{25}$-relation. The mean value of the distribution is -0.16 dex, which is an 8-$\sigma$ deviation from zero based on the standard error in the mean. Our relations also result in an offset, though much smaller and in the opposite direction. The mean of the distribution is 0.04 dex, a 2-$\sigma$ deviation from zero. The width of the Gaussian fit to our values is 0.23 dex, which is an estimate of the 1-$\sigma$ random uncertainty in the value of HI-deficiency, and is also in agreement with the value of the intrinsic scatter fit in the original relations in section \ref{sec:HI_rels}. The deviation of the mean from zero indicates that there is also a systematic error in the estimates of HI-deficiency, of magnitude $\sim$0.05 dex.

\section{Summary}
\label{sec:summary}

We have compiled a database of the global HI properties of 844 isolated galaxies (from the CIG) using our own single-dish observations and spectra from the literature, and a uniform method of profile characterisation. The large size and uniform nature of this dataset has allowed completeness and isolation cuts to be made while still retaining enough sources to perform a statistical analysis. Therefore, our final HI science dataset of 544 galaxies is not simply larger than previous HI datasets of isolated galaxies, it is also more complete and the galaxies are more isolated.

This dataset was used to measure scaling relations between HI mass and optical properties, in order to set an up-to-date baseline of the HI content of galaxies. As the AMIGA project has shown, these galaxies are isolated and represent, as near as possible, a ``nurture free'' sample that has been isolated on average for at least 3 Gyr \citep{Verdes_Montenegro_2005}. Thus, these scaling relations are applicable to evolutionary scenarios addressing the impact of ``nature'' versus ``nurture'' on the neutral gas of a galaxy. 

The regression model used to fit these relations is also more sophisticated than those of previous studies, incorporating measurement uncertainty in both variables, correlated distance errors, and upper limits from non-detections. We find that a galaxy's HI mass is related to either its B-band luminosity or diameter with an intrinsic scatter of about 0.2 dex. With the inclusion of measurement uncertainties this means that the expected HI mass of an individual galaxy (in the absence of interactions) can be predicted with an accuracy of about 0.25 dex. This accuracy is very similar to that found by \citetalias{Haynes_1984}, however, as described throughout the paper, our relations make numerous improvements, including increasing the number of sources, particularly for early-type morphologies, and incorporating HI upper limits and realistic uncertainties into the regression analysis.

Morphology is found to be an important covariate accounting for some of the intrinsic scatter. The trend with morphology indicates that at a given optical size or luminosity, later type galaxies are more HI-rich, and that this difference is most pronounced for low-luminosity galaxies. However, this effect manifests as a simple offset in the optical diameter scaling relations, but as a change to the gradient of luminosity scaling relations. These trends were not apparent in \citetalias{Haynes_1984} due to the small number of detected early-type galaxies.

Our relations were found to differ slightly from those in the literature, but in ways that likely have straightforward explanations. Previous samples were generally even more rich in late types than our sample, which led to those relations being slightly more HI-rich overall. This later average type was either an intentional selection to avoid contamination from galaxies in higher density regions or due to selection effects, but either way meant that the samples were a biased selection of isolated galaxies. Previous relations also typically had shallower gradients which can in part be attributed to the later morphologies, as well as to the lack of accounting for uncertainties in the dependent variable. 

When contrasted with a cluster population from the VCC, we found that although the typical Virgo cluster galaxy was only HI-deficient by a factor of about 2, the tail of the distribution extended to more than an order of magnitude past that for isolated galaxies. Indicating that some cluster galaxies have lost $\sim$90\% of their HI gas. This comparison also revealed that the relations of this paper produce more consistent measures of HI-deficiency when estimated using either the optical diameter or optical luminosity, than the existing relations for isolated galaxies.

In conclusion, to predict the expected HI mass (in the absence of interaction) of a galaxy we recommend either the optical diameter ($D_{25}$) relation 
\begin{equation}
\log M_{\mathrm{HI}}\,h_{70}^{2}/\mathrm{M_{\odot}} = 0.86\log D_{25}^{2}\,h_{70}^{2}/\mathrm{kpc}^{2} + 7.30,
\end{equation}
or the B-band luminosity ($L_{\mathrm{B}}$) relation
\begin{equation}
\log M_{\mathrm{HI}}\,h_{70}^{2}/\mathrm{M_{\odot}} = 0.94\log L_{\mathrm{B}}\,h_{70}^{2}/\mathrm{L_{\odot}} + 0.18,
\end{equation}
should be used in cases where they is no morphology information, or where the sample is dominated by Sb-Sc galaxies. However, as morphology is a strong covariate, if the sample has morphological types that fall well outside this range it will lead to biases in the prediction of their HI content. In such cases it is recommended to use the piece-wise relations listed in Tables \ref{tab:morph_rel_D25} and \ref{tab:morph_rel_LB}.

\begin{acknowledgements}
We thanks E. Battaner, J. Vilchez, E. Perez, and S. Verley for their useful comments, and to staff members of the different telescopes from which data are presented in this paper, especially to those where we have observed: Arecibo, Effelsberg, Nan\c{c}ay, and GBT. We also thank J. R. Fisher for his assistance in obtaining GBT spectra.
MGJ and LVM acknowledge support from the grant AYA2015-65973-C3-1-R (MINECO/FEDER, UE).
DE was supported by a Marie Curie International Fellowship during this work (MOIF-CT-2006-40298) within the 6th European Community Framework Programme. 
This work is supported by DGI Grant AYA 2008-06181-C02 and the Junta de Andaluc\'{i}a (Spain) P08-FQM-4205.
UL acknowledges support by the research projects AYA2014-53506-P from the Spanish Ministerio de Econom\'{i}a y Competitividad, from the European Regional Development Funds (FEDER)and the Junta de Andaluc\'{i}a (Spain) grants FQM108. 
JS is grateful for support from the UK STFC via grant ST/M001229/1. 
DEJ acknowledges support from the National Science Foundation under Grant DMS-1127914 to the Statistical and Applied Mathematical Sciences Institute. This work also received support from the Junta de Andaluc\'{i}a (Spain) grant TIC-114.
This research has made use of the NASA/IPAC Extragalactic Database (NED), which is operated by the Jet Propulsion Laboratory, California Institute of Technology, under contract with the National Aeronautics and Space Administration. We also acknowledge the use of the HyperLeda database.
This research has made use of the SDSS. Funding for SDSS-III has been provided by the Alfred P. Sloan Foundation, the Participating Institutions, the National Science Foundation, and the U.S. Department of Energy Office of Science. The SDSS-III web site is \url{http://www.sdss3.org/}.
\end{acknowledgements}

\bibliographystyle{aa}
\bibliography{refs}

\begin{thebibliography}{96}
\expandafter\ifx\csname natexlab\endcsname\relax\def\natexlab#1{#1}\fi

\bibitem[{{Abramson} {et~al.}(2011){Abramson}, {Kenney}, {Crowl}, {Chung}, {van
  Gorkom}, {Vollmer}, \& {Schiminovich}}]{Abramson_2011}
{Abramson}, A., {Kenney}, J.~D.~P., {Crowl}, H.~H., {et~al.} 2011, \aj, 141,
  164

\bibitem[{{Ahn} {et~al.}(2012){Ahn}, {Alexandroff}, {Allende Prieto},
  {Anderson}, {Anderton}, {Andrews}, {Aubourg}, {Bailey}, {Balbinot}, {Barnes},
  \& et~al.}]{Ahn_2012}
{Ahn}, C.~P., {Alexandroff}, R., {Allende Prieto}, C., {et~al.} 2012, \apjs,
  203, 21

\bibitem[{{Argudo-Fern{\'a}ndez} {et~al.}(2013){Argudo-Fern{\'a}ndez},
  {Verley}, {Bergond}, {Sulentic}, {Sabater}, {Fern{\'a}ndez Lorenzo}, {Leon},
  {Espada}, {Verdes-Montenegro}, {Santander-Vela}, {Ruiz}, \&
  {S{\'a}nchez-Exp{\'o}sito}}]{Argudo_Fernandez_2013}
{Argudo-Fern{\'a}ndez}, M., {Verley}, S., {Bergond}, G., {et~al.} 2013, \aap,
  560, A9

\bibitem[{{Balkowski} \& {Chamaraux}(1981)}]{Balkowski_1981}
{Balkowski}, C. \& {Chamaraux}, P. 1981, \aap, 97, 223

\bibitem[{{Barnes} {et~al.}(2001){Barnes}, {Staveley-Smith}, {de Blok},
  {Oosterloo}, {Stewart}, {Wright}, {Banks}, {Bhathal}, {Boyce}, {Calabretta},
  {Disney}, {Drinkwater}, {Ekers}, {Freeman}, {Gibson}, {Green}, {Haynes}, {te
  Lintel Hekkert}, {Henning}, {Jerjen}, {Juraszek}, {Kesteven}, {Kilborn},
  {Knezek}, {Koribalski}, {Kraan-Korteweg}, {Malin}, {Marquarding}, {Minchin},
  {Mould}, {Price}, {Putman}, {Ryder}, {Sadler}, {Schr{\"o}der}, {Stootman},
  {Webster}, {Wilson}, \& {Ye}}]{Barnes_2001}
{Barnes}, D.~G., {Staveley-Smith}, L., {de Blok}, W.~J.~G., {et~al.} 2001,
  \mnras, 322, 486

\bibitem[{{Bicay} \& {Giovanelli}(1986)}]{Bicay_1986}
{Bicay}, M.~D. \& {Giovanelli}, R. 1986, \aj, 91, 705

\bibitem[{{Binggeli} {et~al.}(1985){Binggeli}, {Sandage}, \&
  {Tammann}}]{Binggeli_1985}
{Binggeli}, B., {Sandage}, A., \& {Tammann}, G.~A. 1985, \aj, 90, 1681

\bibitem[{{Borthakur} {et~al.}(2015){Borthakur}, {Yun}, {Verdes-Montenegro},
  {Heckman}, {Zhu}, \& {Braatz}}]{Borthakur_2015}
{Borthakur}, S., {Yun}, M.~S., {Verdes-Montenegro}, L., {et~al.} 2015, \apj,
  812, 78

\bibitem[{{Bothun} {et~al.}(1985){Bothun}, {Beers}, {Mould}, \&
  {Huchra}}]{Bothun_1985}
{Bothun}, G.~D., {Beers}, T.~C., {Mould}, J.~R., \& {Huchra}, J.~P. 1985, \aj,
  90, 2487

\bibitem[{{Bottinelli} {et~al.}(1993){Bottinelli}, {Durand}, {Fouque},
  {Garnier}, {Gouguenheim}, {Loulergue}, {Paturel}, {Petit}, \&
  {Teerikorpi}}]{Bottinelli_1993}
{Bottinelli}, L., {Durand}, N., {Fouque}, P., {et~al.} 1993, \aaps, 102, 57

\bibitem[{{Bottinelli} {et~al.}(1992){Bottinelli}, {Durand}, {Fouque},
  {Garnier}, {Gouguenheim}, {Paturel}, \& {Teerikorpi}}]{Bottinelli_1992}
{Bottinelli}, L., {Durand}, N., {Fouque}, P., {et~al.} 1992, \aaps, 93, 173

\bibitem[{{Bradford} {et~al.}(2015){Bradford}, {Geha}, \&
  {Blanton}}]{Bradford_2015}
{Bradford}, J.~D., {Geha}, M.~C., \& {Blanton}, M.~R. 2015, \apj, 809, 146

\bibitem[{{Brown} {et~al.}(2017){Brown}, {Catinella}, {Cortese}, {Lagos},
  {Dav{\'e}}, {Kilborn}, {Haynes}, {Giovanelli}, \&
  {Rafieferantsoa}}]{Brown_2017}
{Brown}, T., {Catinella}, B., {Cortese}, L., {et~al.} 2017, \mnras, 466, 1275

\bibitem[{{Catinella} {et~al.}(2010){Catinella}, {Schiminovich}, {Kauffmann},
  {Fabello}, {Wang}, {Hummels}, {Lemonias}, {Moran}, {Wu}, {Giovanelli},
  {Haynes}, {Heckman}, {Basu-Zych}, {Blanton}, {Brinchmann}, {Budav{\'a}ri},
  {Gon{\c c}alves}, {Johnson}, {Kennicutt}, {Madore}, {Martin}, {Rich},
  {Tacconi}, {Thilker}, {Wild}, \& {Wyder}}]{Catinella_2010}
{Catinella}, B., {Schiminovich}, D., {Kauffmann}, G., {et~al.} 2010, \mnras,
  403, 683

\bibitem[{{Chung} {et~al.}(2009){Chung}, {van Gorkom}, {Kenney}, {Crowl}, \&
  {Vollmer}}]{Chung_2009}
{Chung}, A., {van Gorkom}, J.~H., {Kenney}, J.~D.~P., {Crowl}, H., \&
  {Vollmer}, B. 2009, \aj, 138, 1741

\bibitem[{{Cortese} {et~al.}(2011){Cortese}, {Catinella}, {Boissier},
  {Boselli}, \& {Heinis}}]{Cortese_2011}
{Cortese}, L., {Catinella}, B., {Boissier}, S., {Boselli}, A., \& {Heinis}, S.
  2011, \mnras, 415, 1797

\bibitem[{{Courtois} {et~al.}(2009){Courtois}, {Tully}, {Fisher}, {Bonhomme},
  {Zavodny}, \& {Barnes}}]{Courtois_2009}
{Courtois}, H.~M., {Tully}, R.~B., {Fisher}, J.~R., {et~al.} 2009, \aj, 138,
  1938

\bibitem[{{Delhaize} {et~al.}(2013){Delhaize}, {Meyer}, {Staveley-Smith}, \&
  {Boyle}}]{Delhaize_2013}
{Delhaize}, J., {Meyer}, M.~J., {Staveley-Smith}, L., \& {Boyle}, B.~J. 2013,
  \mnras, 433, 1398

\bibitem[{{D{\'e}nes} {et~al.}(2014){D{\'e}nes}, {Kilborn}, \&
  {Koribalski}}]{Denes_2014}
{D{\'e}nes}, H., {Kilborn}, V.~A., \& {Koribalski}, B.~S. 2014, \mnras, 444,
  667

\bibitem[{{Elson} {et~al.}(2016){Elson}, {Blyth}, \& {Baker}}]{Elson_2016}
{Elson}, E.~C., {Blyth}, S.~L., \& {Baker}, A.~J. 2016, \mnras, 460, 4366

\bibitem[{{Espada} {et~al.}(2011){Espada}, {Verdes-Montenegro}, {Huchtmeier},
  {Sulentic}, {Verley}, {Leon}, \& {Sabater}}]{Espada_2011}
{Espada}, D., {Verdes-Montenegro}, L., {Huchtmeier}, W.~K., {et~al.} 2011,
  \aap, 532, A117

\bibitem[{{Fern{\'a}ndez Lorenzo} {et~al.}(2012){Fern{\'a}ndez Lorenzo},
  {Sulentic}, {Verdes-Montenegro}, {Ruiz}, {Sabater}, \&
  {S{\'a}nchez}}]{Fernandez_Lorenzo_2012}
{Fern{\'a}ndez Lorenzo}, M., {Sulentic}, J., {Verdes-Montenegro}, L., {et~al.}
  2012, \aap, 540, A47

\bibitem[{{Fouqu{\'e}} {et~al.}(1990){Fouqu{\'e}}, {Durand}, {Bottinelli},
  {Gouguenheim}, \& {Paturel}}]{Fouque_1990}
{Fouqu{\'e}}, P., {Durand}, N., {Bottinelli}, L., {Gouguenheim}, L., \&
  {Paturel}, G. 1990, \aaps, 86, 473

\bibitem[{{Giovanelli} {et~al.}(2005){Giovanelli}, {Haynes}, {Kent},
  {Perillat}, {Saintonge}, {Brosch}, {Catinella}, {Hoffman}, {Stierwalt},
  {Spekkens}, {Lerner}, {Masters}, {Momjian}, {Rosenberg}, {Springob},
  {Boselli}, {Charmandaris}, {Darling}, {Davies}, {Garcia Lambas}, {Gavazzi},
  {Giovanardi}, {Hardy}, {Hunt}, {Iovino}, {Karachentsev}, {Karachentseva},
  {Koopmann}, {Marinoni}, {Minchin}, {Muller}, {Putman}, {Pantoja}, {Salzer},
  {Scodeggio}, {Skillman}, {Solanes}, {Valotto}, {van Driel}, \& {van
  Zee}}]{Giovanelli_2005}
{Giovanelli}, R., {Haynes}, M.~P., {Kent}, B.~R., {et~al.} 2005, \aj, 130, 2598

\bibitem[{Haynes \& Giovanelli(1980)}]{Haynes_1980}
Haynes, M.~P. \& Giovanelli, R. 1980, The Astrophysical Journal, 240, L87

\bibitem[{{Haynes} \& {Giovanelli}(1984)}]{Haynes_1984}
{Haynes}, M.~P. \& {Giovanelli}, R. 1984, \aj, 89, 758

\bibitem[{{Haynes} \& {Giovanelli}(1991)}]{Haynes_1991}
{Haynes}, M.~P. \& {Giovanelli}, R. 1991, \apjs, 77, 331

\bibitem[{{Haynes} {et~al.}(1984){Haynes}, {Giovanelli}, \&
  {Chincarini}}]{Haynes_1984b}
{Haynes}, M.~P., {Giovanelli}, R., \& {Chincarini}, G.~L. 1984, \araa, 22, 445

\bibitem[{{Haynes} {et~al.}(2011){Haynes}, {Giovanelli}, {Martin}, {Hess},
  {Saintonge}, {Adams}, {Hallenbeck}, {Hoffman}, {Huang}, {Kent}, {Koopmann},
  {Papastergis}, {Stierwalt}, {Balonek}, {Craig}, {Higdon}, {Kornreich},
  {Miller}, {O'Donoghue}, {Olowin}, {Rosenberg}, {Spekkens}, {Troischt}, \&
  {Wilcots}}]{Haynes_2011}
{Haynes}, M.~P., {Giovanelli}, R., {Martin}, A.~M., {et~al.} 2011, \aj, 142,
  170

\bibitem[{{Hess} {et~al.}(2017){Hess}, {Cluver}, {Yahya}, {Leisman}, {Serra},
  {Lucero}, {Passmoor}, \& {Carignan}}]{Hess_2017}
{Hess}, K.~M., {Cluver}, M.~E., {Yahya}, S., {et~al.} 2017, \mnras, 464, 957

\bibitem[{{Hess} \& {Wilcots}(2013)}]{Hess_2013}
{Hess}, K.~M. \& {Wilcots}, E.~M. 2013, \aj, 146, 124

\bibitem[{{Hewitt} {et~al.}(1983){Hewitt}, {Haynes}, \&
  {Giovanelli}}]{Hewitt_1983}
{Hewitt}, J.~N., {Haynes}, M.~P., \& {Giovanelli}, R. 1983, \aj, 88, 272

\bibitem[{{Hickson}(1982)}]{Hickson_1982}
{Hickson}, P. 1982, \apj, 255, 382

\bibitem[{{Hickson} {et~al.}(1992){Hickson}, {Mendes de Oliveira}, {Huchra}, \&
  {Palumbo}}]{Hickson_1992}
{Hickson}, P., {Mendes de Oliveira}, C., {Huchra}, J.~P., \& {Palumbo}, G.~G.
  1992, \apj, 399, 353

\bibitem[{{Hinshaw} {et~al.}(2013){Hinshaw}, {Larson}, {Komatsu}, {Spergel},
  {Bennett}, {Dunkley}, {Nolta}, {Halpern}, {Hill}, {Odegard}, {Page}, {Smith},
  {Weiland}, {Gold}, {Jarosik}, {Kogut}, {Limon}, {Meyer}, {Tucker}, {Wollack},
  \& {Wright}}]{Hinshaw_2013}
{Hinshaw}, G., {Larson}, D., {Komatsu}, E., {et~al.} 2013, \apjs, 208, 19

\bibitem[{{Huang} {et~al.}(2012){Huang}, {Haynes}, {Giovanelli}, \&
  {Brinchmann}}]{Huang_2012}
{Huang}, S., {Haynes}, M.~P., {Giovanelli}, R., \& {Brinchmann}, J. 2012, \apj,
  756, 113

\bibitem[{{Huchtmeier} \& {Richter}(1989)}]{Huchtmeier_1989}
{Huchtmeier}, W.~K. \& {Richter}, O.-G. 1989, \aap, 210, 1

\bibitem[{{Huchtmeier} {et~al.}(1995){Huchtmeier}, {Sage}, \&
  {Henkel}}]{Huchtmeier_1995}
{Huchtmeier}, W.~K., {Sage}, L.~J., \& {Henkel}, C. 1995, \aap, 300, 675

\bibitem[{{Jones} {et~al.}(2016){Jones}, {Haynes}, {Giovanelli}, \&
  {Papastergis}}]{Jones_2016}
{Jones}, M.~G., {Haynes}, M.~P., {Giovanelli}, R., \& {Papastergis}, E. 2016,
  \mnras, 455, 1574

\bibitem[{{Jones} {et~al.}(2015){Jones}, {Papastergis}, {Haynes}, \&
  {Giovanelli}}]{Jones_2015}
{Jones}, M.~G., {Papastergis}, E., {Haynes}, M.~P., \& {Giovanelli}, R. 2015,
  \mnras, 449, 1856

\bibitem[{{Karachentsev}(1972)}]{Karachentsev_1972}
{Karachentsev}, I.~D. 1972, Soobshcheniya Spetsial'noj Astrofizicheskoj
  Observatorii, 7

\bibitem[{{Karachentseva}(1973)}]{Karachentseva_1973}
{Karachentseva}, V.~E. 1973, Soobshcheniya Spetsial'noj Astrofizicheskoj
  Observatorii, 8

\bibitem[{{Kenney} {et~al.}(2004){Kenney}, {van Gorkom}, \&
  {Vollmer}}]{Kenney_2004}
{Kenney}, J.~D.~P., {van Gorkom}, J.~H., \& {Vollmer}, B. 2004, \aj, 127, 3361

\bibitem[{{Lah} {et~al.}(2009){Lah}, {Pracy}, {Chengalur}, {Briggs}, {Colless},
  {de Propris}, {Ferris}, {Schmidt}, \& {Tucker}}]{Lah_2009}
{Lah}, P., {Pracy}, M.~B., {Chengalur}, J.~N., {et~al.} 2009, \mnras, 399, 1447

\bibitem[{{Leisman} {et~al.}(2016){Leisman}, {Haynes}, {Giovanelli},
  {J{\'o}zsa}, {Adams}, \& {Hess}}]{Leisman_2016}
{Leisman}, L., {Haynes}, M.~P., {Giovanelli}, R., {et~al.} 2016, \mnras, 463,
  1692

\bibitem[{{Leon} \& {Verdes-Montenegro}(2003)}]{Leon_2003}
{Leon}, S. \& {Verdes-Montenegro}, L. 2003, \aap, 411, 391

\bibitem[{{Leon} {et~al.}(2008){Leon}, {Verdes-Montenegro}, {Sabater},
  {Espada}, {Lisenfeld}, {Ballu}, {Sulentic}, {Verley}, {Bergond}, \&
  {Garc{\'{\i}}a}}]{Leon_2008}
{Leon}, S., {Verdes-Montenegro}, L., {Sabater}, J., {et~al.} 2008, \aap, 485,
  475

\bibitem[{{Lewis}(1983)}]{Lewis_1983}
{Lewis}, B.~M. 1983, \aj, 88, 1695

\bibitem[{{Lewis} {et~al.}(1985){Lewis}, {Helou}, \& {Salpeter}}]{Lewis_1985}
{Lewis}, B.~M., {Helou}, G., \& {Salpeter}, E.~E. 1985, \apjs, 59, 161

\bibitem[{{Lisenfeld} {et~al.}(2011){Lisenfeld}, {Espada}, {Verdes-Montenegro},
  {Kuno}, {Leon}, {Sabater}, {Sato}, {Sulentic}, {Verley}, \&
  {Yun}}]{Lisenfeld_2011}
{Lisenfeld}, U., {Espada}, D., {Verdes-Montenegro}, L., {et~al.} 2011, \aap,
  534, A102

\bibitem[{{Lisenfeld} {et~al.}(2007){Lisenfeld}, {Verdes-Montenegro},
  {Sulentic}, {Leon}, {Espada}, {Bergond}, {Garc{\'{\i}}a}, {Sabater},
  {Santander-Vela}, \& {Verley}}]{Lisenfeld_2007}
{Lisenfeld}, U., {Verdes-Montenegro}, L., {Sulentic}, J., {et~al.} 2007, \aap,
  462, 507

\bibitem[{{Lu} {et~al.}(1993){Lu}, {Hoffman}, {Groff}, {Roos}, \&
  {Lamphier}}]{Lu_1993}
{Lu}, N.~Y., {Hoffman}, G.~L., {Groff}, T., {Roos}, T., \& {Lamphier}, C. 1993,
  \apjs, 88, 383

\bibitem[{{Lucero} {et~al.}(2005){Lucero}, {Young}, \& {van
  Gorkom}}]{Lucero_2005}
{Lucero}, D.~M., {Young}, L.~M., \& {van Gorkom}, J.~H. 2005, \aj, 129, 647

\bibitem[{{Maddox} {et~al.}(2015){Maddox}, {Hess}, {Obreschkow}, {Jarvis}, \&
  {Blyth}}]{Maddox_2015}
{Maddox}, N., {Hess}, K.~M., {Obreschkow}, D., {Jarvis}, M.~J., \& {Blyth},
  S.-L. 2015, \mnras, 447, 1610

\bibitem[{{Makarov} {et~al.}(2014){Makarov}, {Prugniel}, {Terekhova},
  {Courtois}, \& {Vauglin}}]{Makarov_2014}
{Makarov}, D., {Prugniel}, P., {Terekhova}, N., {Courtois}, H., \& {Vauglin},
  I. 2014, \aap, 570, A13

\bibitem[{{Masters}(2005)}]{Masters_2005}
{Masters}, K.~L. 2005, PhD thesis, Cornell University, New York, USA

\bibitem[{{Masters} {et~al.}(2014){Masters}, {Crook}, {Hong}, {Jarrett},
  {Koribalski}, {Macri}, {Springob}, \& {Staveley-Smith}}]{Masters_2014}
{Masters}, K.~L., {Crook}, A., {Hong}, T., {et~al.} 2014, \mnras, 443, 1044

\bibitem[{{Meyer} {et~al.}(2004){Meyer}, {Zwaan}, {Webster}, {Staveley-Smith},
  {Ryan-Weber}, {Drinkwater}, {Barnes}, {Howlett}, {Kilborn}, {Stevens},
  {Waugh}, {Pierce}, {Bhathal}, {de Blok}, {Disney}, {Ekers}, {Freeman},
  {Garcia}, {Gibson}, {Harnett}, {Henning}, {Jerjen}, {Kesteven}, {Knezek},
  {Koribalski}, {Mader}, {Marquarding}, {Minchin}, {O'Brien}, {Oosterloo},
  {Price}, {Putman}, {Ryder}, {Sadler}, {Stewart}, {Stootman}, \&
  {Wright}}]{Meyer_2004}
{Meyer}, M.~J., {Zwaan}, M.~A., {Webster}, R.~L., {et~al.} 2004, \mnras, 350,
  1195

\bibitem[{{Mirabel} \& {Sanders}(1988)}]{Mirabel_1988}
{Mirabel}, I.~F. \& {Sanders}, D.~B. 1988, \apj, 335, 104

\bibitem[{{Mould} {et~al.}(2000){Mould}, {Huchra}, {Freedman}, {Kennicutt},
  {Ferrarese}, {Ford}, {Gibson}, {Graham}, {Hughes}, {Illingworth}, {Kelson},
  {Macri}, {Madore}, {Sakai}, {Sebo}, {Silbermann}, \& {Stetson}}]{Mould_2000}
{Mould}, J.~R., {Huchra}, J.~P., {Freedman}, W.~L., {et~al.} 2000, \apj, 529,
  786

\bibitem[{{Nilson}(1973)}]{Nilson_1973}
{Nilson}, P. 1973, {Uppsala general catalogue of galaxies}

\bibitem[{{Odekon} {et~al.}(2016){Odekon}, {Koopmann}, {Haynes}, {Finn},
  {McGowan}, {Micula}, {Reed}, {Giovanelli}, \& {Hallenbeck}}]{Odekon_2016}
{Odekon}, M.~C., {Koopmann}, R.~A., {Haynes}, M.~P., {et~al.} 2016, \apj, 824,
  110

\bibitem[{{Papastergis} {et~al.}(2013){Papastergis}, {Giovanelli}, {Haynes},
  {Rodr{\'{\i}}guez-Puebla}, \& {Jones}}]{Papastergis_2013}
{Papastergis}, E., {Giovanelli}, R., {Haynes}, M.~P.,
  {Rodr{\'{\i}}guez-Puebla}, A., \& {Jones}, M.~G. 2013, \apj, 776, 43

\bibitem[{{Patton} {et~al.}(2013){Patton}, {Torrey}, {Ellison}, {Mendel}, \&
  {Scudder}}]{Patton_2013}
{Patton}, D.~R., {Torrey}, P., {Ellison}, S.~L., {Mendel}, J.~T., \& {Scudder},
  J.~M. 2013, \mnras, 433, L59

\bibitem[{{Planck Collaboration} {et~al.}(2016){Planck Collaboration}, {Ade},
  {Aghanim}, {Arnaud}, {Ashdown}, {Aumont}, {Baccigalupi}, {Banday},
  {Barreiro}, {Bartlett}, \& et~al.}]{Ade_2016}
{Planck Collaboration}, {Ade}, P.~A.~R., {Aghanim}, N., {et~al.} 2016, \aap,
  594, A13

\bibitem[{{Rasmussen} {et~al.}(2008){Rasmussen}, {Ponman}, {Verdes-Montenegro},
  {Yun}, \& {Borthakur}}]{Rasmussen_2008}
{Rasmussen}, J., {Ponman}, T.~J., {Verdes-Montenegro}, L., {Yun}, M.~S., \&
  {Borthakur}, S. 2008, \mnras, 388, 1245

\bibitem[{{Reid} {et~al.}(1991){Reid}, {Brewer}, {Brucato}, {McKinley},
  {Maury}, {Mendenhall}, {Mould}, {Mueller}, {Neugebauer}, {Phinney},
  {Sargent}, {Schombert}, \& {Thicksten}}]{Reid_1991}
{Reid}, I.~N., {Brewer}, C., {Brucato}, R.~J., {et~al.} 1991, \pasp, 103, 661

\bibitem[{{Richter} \& {Huchtmeier}(1987)}]{Richter_1987}
{Richter}, O.-G. \& {Huchtmeier}, W.~K. 1987, \aaps, 68, 427

\bibitem[{{Rubin} {et~al.}(1976){Rubin}, {Roberts}, {Graham}, {Ford}, \&
  {Thonnard}}]{Rubin_1976}
{Rubin}, V.~C., {Roberts}, M.~S., {Graham}, J.~A., {Ford}, Jr., W.~K., \&
  {Thonnard}, N. 1976, \aj, 81, 687

\bibitem[{{Sabater} {et~al.}(2008){Sabater}, {Leon}, {Verdes-Montenegro},
  {Lisenfeld}, {Sulentic}, \& {Verley}}]{Sabater_2008}
{Sabater}, J., {Leon}, S., {Verdes-Montenegro}, L., {et~al.} 2008, \aap, 486,
  73

\bibitem[{{Sabater} {et~al.}(2012){Sabater}, {Verdes-Montenegro}, {Leon},
  {Best}, \& {Sulentic}}]{Sabater_2012}
{Sabater}, J., {Verdes-Montenegro}, L., {Leon}, S., {Best}, P., \& {Sulentic},
  J. 2012, \aap, 545, A15

\bibitem[{{Schneider} {et~al.}(1992){Schneider}, {Thuan}, {Mangum}, \&
  {Miller}}]{Schneider_1992}
{Schneider}, S.~E., {Thuan}, T.~X., {Mangum}, J.~G., \& {Miller}, J. 1992,
  \apjs, 81, 5

\bibitem[{{Serra} {et~al.}(2013){Serra}, {Koribalski}, {Duc}, {Oosterloo},
  {McDermid}, {Michel-Dansac}, {Emsellem}, {Cuillandre}, {Alatalo}, {Blitz},
  {Bois}, {Bournaud}, {Bureau}, {Cappellari}, {Crocker}, {Davies}, {Davis}, {de
  Zeeuw}, {Khochfar}, {Krajnovi{\'c}}, {Kuntschner}, {Lablanche}, {Morganti},
  {Naab}, {Sarzi}, {Scott}, {Weijmans}, \& {Young}}]{Serra_2013}
{Serra}, P., {Koribalski}, B., {Duc}, P.-A., {et~al.} 2013, \mnras, 428, 370

\bibitem[{{Serra} {et~al.}(2012){Serra}, {Oosterloo}, {Morganti}, {Alatalo},
  {Blitz}, {Bois}, {Bournaud}, {Bureau}, {Cappellari}, {Crocker}, {Davies},
  {Davis}, {de Zeeuw}, {Duc}, {Emsellem}, {Khochfar}, {Krajnovi{\'c}},
  {Kuntschner}, {Lablanche}, {McDermid}, {Naab}, {Sarzi}, {Scott}, {Trager},
  {Weijmans}, \& {Young}}]{Serra_2012}
{Serra}, P., {Oosterloo}, T., {Morganti}, R., {et~al.} 2012, \mnras, 422, 1835

\bibitem[{{Shostak}(1978)}]{Shostak_1978}
{Shostak}, G.~S. 1978, \aap, 68, 321

\bibitem[{{Solanes} {et~al.}(1996){Solanes}, {Giovanelli}, \&
  {Haynes}}]{Solanes_1996}
{Solanes}, J.~M., {Giovanelli}, R., \& {Haynes}, M.~P. 1996, \apj, 461, 609

\bibitem[{{Solanes} {et~al.}(2001){Solanes}, {Manrique},
  {Garc{\'{\i}}a-G{\'o}mez}, {Gonz{\'a}lez-Casado}, {Giovanelli}, \&
  {Haynes}}]{Solanes_2001}
{Solanes}, J.~M., {Manrique}, A., {Garc{\'{\i}}a-G{\'o}mez}, C., {et~al.} 2001,
  \apj, 548, 97

\bibitem[{{Springob} {et~al.}(2005){Springob}, {Haynes}, {Giovanelli}, \&
  {Kent}}]{Springob_2005}
{Springob}, C.~M., {Haynes}, M.~P., {Giovanelli}, R., \& {Kent}, B.~R. 2005,
  \apjs, 160, 149

\bibitem[{{Staveley-Smith} \& {Davies}(1987)}]{Staveley-Smith_1987}
{Staveley-Smith}, L. \& {Davies}, R.~D. 1987, \mnras, 224, 953

\bibitem[{{Sulentic} {et~al.}(2006){Sulentic}, {Verdes-Montenegro}, {Bergond},
  {Lisenfeld}, {Durbala}, {Espada}, {Garcia}, {Leon}, {Sabater}, {Verley},
  {Casanova}, \& {Sota}}]{Sulentic_2006}
{Sulentic}, J.~W., {Verdes-Montenegro}, L., {Bergond}, G., {et~al.} 2006, \aap,
  449, 937

\bibitem[{{Theureau} {et~al.}(1998){Theureau}, {Bottinelli}, {Coudreau-Durand},
  {Gouguenheim}, {Hallet}, {Loulergue}, {Paturel}, \&
  {Teerikorpi}}]{Theureau_1998}
{Theureau}, G., {Bottinelli}, L., {Coudreau-Durand}, N., {et~al.} 1998, \aaps,
  130, 333

\bibitem[{{Theureau} {et~al.}(2005){Theureau}, {Coudreau}, {Hallet}, {Hanski},
  {Alsac}, {Bottinelli}, {Gouguenheim}, {Martin}, \& {Paturel}}]{Theureau_2005}
{Theureau}, G., {Coudreau}, N., {Hallet}, N., {et~al.} 2005, \aap, 430, 373

\bibitem[{{Theureau} {et~al.}(2007){Theureau}, {Hanski}, {Coudreau}, {Hallet},
  \& {Martin}}]{Theureau_2007}
{Theureau}, G., {Hanski}, M.~O., {Coudreau}, N., {Hallet}, N., \& {Martin},
  J.-M. 2007, \aap, 465, 71

\bibitem[{{Tifft} \& {Cocke}(1988)}]{Tifft_1988}
{Tifft}, W.~G. \& {Cocke}, W.~J. 1988, \apjs, 67, 1

\bibitem[{{Toribio} {et~al.}(2011{\natexlab{a}}){Toribio}, {Solanes},
  {Giovanelli}, {Haynes}, \& {Martin}}]{Toribio_2011}
{Toribio}, M.~C., {Solanes}, J.~M., {Giovanelli}, R., {Haynes}, M.~P., \&
  {Martin}, A.~M. 2011{\natexlab{a}}, \apj, 732, 93

\bibitem[{{Toribio} {et~al.}(2011{\natexlab{b}}){Toribio}, {Solanes},
  {Giovanelli}, {Haynes}, \& {Masters}}]{Toribio_2011a}
{Toribio}, M.~C., {Solanes}, J.~M., {Giovanelli}, R., {Haynes}, M.~P., \&
  {Masters}, K.~L. 2011{\natexlab{b}}, \apj, 732, 92

\bibitem[{{Torres-Flores} {et~al.}(2010){Torres-Flores}, {Mendes de Oliveira},
  {Amram}, {Plana}, {Epinat}, {Carignan}, \& {Balkowski}}]{Torres-Flores+2010}
{Torres-Flores}, S., {Mendes de Oliveira}, C., {Amram}, P., {et~al.} 2010,
  \aap, 521, A59

\bibitem[{{van Driel} {et~al.}(2016){van Driel}, {Butcher}, {Schneider},
  {Lehnert}, {Minchin}, {Blyth}, {Chemin}, {Hallet}, {Joseph}, {Kotze},
  {Kraan-Korteweg}, {Olofsson}, \& {Ramatsoku}}]{vanDriel_2016}
{van Driel}, W., {Butcher}, Z., {Schneider}, S., {et~al.} 2016, \aap, 595, A118

\bibitem[{{van Driel} {et~al.}(1995){van Driel}, {van den Broek}, \&
  {Baan}}]{vanDriel_1995}
{van Driel}, W., {van den Broek}, A.~C., \& {Baan}, W. 1995, \apj, 444, 80

\bibitem[{{Verdes-Montenegro} {et~al.}(2005){Verdes-Montenegro}, {Sulentic},
  {Lisenfeld}, {Leon}, {Espada}, {Garcia}, {Sabater}, \&
  {Verley}}]{Verdes_Montenegro_2005}
{Verdes-Montenegro}, L., {Sulentic}, J., {Lisenfeld}, U., {et~al.} 2005, \aap,
  436, 443

\bibitem[{{Verdes-Montenegro} {et~al.}(2001){Verdes-Montenegro}, {Yun},
  {Williams}, {Huchtmeier}, {Del Olmo}, \& {Perea}}]{Verdes-Montenegro_2001}
{Verdes-Montenegro}, L., {Yun}, M.~S., {Williams}, B.~A., {et~al.} 2001, \aap,
  377, 812

\bibitem[{{Verley} {et~al.}(2007{\natexlab{a}}){Verley}, {Leon},
  {Verdes-Montenegro}, {Combes}, {Sabater}, {Sulentic}, {Bergond}, {Espada},
  {Garc{\'{\i}}a}, {Lisenfeld}, \& {Odewahn}}]{Verley_2007b}
{Verley}, S., {Leon}, S., {Verdes-Montenegro}, L., {et~al.} 2007{\natexlab{a}},
  \aap, 472, 121

\bibitem[{{Verley} {et~al.}(2007{\natexlab{b}}){Verley}, {Odewahn},
  {Verdes-Montenegro}, {Leon}, {Combes}, {Sulentic}, {Bergond}, {Espada},
  {Garc{\'{\i}}a}, {Lisenfeld}, \& {Sabater}}]{Verley_2007a}
{Verley}, S., {Odewahn}, S.~C., {Verdes-Montenegro}, L., {et~al.}
  2007{\natexlab{b}}, \aap, 470, 505

\bibitem[{{Walker} {et~al.}(2016){Walker}, {Johnson}, {Gallagher}, {Privon},
  {Kepley}, {Whelan}, {Desjardins}, \& {Zabludoff}}]{Walker_2016}
{Walker}, L.~M., {Johnson}, K.~E., {Gallagher}, S.~C., {et~al.} 2016, \aj, 151,
  30

\bibitem[{{Zasov} \& {Sulentic}(1994)}]{Zasov_1994}
{Zasov}, A.~V. \& {Sulentic}, J.~W. 1994, \apj, 430, 179

\bibitem[{{Zwicky} {et~al.}(1961){Zwicky}, {Herzog}, {Wild}, {Karpowicz}, \&
  {Kowal}}]{Zwicky_1961}
{Zwicky}, F., {Herzog}, E., {Wild}, P., {Karpowicz}, M., \& {Kowal}, C.~T.
  1961, {Catalogue of galaxies and of clusters of galaxies, Vol. I}

\end{thebibliography}

\appendix

\section{Predicting HI mass}

The relations calculated throughout this paper are designed to approximate the scaling between the ``true'' values of an isolated galaxy's optical linear size or optical luminosity and its ``true'' HI mass. While the word ``true'' may seem redundant, but it makes the important distinction between the observed values of each of these properties and their actual physical values, which can never be known perfectly. Although the true values of these properties are unknown, by incorporating estimates of the uncertainties in their measured values into the fitting procedure, we have obtained the maximum likelihood estimate of the underlying physical relations (between the true values) for our dataset. This differs fundamentally from OLS regression, or similar methods that do not account for uncertainties in both variables, which find a relation between the \textit{observed} values.

Typically OLS regression is used when one wishes to use observations of $x$ to predict $y$, as this is the optimal linear predictor (in the least squares sense). Similarly, if each $y$ value has a different variance then a form of weighted least squares is optimal. However, both approaches are only optimal for predicting the value of $y$ (observed or true) from the observed value of $x$, under the assumption that $x$ is observed either without error or that the errors in observing $x$ have the same variance for every observation (and zero mean), and that the errors in future observations of $x$ (to be used to predict $y$) also have the same variance as in the original dataset. If either assumption is violated then OLS is non-optimal. If the first assumption is violated then the physical interpretation of the relation is lost because the regression coefficients become dependent on the magnitude of the $x$ error variance, which in turn can depend on the observation and reduction methods, which are unconnected with the physical properties of the galaxies. For our dataset neither of these assumptions hold. Therefore, the OLS relation (or more appropriately, the weighted least squares relation) is not optimal in our case. By taking the magnitude of the errors in each measurement into account we can construct a better method to predict $y$. Similarly, in quantifying the uncertainty in predictions of $y$ from newly observed values of $x$, it is important to take into account the $x$ error variance, see Equation \ref{eqn:prediction_var}. In addition, failure to take into account the fact that the $x$ and $y$ errors are correlated would reduce prediction accuracy, although this would likely be a minor effect in this case.

By incorporating estimates of measurement and distance errors in the fitting procedure we have made an estimate of what the intrinsic scatter in the relations is (i.e. scatter which is not accounted for within the error budget), and similarly the slope and intercept of the relations are also estimates of their intrinsic values for the true values of $D_{25}$, $L_{\mathrm{B}}$, and $M_{\mathrm{HI}}$. This presents a technical problem because when using the relations to predict $M_{\mathrm{HI}}$ one cannot measure the true value of $D_{25}$ or $L_{\mathrm{B}}$. To overcome this difficulty we consider how the quantity $\Delta y^{\mathrm{est}}$ is distributed, where $\Delta y^{\mathrm{est}}$ is the deviation between the true value of $y$ and the value estimated by treating an observed value of $x$ as the true value (which is what one must do in order to make a prediction of $y$ using the scaling relation), i.e.
\begin{equation}
    \Delta y^{\mathrm{est}} = y^{*} - (\beta_{0} + \beta_{1} x^{\mathrm{obs}}).
\end{equation}
The relation that we fitted was $y^{*} = \beta_{0} + \beta_{1} x^{*} + \xi$ (this is equation \ref{eqn:reg_mod} restated), which can be substituted for $y^{*}$ in the above equation. We can also substitute $x^{\mathrm{obs}} = x^{*} + \eta + \delta$, where $\sigma_{\eta}$ and $\sigma_{\delta}$ are the measurement and distance uncertainties as defined in Section \ref{sec:reg_mod}. This gives $\Delta y^{\mathrm{est}} = \xi - \beta_{1} (\eta + \delta)$, which implies
\begin{equation}
    \label{eqn:prediction_var}
    \Delta y^{\mathrm{est}} \sim \mathrm{N}(0,\sigma^{2}_{\xi} + \beta^{2}_{1} (\sigma^{2}_{\eta} + \sigma^{2}_{\delta})),
\end{equation}
where $N(\mu,\sigma^{2})$ indicates a normally distributed random variable with mean $\mu$ and variance $\sigma^{2}$.

Therefore, by simply using our scaling relation and an observation of $x$ as input, an unbiased prediction of the true value of $y$ can be made and its confidence interval can be estimated from the above normal distribution. For the slopes of our regression fits and typical values of the uncertainties in $\log D_{25}$ and $\log L_{\mathrm{B}}$, this gives estimates of the standard deviation of $\Delta y^{\mathrm{est}}$ as 0.23 and 0.25 dex, respectively. These values match well with the width of the distribution in Figure \ref{fig:HIdef_diff_comp}, showing self-consistency among our estimates of the precision of these relations as predictors of HI mass.

In other words, this approach allows for any future observation of either a galaxy's $D_{25}$ or $L_{\mathrm{B}}$ to be used to make a prediction of the physical value of its HI mass and an estimate of the uncertainty, regardless of whether the uncertainties in those future measurements are similar to those in our dataset or not.

It should also be noted that this approach assumes that the relation is known, i.e. that our fit is both accurate and precise. This is a good approximation near the centre of the data range where there is an abundance of data points and the fit is tightly constrained (see Figure \ref{fig:reg_coeff}). At the extremes of the data the uncertainty in the position of the relation becomes larger, however, the standard deviation of $\Delta y_{\mathrm{est}}$ about the relation is still several times larger. This shortcoming is not unique to our method, but applies to all approaches where the relation is not known exactly, but is treated as so. Ideally the extremes would be constrained with more data, but this is not presently possible.

\section{Data table description}
\label{app:data_tab}

The complete HI dataset is available with the online version of this paper. The data is displayed in three separate tables: 1) the observed HI properties, 2) the derived properties, and 3) various flags. Here we describe in detail the properties listed in each column of these tables.\\\\
Observed HI properties (Table \ref{tab:HI_prop}):\\
\textbf{Column 1} - CIG: ID number corresponding to the assignment in the original catalogue of isolated galaxies \citep{Karachentseva_1973}.\\
\textbf{Column 2} - $S_{\mathrm{p}}$: The peak flux density (mJy) of the emission profile. This is measured in the smoothed spectrum (see Column 6).\\
\textbf{Column 3} - $\sigma_{\mathrm{rms}}$: The root mean squared noise (mJy) in the emission free parts of the spectrum. This is measured in the smoothed spectrum (see Column 6).\\
\textbf{Column 4} - $\sigma_{\mathrm{rms},10}$: The root mean squared noise (mJy) if the spectrum had channel widths of 10 \kms \ (see equation \ref{eqn:rms10}).\\
\textbf{Column 5} - $\delta v$: The average channel width across the emission profile in \kms.\\
\textbf{Column 6} - $n_{\mathrm{smo}}$: The number of channels the spectrum is smoothed over using a Hanning window.\\
\textbf{Column 7} - snr$_{\mathrm{p}}$: The peak signal-to-noise ratio of the profile (S$_{\mathrm{p}}$/N  $= S_{\mathrm{p}}/\sigma_{\mathrm{rms}}$).\\
\textbf{Column 8} - snr$_{\mathrm{i}}$: The integrated signal-to-noise ratio of the profile as calculated by equation \ref{eqn:snr}.\\
\textbf{Column 9} - $S_{\mathrm{int}}$: The integrated flux (Jy \kms) of the profile. This is measured between the two zero-crossings of the straight line fits to the left and right sides of the profile (see section \ref{sec:data_red}).\\
\textbf{Column 10} - $\sigma_{S_{\mathrm{int}}}$: Measurement error in $S_{\mathrm{int}}$ (Jy \kms) as calculated in equation \ref{eqn:Sint_err}.\\
\textbf{Column 11} - $c_{\mathrm{beam}}$: The beam correction factor accounting for the beam response and flux that is outside the beam. This assumes a Gaussian beam response and a distribution of HI within the galaxy. See section \ref{sec:corrections} for further details.\\
\textbf{Column 12} - $S_{\mathrm{int,c}}$: The corrected integrated flux (Jy \kms). $S_{\mathrm{int,c}} = c_{\mathrm{beam}} S_{\mathrm{int}}$.\\
\textbf{Column 13} - $V_{50}$: The heliocentric velocity of the mid-point of the profile at the 50\% level (\kms). See section \ref{sec:data_red}.\\
\textbf{Column 14} - $\sigma_{V_{50}}$: The measurement error of the heliocentric velocity (\kms) as estimated in equation \ref{eqn:W50_err}.\\
\textbf{Column 15} - $z_{50}$: The heliocentric redshift of the source. Here we assume the velocities are small relative to $c$, i.e. $z_{50} = V_{50}/c$.\\
\textbf{Column 16} - $W_{50}$: The full velocity width of the emission profile at the 50\% level (\kms). The half point velocity is identified independently on either side of the profile by fitting a straight line to the profile edge and taking the velocity where the line equals half the peak value (minus the rms noise) on that side. For further details see section \ref{sec:data_red}.\\
\textbf{Column 17} - $\sigma_{W_{50}}$: The measurement error of the velocity width (\kms). Assumed to be $\sqrt{2}$ times the error in $V_{50}$ (Column 13).\\
\textbf{Column 18} - $c_{\mathrm{inst}}$: Correction for instrumental broadening (\kms). This correction is calculated exactly as in \citet{Springob_2005} (equations 3, 5-7, and table 2) except that we replace the channel width with the channel width times $n_{\mathrm{smo}}-2$, to account for any smoothing in addition to the 2 channels that method assumes have been smoothed over.\\
\textbf{Column 19} - $c_{\mathrm{cosmo}}$: Correction for the broadening of the velocity width due to cosmological expansion. $c_{\mathrm{cosmo}} = 1/(1+z_{50})$.\\
\textbf{Column 20} - $W_{\mathrm{50,c}}$: The velocity width corrected for instrumental and cosmic broadening (\kms). $W_{\mathrm{50,c}} = (W_{50} - c_{\mathrm{inst}}) c_{\mathrm{cosmo}}$.\\
\textbf{Column 21} - Detection code: A code indicating whether the source was identified by eye to be confidently detected (0), not detected (1), or marginally detected (2).\\
\textbf{Column 22} - Quality code: A code to identify spectra with features such as potentially spurious flux spikes that likely introduce major uncertainty into the measured parameters. 1 corresponds to a spectrum with complications, 0 to a good spectrum.\\
\textbf{Column 23} - Telescope code: A three character code to identify the telescope that the spectrum was observed with. See Table \ref{tab:beam_widths} for details.\\
\textbf{Column 24} - Reference code: A four character code to identify the original article the spectrum was taken from (codes are matched to references in table \ref{tab:refs}). If blank then the observations were performed as part of the AMIGA project.\\\\
Derived properties (Table \ref{tab:derive_prop}):\\
\textbf{Column 1} - CIG: ID number corresponding to the assignment in the original catalogue of isolated galaxies \citep{Karachentseva_1973}.\\
\textbf{Column 2} - $V_{\mathrm{helio}}$: The heliocentric velocity chosen to be used to estimate the source distance (\kms). This may or may not be the velocity measured from the HI spectrum. If $V_{50}$ agreed within 2-$\sigma$ of the existing AMIGA preferred velocity then the one with the smaller measurement error was chosen. Otherwise the velocities were inspected by hand. Typically in these cases the measurement with the smaller uncertainty was chosen, except in cases where it disagreed with the majority of other measurements in the literature.\\
\textbf{Column 3} - $V_{\mathrm{mod}}$: The flow model corrected recession velocity of the source based on the \citet{Mould_2000} flow model (\kms).\\
\textbf{Column 4} - $D_{\mathrm{mod}}$: The flow model distance to the source ($h_{70}\,$Mpc).\\
\textbf{Column 5} - $\sigma_{D}$: Estimate of the uncertainty in the flow model distance (Mpc). Calculated as described in section \ref{sec:dists}.\\
\textbf{Column 6} - $\log M_{\mathrm{HI}}$: The log of the HI mass of the source ($\mathrm{M_{\odot}}\,h_{70}^{2}$), as calculated in equation \ref{eqn:HI_mass}.\\
\textbf{Column 7} - $\sigma_{\log M_{\mathrm{HI}} }$: The error in $\log M_{\mathrm{HI}}$ (dex). This was estimated as
\begin{equation}
\log \left( 1 + \sqrt{ \left( \frac{\sigma_{S_{\mathrm{int}}}}{S_{\mathrm{int}}} \right)^{2} + \left( \frac{\sigma_{D}}{D_{\mathrm{mod}}} \right)^{2}} \right).
\end{equation}
However, the considerable scatter between the ALFALFA and AMIGA measures of $S_{\mathrm{int}}$ for the same sources suggests that the largest contribution to this error is in fact the absolute calibration, not the noise in the spectrum or the distance uncertainty. Therefore, this error was assigned a minimum value of 0.12 dex corresponding to the scatter between ALFALFA and AMIGA (see section \ref{sec:alfa_comp}). In practice this is the relevant value for almost all the sources.
\textbf{Column 8} - $W_{\mathrm{TFR}}$: The estimate of the galaxy's HI velocity width (\kms) based on the Tully-Fisher relation of \citet{Torres-Flores+2010}, as described in section \ref{sec:HI_lims}.
\textbf{Column 9} - $\log M_{\mathrm{HI-lim}}$: The 5-$\sigma$ limit on the HI mass of the source ($\mathrm{M_{\odot}}\,h_{70}^{2}$), based on the velocity width above and the sensitivity of the spectrum. See section \ref{sec:alfa_comp} for more details.\\
\textbf{Column 10} - Limit code: This indicates whether the limit should be used. The limits should be used when this code is 1. Note that for marginal detections $\log M_{\mathrm{HI}}$ is still calculated, but it should not be used.\\\\
Flag table (Table \ref{tab:flags}):\\
\textbf{Column 1} - CIG: ID number corresponding to the assignment in the original catalogue of isolated galaxies \citep{Karachentseva_1973}.\\
\textbf{Column 2} - Isolation: If set to 1 the galaxy is not considered to be well isolated.\\
\textbf{Column 3} - Completeness: If set to 1 the galaxy does not meet the completeness requirements.\\
\textbf{Column 4} - Interpolation: If set to 1 this means that due to the low resolution of high noise of the spectrum some interpolation was necessary to obtain a meaningful fit to the edges of the HI profile.\\
\textbf{Column 5} - Single peak: If set to 1 the galaxy does not have a typical double horn profile, but is closer to a Gaussian.\\
\textbf{Column 6} - Middle peak: If set to 1 the galaxy HI profile is not peaked in on of the horned, but instead the maximum is somewhere between the horns.\\
\textbf{Column 7} - Offset spike: If set to 1 there is a sharp spike in the spectrum that may be contamination. In these cases the velocity widths are likely highly uncertain.\\
\textbf{Column 8} - Radio velocity: If set to 1 then the original spectrum was thought to be recorded in radio velocity and was converted to optical velocity accordingly.\\
\textbf{Column 9} - Digitised: If set to 1 then no digital version of the spectrum was available and a published image of the spectrum was digitised.

\begin{table*}
\centering
\caption{Observed properties of the full HI sample}
\label{tab:HI_prop}
\begin{tabular}{ccccccccccccccc}
\hline
\hline
CIG  & $S_{\mathrm{p}}$ & $\sigma_{\mathrm{rms}}$ & $\sigma_{\mathrm{rms},10}$ & $\delta v$ & $n_{\mathrm{smo}}$ & snr$_{\mathrm{p}}$ & snr$_{\mathrm{i}}$ & $S_{\mathrm{int}}$ & $\sigma_{S_{\mathrm{int}}}$ & $c_{\mathrm{beam}}$ & $S_{\mathrm{int,c}}$ & $V_{50}$ & $\sigma_{V_{50}}$ & $z_{50}$ \\
\hline
1    & 19.3         & 2.8           & 7.3               & 11.1          & 6.0    & 6.8    & 14.7   & 5.89               & 0.48                    & 1.05    & 6.17                  & 7275.1         & 13.1                & 0.02 \\
2    & 17.2         & 1.6           & 1.8               & 4.3           & 3.0    & 10.6   & 33.4   & 3.1                & 0.13                    & 1.05    & 3.25                  & 6994.5         & 4.9                 & 0.02 \\
4    & 79.8         & 6.0           & 12.3              & 4.2           & 10.0   & 13.2   & 29.0   & 19.61              & 0.59                    & 1.02    & 19.91                 & 2313.9         & 10.4                & 0.01 \\
5    & 2.8          & 0.4           & 1.3               & 8.7           & 10.0   & 6.6    & 13.1   & 0.91               & 0.06                    & 1.05    & 0.96                  & 7853.2         & 40.7                & 0.03 \\
6    & 21.1         & 1.2           & 1.3               & 4.1           & 3.0    & 17.4   & 54.8   & 2.74               & 0.07                    & 1.03    & 2.83                  & 4527.4         & 5.1                 & 0.02 \\
7    & 6.4          & 1.5           & 2.0               & 9.0           & 2.0    & 4.4    & 15.2   & 1.63               & 0.22                    & 1.04    & 1.69                  & 12760.7        & 13.4                & 0.04 \\
8    & 24.0         & 1.4           & 3.3               & 8.6           & 6.0    & 16.8   & 25.7   & 4.19               & 0.16                    & 1.04    & 4.34                  & 6355.1         & 6.0                 & 0.02 \\
9    & 20.4         & 1.0           & 2.4               & 8.7           & 6.0    & 19.5   & 39.9   & 5.24               & 0.14                    & 1.05    & 5.51                  & 8484.2         & 6.6                 & 0.03 \\
10   & 14.2         & 1.8           & 1.9               & 5.2           & 2.0    & 7.7    & 22.4   & 2.16               & 0.16                    & 1.01    & 2.18                  & 4994.3         & 3.9                 & 0.02 \\
11   & 79.3         & 5.9           & 8.6               & 10.6          & 2.0    & 13.4   & 31.0   & 14.27              & 0.77                    & 1.09    & 15.53                 & 3963.7         & 6.2                 & 0.01 \\
12   & 15.5         & 1.2           & 2.7               & 8.6           & 6.0    & 12.9   & 22.5   & 3.28               & 0.14                    & 1.04    & 3.42                  & 5478.0         & 8.4                 & 0.02 \\
13   &   -           & 0.2           & 0.1               & 1.3           &   -     &   -     &   -     &   -                 &   -                      &   -      &   -                    &   -             &   -                  &   -   \\
14   & 3.3          & 0.7           & 1.5               & 5.2           & 10.0   & 5.0    & 10.1   & 0.84               & 0.06                    & 1.02    & 0.85                  & 5194.5         & 36.0                & 0.02 \\
15   & 4.1          & 0.7           & 2.0               & 8.9           & 10.0   & 6.0    & 8.2    & 0.91               & 0.09                    & 1.03    & 0.94                  & 11673.1        & 18.4                & 0.04 \\
16   &   -           & 6.5           & 4.7               & 2.6           &   -     &   -     &   -     &   -                 &   -                      &   -      &   -                    &   -             &   -                  &   -   \\
18   & 13.7         & 1.2           & 1.4               & 4.1           & 3.0    & 11.2   & 40.1   & 2.8                & 0.1                     & 1.04    & 2.91                  & 7215.8         & 5.2                 & 0.02 \\
19   &   -           & 8.0           & 4.0               & 1.3           &   -     &   -     &   -     &   -                 &   -                      &   -      &   -                    &   -             &   -                  &   -   \\
20   &   -           & 4.0           & 4.0               & 5.2           &   -     &   -     &   -     &   -                 &   -                      &   -      &   -                    &   -             &   -                  &   -   \\
21   &   -           & 2.0           & 0.7               & 0.7           &   -     &   -     &   -     &   -                 &   -                      &   -      &   -                    &   -             &   -                  &   -   \\
  (...) & (...) & (...) & (...) & (...) & (...) & (...) & (...) & (...) & (...) & (...) & (...) & (...) & (...) & (...) \\
\hline
\end{tabular}

\vspace{20mm}

\begin{tabular}{ccccccccccc}
\hline
\hline
CIG  & (...) & $W_{50}$ & $\sigma_{W_{50}}$ & $c_{\mathrm{inst}}$ & $c_{\mathrm{cosmo}}$ & $W_{\mathrm{50,c}}$ & det\_code & qual\_code & tele\_code & ref\_code \\
\hline
1    & (...)          & 469.5          & 18.5                & 18.8               & 0.98     & 440.1             & 0         & 0          & NRT        & Th98      \\
2    & (...)          & 253.0          & 6.9                 & 0.0                & 0.98     & 247.3             & 0         & 0          & AOL        & HG84      \\
4    & (...)          & 403.6          & 14.8                & 28.8               & 0.99     & 371.9             & 0         & 0          & G43        & Sp05      \\
5    & (...)          & 419.0          & 57.6                & 27.7               & 0.97     & 381.3             & 0         & 1          & AOL        & Sp05      \\
6    & (...)          & 137.8          & 7.3                 & 0.0                & 0.99     & 135.8             & 0         & 0          & AOL        & HG84      \\
7    & (...)          & 432.2          & 19.0                & 0.0                & 0.96     & 414.5             & 0         & 0          & AOL        & Sp05      \\
8    & (...)          & 251.2          & 8.4                 & 29.6               & 0.98     & 217.0             & 0         & 0          & AOL        & Sp05      \\
9    & (...)          & 356.0          & 9.3                 & 30.0               & 0.97     & 317.1             & 0         & 0          & AOL        & Sp05      \\
10   & (...)          & 259.6          & 5.5                 & 0.0                & 0.98     & 255.3             & 0         & 0          & ERT        &           \\
11   & (...)          & 289.0          & 8.8                 & 0.0                & 0.99     & 285.2             & 0         & 0          & NRT        & KLUN      \\
12   & (...)          & 289.7          & 11.9                & 29.4               & 0.98     & 255.7             & 0         & 0          & AOL        & Sp05      \\
13   & (...)          &   -            &   -                 &   -                &   -      &   -               & 1         &   -        & NRT        &           \\
14   & (...)          & 321.6          & 50.9                & 8.9                & 0.98     & 307.4             & 0         & 0          & ERT        &           \\
15   & (...)          & 387.9          & 26.0                & 23.8               & 0.96     & 350.5             & 0         & 0          & AOL        & Sp05      \\
16   & (...)          &   -            &   -                 &   -                &   -      &   -               & 1         &   -        & NRT        &           \\
18   & (...)          & 263.8          & 7.4                 & 0.0                & 0.98     & 257.6             & 0         & 0          & AOL        & HG84      \\
19   & (...)          &   -            &   -                 &   -                &   -      &   -               & 1         &   -        & NRT        &           \\
20   & (...)          &   -            &   -                 &   -                &   -      &   -               & 1         &   -        & ERT        &           \\
21   & (...)          &   -            &   -                 &   -                &   -      &   -               & 1         &   -        & AOG        &           \\
  (...) & (...) & (...) & (...) & (...) & (...) & (...) & (...) & (...) & (...) & (...) \\
\hline
\end{tabular}

\end{table*}

\begin{table*}
\centering
\caption{Derived properties of the full HI sample}
\label{tab:derive_prop}
\begin{tabular}{cccccccccc}
\hline
\hline
CIG  & $V_{\mathrm{helio}}$ & $V_{\mathrm{mod}}$ & $D_{\mathrm{mod}}$ & $\sigma_{D}$ & $\log M_{\mathrm{HI}}$ & $\sigma_{\log M_{\mathrm{HI}}}$ & $W_{\mathrm{TFR}}$ & $\log M_{\mathrm{HI-lim}}$ & Limit\\
\hline
  1 & 7299 & 7180 & 102.6 & 4.2 & 10.08 & 0.12 & 560 & 10.71 & 0\\
  2 & 6995 & 6903 & 98.6 & 4.1 & 9.87 & 0.12 & 251 & 9.72 & 0\\
  4 & 2310 & 2340 & 33.4 & 3.0 & 9.72 & 0.12 & 483 & 9.89 & 0\\
  5 & 7865 & 7737 & 110.5 & 4.3 & 9.44 & 0.12 & 365 & 9.82 & 0\\
  6 & 4527 & 4475 & 63.9 & 3.4 & 9.44 & 0.12 & 300 & 9.29 & 0\\
  7 & 12761 & 12566 & 179.5 & 5.9 & 10.11 & 0.12 & 429 & 10.5 & 1\\
  8 & 6355 & 6242 & 89.2 & 3.9 & 9.91 & 0.12 & 390 & 10.07 & 0\\
  9 & 8484 & 8333 & 119.0 & 4.5 & 10.27 & 0.12 & 416 & 10.22 & 0\\
  10 & 4994 & 4967 & 71.0 & 3.5 & 9.26 & 0.12 & 308 & 9.54 & 0\\
  11 & 3963 & 3906 & 55.8 & 3.2 & 9.87 & 0.12 & 342 & 10.03 & 0\\
  12 & 5478 & 5382 & 76.9 & 3.6 & 9.68 & 0.12 & 322 & 9.78 & 0\\
  13 & 5240 & 5147 & 73.5 & 3.6 &    -   &    -   & 486 & 8.5 & 1\\
  14 & 5195 & 5101 & 72.9 & 3.6 & 8.88 & 0.12 & 336 & 9.5 & 0\\
  15 & 11665 & 11491 & 164.2 & 5.6 & 9.78 & 0.12 & 496 & 10.5 & 0\\
  16 & 5485 & 5383 & 76.9 & 3.7 &    -   &    -   & 254 & 9.92 & 1\\
  18 & 7216 & 7102 & 101.5 & 4.1 & 9.85 & 0.12 & 309 & 9.71 & 0\\
  19 & 5390 & 5290 & 75.6 & 3.6 &    -   &    -   & 309 & 9.92 & 1\\
  20 & 5104 & 5083 & 72.6 & 3.5 &    -   &    -   & 158 & 9.6 & 1\\
  21 & 7969 & 7822 & 111.7 & 4.3 &    -   &    -   & 297 & 9.5 & 1\\
  (...) & (...) & (...) & (...) & (...) &  (...) &  (...) & (...) & (...) & (...)\\
\hline
\end{tabular}
\end{table*}

\begin{table*}
\centering
\caption{Flags for the full HI sample}
\label{tab:flags}
\begin{tabular}{ccccccccc}
\hline
\hline
CIG & Isolate & Complete & Interpolate & Single Peak& Middle Peak & Offset Spike & Radio Vel. & Digitised\\
\hline
1    & 1       & 0        & 1        & 0            & 0            & 0             & 0          & 0         \\
2    & 0       & 0        & 0        & 0            & 0            & 0             & 0          & 0         \\
4    & 0       & 0        & 0        & 0            & 0            & 0             & 0          & 0         \\
5    & 0       & 0        & 0        & 0            & 0            & 1             & 0          & 0         \\
6    & 0       & 0        & 0        & 0            & 0            & 0             & 0          & 0         \\
7    & 0       & 0        & 1        & 0            & 0            & 0             & 0          & 0         \\
8    & 1       & 0        & 0        & 0            & 0            & 0             & 0          & 0         \\
9    & 0       & 0        & 0        & 0            & 0            & 0             & 0          & 0         \\
10   & 0       & 0        & 0        & 0            & 0            & 0             & 0          & 0         \\
11   & 1       & 0        & 0        & 0            & 0            & 0             & 1          & 0         \\
12   & 0       & 0        & 0        & 0            & 0            & 0             & 0          & 0         \\
13   & 0       & 0        &   -      &   -          &   -          &   -           &   -        & 0         \\
14   & 0       & 0        & 1        & 0            & 1            & 0             & 0          & 0         \\
15   & 0       & 0        & 0        & 0            & 0            & 0             & 0          & 0         \\
16   & 1       & 1        &   -      &   -          &   -          &   -           &   -        & 0         \\
18   & 0       & 0        & 0        & 0            & 0            & 0             & 0          & 0         \\
19   & 1       & 0        &   -      &   -          &   -          &   -           &   -        & 0         \\
20   & 0       & 0        &   -      &   -          &   -          &   -           &   -        & 0         \\
21   & 0       & 0        &   -      &   -          &   -          &   -           &   -        & 0         \\
  (...) &(...) &(...) &(...) &(...) &(...) &(...) &(...) &(...)\\
\hline
\end{tabular}
\end{table*}

\section{Conversion of previous scaling relations}
\label{sec:rel_conv}

We compare our trend lines between $D_{25}$, $L_{\mathrm{B}}$ and $M_{\mathrm{HI}}$ with those of \citet{Haynes_1984,Solanes_1996,Denes_2014}. However, these papers all use slightly different units systems or measures of galaxy diameter, luminosity or mass. Therefore, we must make corrections to facilitate a fair comparison.

\citetalias{Haynes_1984} correct their HI masses for internal absorption, which we make no correction for as it may introduce a bias depending on the galaxy type. Unfortunately as this correction was applied with a piece wise function that varied across different morphologies we cannot make a simple conversion factor to account for it. Therefore, no conversion is made for the HI masses. As the typical correction for internal absorption should be small ($<$10\%) this is not expected to make a significant difference to the comparison.

The isophotal diameters they used were edited from the UGC values depending on the galaxy's apparent surface magnitude. We took the overlapping 323 sources in the UGC and CIG and recalculated the \citetalias{Haynes_1984} correction to the UGC B-band diameters. An OLS linear trend line was then fit between these corrected UGC diameters and our B-band $D_{25}$ values, giving the relation $\log (D_{25}\,h_{70} / \mathrm{kpc}) = 0.96 \log (a_{\mathrm{c}}\,h / \mathrm{kpc}) + 0.12$, where $a_{\mathrm{c}}$ is the corrected UGC B-band diameter. \citetalias{Haynes_1984} also use $h=1$ cosmology whereas we assume $h=0.7$.

The original relation from \citetalias{Haynes_1984} is
\begin{equation}
\log \frac{M_{\mathrm{HI}}\,h^{2}}{\mathrm{M_{\odot}}} = 0.88 \times 2\log \frac{a_{\mathrm{c}}\,h}{\mathrm{kpc}} + 7.12.
\end{equation}
After conversion to our unit system the gradient becomes $0.88/0.96$ and the intercept $7.12 - 2\log 0.7 - (2\times0.88\times0.12)/0.96$, making the final relation
\begin{equation}
\log \frac{M_{\mathrm{HI}}\,h_{70}^{2}}{\mathrm{M_{\odot}}} = 0.92 \times 2 \log \frac{D_{25}\,h_{70}}{\mathrm{kpc}} + 7.21.
\end{equation}

\citetalias{Haynes_1984} also fit a relation based on $L_{\mathrm{B}}$. Their magnitudes are taken from the UGC and then corrected internal extinction, Galactic extinction, the K-correction and some systematic errors in the magnitude system used at the time. We do not duplicate their exact corrections, but make similar ones of our own. Also when our raw magnitudes were compared to those from the UGC for the 323 objects that overlap with the CIG, we found excellent agreement. Therefore, we do not make any conversion between the two papers' magnitude systems. However, the difference in Hubble constant must be accounted for and \citetalias{Haynes_1984} used 5.37 as the bolometric absolute magnitude of the Sun, whereas we adopt the value 4.88. The original published relations was 
\begin{equation}
\log \frac{M_{\mathrm{HI}}\,h^{2}}{\mathrm{M_{\odot}}} = 0.66 \log \frac{L_{\mathrm{B}}\,h^{2}}{\mathrm{L_{\odot}}} + 2.94,
\end{equation}
which after conversion has the same gradient, but an intercept given by $2.94 - 2 \times (1-0.66) \log 0.7 - (0.4 \times 0.66)(4.88 - 5.37)$, making the final comparison relation
\begin{equation}
\log \frac{M_{\mathrm{HI}}\,h_{70}^{2}}{\mathrm{M_{\odot}}} = 0.66 \log \frac{L_{\mathrm{B}}\,h_{70}^{2}}{\mathrm{L_{\odot}}} + 3.17.
\end{equation}

\citet{Solanes_1996} fit a relation based on the optical size of field galaxies based on by eye measurements from POSS blue prints. They state that these are in good agreement with the UGC diameters and that no corrections were made. We therefore again compare the matched sources between the UGC and CIG, but now do not make any corrections to the diameters. This gives the relation $\log (a\,h / \mathrm{kpc}) = 0.87 \log (D_{25}\,h_{70} / \mathrm{kpc}) + 0.12$. The original relation from \citet{Solanes_1996} is
\begin{equation}
\log \frac{M_{\mathrm{HI}}\,h^{2}}{\mathrm{M_{\odot}}} = 0.73 \times 2\log \frac{a_{\mathrm{c}}\,h}{\mathrm{kpc}} + 7.51,
\end{equation}
which combined with the relation between the two diameter measurements gives the gradient as $0.73 \times 0.87$ and the intercept as $7.51 - 2 \log 0.7 + 2 \times 0.73 \times 0.12 $, making the final relation
\begin{equation}
\log \frac{M_{\mathrm{HI}}\,h_{70}^{2}}{\mathrm{M_{\odot}}} = 0.64 \times 2 \log \frac{D_{25}\,h_{70}}{\mathrm{kpc}} + 8.00.
\end{equation}

Finally, we considered the two relations of \citet{Denes_2014}, which are based on B-band magnitudes and diameters at the 25 mag arcsec$^{-2}$ (as in this work), and HI masses from HIPASS. No conversion to a different unit system is required in this case and the scaling with size can be used directly,
\begin{equation}
\log \frac{M_{\mathrm{HI}}\,h_{70}^{2}}{\mathrm{M_{\odot}}} = 0.64 \times 2 \log \frac{D_{25}\,h_{70}}{\mathrm{kpc}} + 8.21,
\end{equation}
while the B-band magnitude relation only requires conversion from a magnitude to a luminosity scale (using equation \ref{eqn:luminosity}). Which makes that relation
\begin{equation}
\log \frac{M_{\mathrm{HI}}\,h_{70}^{2}}{\mathrm{M_{\odot}}} = 0.85 \log \frac{L_{\mathrm{B}}\,h_{70}^{2}}{\mathrm{L_{\odot}}} + 1.23.
\end{equation}

\section{Scaling relations for SDSS-based isolation revision}
\label{sec:isolation_regression}

\citet{Argudo_Fernandez_2013} revised the AMIGA isolation criteria for those sources which fall within the SDSS footprint. Their photometric analysis produced a sample of well isolated galaxies that was 67\% of the input sample of CIG objects. This represented a reduction from the 83\% that \citet{Verley_2007b} found to be isolated, with the difference arising due to faint neighbours being identified in SDSS that were not apparent in POSS I \& II. However, there is considerable scatter in the estimates of neighbour density and tidal forces between the two works, therefore, a galaxy that is identified as isolated by the photometric criteria of \citet{Argudo_Fernandez_2013} may or may not be isolated according to the criteria of \citet{Verley_2007b}, although there is a definite correlation, and vice versa. When considering the SDSS spectroscopic sample \citet{Argudo_Fernandez_2013} found that 84\% of the sources with good spectroscopic coverage were found to be isolated when neighbours separated by more than 500 \kms \ from the central object were excluded, which may suggest, contrary to the photometric findings, that the full CIG sample could contain slightly more well isolated galaxies than found by \citet{Verley_2007b}. 

Regardless of which of the criteria from \citet{Argudo_Fernandez_2013} we choose to use it represents a major reduction in our sample size simply because their work is restricted to the SDSS footprint. This is the primary motivation for choosing the \citet{Verley_2007b} criteria.

When we do employ the SDSS-based photometric criteria it produces a sample of 309 isolated objects with HI observations, a decrease of about 45\% compared to using the \citet{Verley_2007b} criteria. This more restricted sample results in the relations $\log M_{\mathrm{HI}} = 0.79 \times 2 \log D_{25}/\mathrm{kpc} + 7.56$ and $\log M_{\mathrm{HI}} = 0.88 \log L_{\mathrm{B}}/\mathrm{L_{\odot}} + 0.90$, both with intrinsic scatter estimates of 0.17 dex. In the case of the SDSS spectroscopic isolation criteria, the sample available to us is only 243 objects. Here the relation fits become $\log M_{\mathrm{HI}} = 0.82 \times 2 \log D_{25}/\mathrm{kpc} + 7.44$ and $\log M_{\mathrm{HI}} = 0.97 \log L_{\mathrm{B}}/\mathrm{L_{\odot}} - 0.05$, with intrinsic scatters of 0.18 and 0.17, respectively. The parameter values and error estimates are given in Tables \ref{tab:SDSS_iso_D25_rel} \& \ref{tab:SDSS_iso_LB_rel}, and the relations are plotted in Figure \ref{fig:reg_coeff}.

It is clear from Figure \ref{fig:reg_coeff} that for both the $D_{25}$ and $L_{\mathrm{B}}$ relations that the relations based on the isolated SDSS spectroscopic sample are almost entirely consistent with the relations from the HI science sample. The relations from the isolated SDSS photometric sample appears to have a slightly flatter gradient and larger intercept, but agrees at high HI masses (for both relations). This might suggest that some of the smaller and less HI-rich AMIGA galaxies are not as well isolated when measured using SDSS rather than POSS I \& II. However, this is a marginal result, and in fact the $L_\mathrm{B}$-relations are consistent at 1-$\sigma$ for the lowest HI masses.

A point to note is that the intrinsic scatter estimates for these fits are smaller by about 0.05 dex which could indicate that the \citet{Argudo_Fernandez_2013} criteria have removed the non-isolated galaxies more reliably and thus reduced scatter due to contamination by non-isolated objects. However, considerable caution is required in interpreting this result because there are many possible explanations. For example, not only is the sample size smaller, which will reduce the accuracy of the intrinsic scatter estimate, but also the SDSS footprint restricts the available sky area and the CIG is a local Universe sample, therefore, it is possible that some of the intrinsic scatter is related to the larger scale environment that the isolated galaxies reside in, but the range of possible environments has been restricted by not looking at the full northern sky.

\begin{table*}
\centering
\caption{Regression fits between $2 \log D_{25}/\mathrm{kpc}$ and $\log M_{\mathrm{HI}}/\mathrm{M_{\odot}}$ for the SDSS-defined isolated sample}
\label{tab:SDSS_iso_D25_rel}
\begin{tabular}{c c c c c c}
\hline\hline
SDSS Isolation & Method & Sample     & Gradient & Intercept & Intrinsic Scatter (dex) \\ \hline
Spectroscopic     & MLE    & All        & 0.82 $\pm$ 0.06     & 7.44 $\pm$ 0.19      & 0.18 $\pm$ 0.02    \\ 
Spectroscopic     & OLS    & Detections & 0.77 $\pm$ 0.05     & 7.59 $\pm$ 0.14      & -    \\
Photometric       & MLE    & All        & 0.79 $\pm$ 0.05     & 7.56 $\pm$ 0.14      & 0.17 $\pm$ 0.02    \\ 
Photometric       & OLS    & Detections & 0.74 $\pm$ 0.04     & 7.69 $\pm$ 0.10      & -    \\
\hline
\end{tabular}
\end{table*}

\begin{table*}
\centering
\caption{Regression fits between $\log L_{\mathrm{B}}/\mathrm{L_{\odot}}$ and $\log M_{\mathrm{HI}}/\mathrm{M_{\odot}}$ for the SDSS-defined isolated sample}
\label{tab:SDSS_iso_LB_rel}
\begin{tabular}{c c c c c c}
\hline\hline
SDSS Isolation & Method & Sample     & Gradient & Intercept & Intrinsic Scatter (dex) \\ \hline
Spectroscopic     & MLE    & All        & 0.97 $\pm$ 0.18     &-0.05 $\pm$ 1.82      & 0.17 $\pm$ 0.07    \\ 
Spectroscopic     & OLS    & Detections & 0.73 $\pm$ 0.06     & 2.34 $\pm$ 0.60      & -    \\
Photometric       & MLE    & All        & 0.88 $\pm$ 0.10     & 0.90 $\pm$ 0.98      & 0.17 $\pm$ 0.03    \\ 
Photometric       & OLS    & Detections & 0.71 $\pm$ 0.05     & 2.67 $\pm$ 0.46      & -    \\
\hline
\end{tabular}
\end{table*}

\section{Isolation comparison with Solanes et al. 1996}
\label{Solanes_iso_comp}

\citet{Solanes_1996} compiled a sample of field galaxies from the CGCG \citep[Catalog of Galaxies and Galaxy Clusters,][]{Zwicky_1961} in the direction of Pisces-Perseus. In this paper we point out that this is a field sample which is quantitatively different from an isolated sample. To compare the degree of isolation we selected the $\sim$2500 most isolated galaxies (from other CGCG sources) in the CGCG in the same area as \citet{Solanes_1996}. These sources were then cross matched with SDSS DR9 \citep{Ahn_2012} photometric sources with clean photometry that were identified as galaxies. A positional match of less than 5 arcsec was required, and if there were multiple matches within that radius the CGCG source was discarded. Next, all neighbouring SDSS photometric galaxies that had Petrosian radii (in r-band) within a factor of 4 of the Petrosian radius of the central object were selected from DR9. Using these neighbouring sources the dimensionless local number density was calculated \citep[following][]{Verley_2007b} and the Karachentseva criteria were evaluated: a galaxy is considered isolated if there are no neighbours within an angular separation of 20 times their optical diameter, that have optical diameters within a factor of 4 of the central object. Originally the B-band diameter was used, however, for this approximate comparison we use the SDSS Petrosian diameters in r-band for simplicity. Due to the minimal SDSS DR9 coverage in this region, this process only returned usable results for 67 galaxies, therefore, we make our estimates based on this small sub-population. An equivalent exercise was performed with all of the CIG galaxies for which it was possible (451 sources). 

It was found that while ~20\% of the CIG objects meet these adapted Karachentseva criteria, none of the Solanes-like sample do. Furthermore, the average projected neighbour density for the Solanes-like sample was found to be more than double that of the CIG sample. We therefore conclude that these two samples cover different environments and that almost all of the \citet{Solanes_1996} galaxies would not be considered isolated by the AMIGA criteria, and thus are referred to a field objects rather than isolated objects.

\section{Impact of NRT and ERT flux discrepancies}
\label{sec:flux_disc}

\begin{figure*}
    \centering
    \includegraphics[width=\columnwidth]{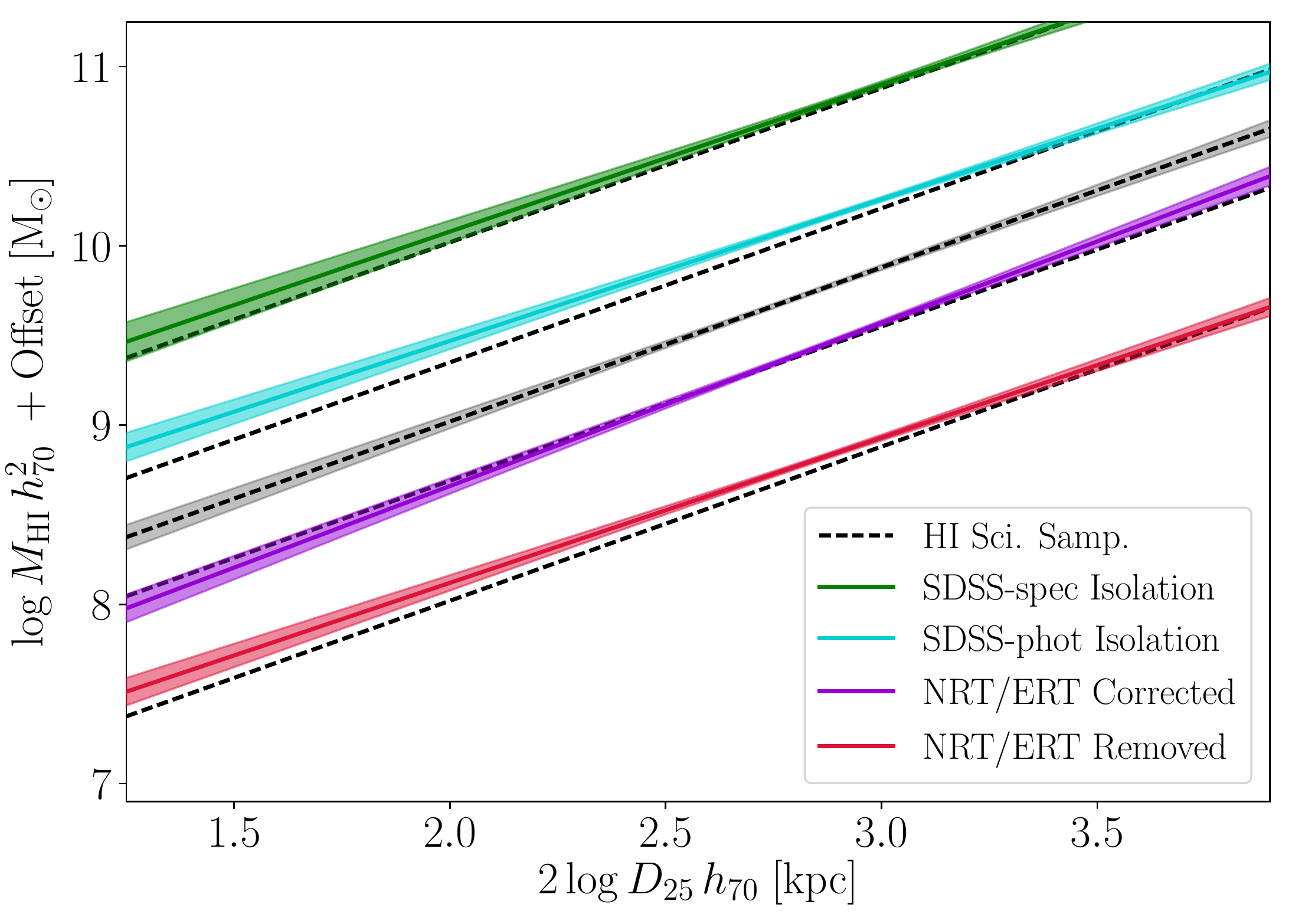}
    \includegraphics[width=\columnwidth]{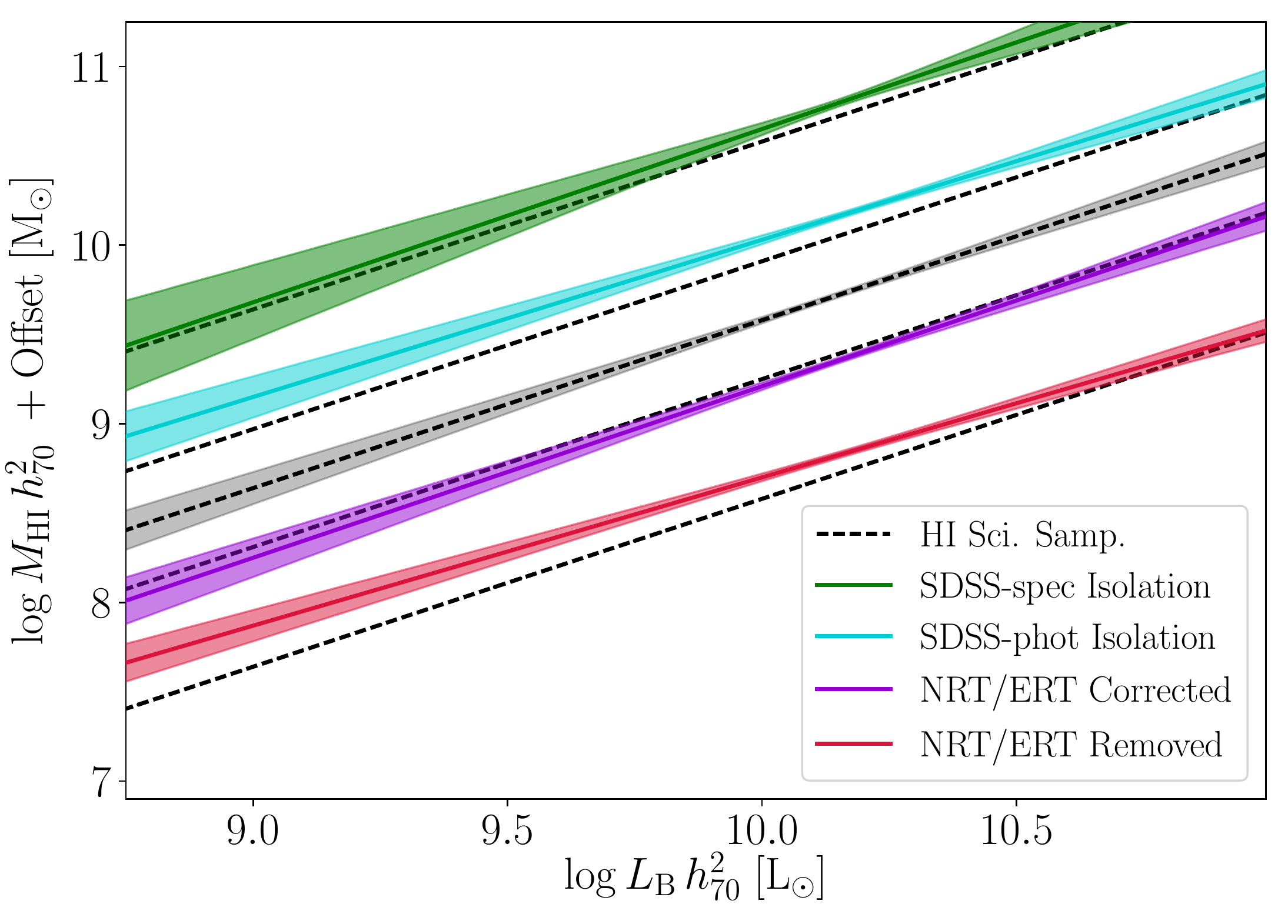}
    \caption{The $D_{25}$ (left) and $L_{\mathrm{B}}$ (right) relations along with 1-$\sigma$ uncertainties (shaded regions) for the HI science sample (dashed black line), the sample with SDSS spectroscopic isolation criteria (light blue), the sample with SDSS photometric isolation criteria (green), the HI science sample with corrected NRT and ERT fluxes (purple), and with NRT and ERT observations removed (red). Offsets have been applied to the relations (except the central relation, the HI science sample) to aid readability. The HI science sample relation is duplicated each time for comparison.}
    \label{fig:reg_coeff}
\end{figure*}

As detailed in Section \ref{sec:alfa_comp} the data observed with NRT and ERT appear to have a frequency dependent discrepancy in their absolute flux calibration. We were unable to determine the source of this difference and decided to use the data unaltered. Here we describe what impact that decision has on the final relations which we calculate.

A straight line was fit using weighted least squares regression between the redshift ($z_{50}$) and the logarithmic flux offset between the NRT and ERT observations and the ALFALFA fluxes for the same objects (considering only the uncertainty in the flux offset). This fit also included comparisons between our the NRT and ERT data and fluxes from \citet{Springob_2005} as this doubled the number of available sources for the fit, making the total 64. The \citet{Springob_2005} data were not used for comparisons when these were the same data as in our compilation. In this scenario OLS is a reasonable method as the uncertainties in $z_{50}$ are negligible in comparison to those in the flux. This fit was then used to alter all the NRT and ERT fluxes such that they fall in line with the ALFALFA and Springob et al. measurements. The resulting relations are listed in Tables \ref{tab:NRT_ERT_D25_rel} \& \ref{tab:NRT_ERT_LB_rel}, and plotted in Figure \ref{fig:reg_coeff}.

For both the $D_{25}$ and $L_{\mathrm{B}}$ relations, when the intercept and gradient are considered separately, they consistent (at 1-$\sigma$) with the relations that we calculated in Tables \ref{tab:D25_rel} \& \ref{tab:LB_rel}. However, in all cases (including the OLS fits) the relations with the correction applied to the NRT and ERT fluxes are marginally steeper. This suggests that by not applying a correction to the fluxes in the main part of the paper the gradients may have been flattened by approximately 5\%. Having said this, there is clearly substantial overlap between the 1-$\sigma$ uncertainties of the HI science sample relations with and without the flux corrections (Figure \ref{fig:reg_coeff}). In addition, the correction is both entirely empirical and quite uncertain. Therefore, we recommend using the relation given in Tables \ref{tab:D25_rel} \& \ref{tab:LB_rel}, not those in this Section.

The other approach is to simply remove the spectra observed with NRT and ERT. This has the advantage of not necessitating the calculation of an uncertain and empirical relation to make a correction, however, it also removes almost half of the dataset. The relations calculated for this reduced dataset are also shown in Tables \ref{tab:NRT_ERT_D25_rel} \& \ref{tab:NRT_ERT_LB_rel}, and Figure \ref{fig:reg_coeff}. 

In both relations removing the NRT and ERT sources caused a marginal flattening of the slope (and increase in the intercept), as opposed to the steepening found when the flux correction was applied. We suspect that, at least in part, this is caused by the different morphological distribution of the sources observed with NRT and ERT relative to the full population. For the detections approximately half of the sources that are S0 or earlier types were removed, whereas less than a third of sources later than S0 were removed. Hence, the remaining sample is richer in late types than the original sample. As the relations are found to be flatter for later types (which are also more HI-rich), this must contribute to the resulting flattening (and increase in the intercept). In addition, all but 17 of the marginal and non-detections were observed with NRT or ERT. These are also all removed, but this likely has minimal effects on the relations because the upper limits do not contain a lot of information.

\begin{table*}
\centering
\caption{Regression fits between $2 \log D_{25}/\mathrm{kpc}$ and $\log M_{\mathrm{HI}}/\mathrm{M_{\odot}}$ with NRT and ERT fluxes corrected or removed}
\label{tab:NRT_ERT_D25_rel}
\begin{tabular}{c c c c c c}
\hline\hline
NRT \& ERT Fluxes & Method & Sample     & Gradient & Intercept & Intrinsic Scatter (dex) \\ \hline
Corrected         & MLE    & All        & 0.91 $\pm$ 0.05     & 7.17 $\pm$ 0.14      & 0.23 $\pm$ 0.02    \\ 
Corrected         & OLS    & Detections & 0.81 $\pm$ 0.04     & 7.45 $\pm$ 0.11      & -    \\
Removed           & MLE    & All        & 0.81 $\pm$ 0.05     & 7.50 $\pm$ 0.13      & 0.21 $\pm$ 0.02    \\ 
Removed           & OLS    & Detections & 0.73 $\pm$ 0.04     & 7.71 $\pm$ 0.11      & -    \\
\hline
\end{tabular}
\end{table*}

\begin{table*}
\centering
\caption{Regression fits between $\log L_{\mathrm{B}}/\mathrm{L_{\odot}}$ and $\log M_{\mathrm{HI}}/\mathrm{M_{\odot}}$ with NRT and ERT fluxes corrected or removed}
\label{tab:NRT_ERT_LB_rel}
\begin{tabular}{c c c c c c}
\hline\hline
NRT \& ERT Fluxes & Method & Sample     & Gradient & Intercept & Intrinsic Scatter (dex) \\ \hline
Corrected         & MLE    & All        & 0.96 $\pm$ 0.09     &-0.06 $\pm$ 0.94      & 0.23 $\pm$ 0.03    \\ 
Corrected         & OLS    & Detections & 0.77 $\pm$ 0.04     & 1.86 $\pm$ 0.43      & -    \\
Removed           & MLE    & All        & 0.83 $\pm$ 0.07     & 1.40 $\pm$ 0.75      & 0.22 $\pm$ 0.02    \\ 
Removed           & OLS    & Detections & 0.67 $\pm$ 0.04     & 2.92 $\pm$ 0.44      & -    \\
\hline
\end{tabular}
\end{table*}

\end{document}